\documentclass[showpacs,twocolumn,floatfix]{revtex4}

\usepackage{amsmath,amssymb,mathrsfs}
\usepackage{latexsym}
\usepackage{graphicx,psfrag} 
\usepackage[dvips]{color}

\newcommand{\bs}[1]{\boldsymbol{#1}}
\newcommand{\nn}{\nonumber}
\newcommand{\mb}[1]{\mathbf{#1}}
\newcommand{\half}{$\frac{1}{2}$ }
\newcommand{\comm}[2]{\left[#1,#2\right]}

\newcommand{\vac}{\left|\,0\,\right\rangle}
\newcommand{\ket}[1]{\left|#1\right\rangle}

\newcommand{\up}{\uparrow}
\newcommand{\dw}{\downarrow}

\newcommand{\casi}[2]{\mathcal{C}^2_{\text{SU(3)}}(#1,#2)}
\newcommand{\casin}[1]{\mathcal{C}^2_{\text{SU($n$)}}(#1)}
\newcommand{\casinn}[2]{\mathcal{C}^2_{\text{SU($#1$)}}(#2)}
\newcommand{\dimn}[2]{\text{dim}_{\text{SU($#1$)}}(#2)}

\def\ie{\emph{i.e.},\ }
\def\eg{\emph{e.g.}\ }
\def\ea{\emph{et al.}}

\def\b{\text{b}}
\def\r{\text{r}}
\def\g{\text{g}}
\def\y{\text{y}}
\def\c{\text{c}}
\def\m{\text{m}}

\allowdisplaybreaks[1]\begin{document}
\title{Valence bond solids for SU($\bs{n}$) spin chains:\\
exact models, spinon confinement, and the Haldane gap}
\author{Martin Greiter and Stephan Rachel}
\affiliation{Institut f\"ur Theorie der Kondensierten Materie,\\
  Universit\"at Karlsruhe, Postfach 6980, 76128 Karlsruhe, Germany}
\pagestyle{plain}

\begin{abstract}
  To begin with, we introduce several exact models for SU(3) spin
  chains: (1) a translationally invariant parent Hamiltonian involving
  four-site interactions for the trimer chain, with a three-fold
  degenerate ground state.  We provide numerical evidence that the
  elementary excitations of this model transform under representation
  $\bs{\bar 3}$ of SU(3) if the original spins of the model transform
  under rep.\ $\bs{3}$.  (2) a family of parent Hamiltonians for
  valence bond solids of SU(3) chains with spin reps.\ $\bs{6}$,
  $\bs{10}$, and $\bs{8}$ on each lattice site.  We argue that of
  these three models, only the latter two exhibit spinon confinement
  and hence a Haldane gap in the excitation spectrum.  We generalize
  some of our models to SU($n$).  Finally, we use the emerging rules
  for the construction of VBS states to argue that models of
  antiferromagnetic chains of SU($n$) spins in general possess a
  Haldane gap if the spins transform under a representation
  corresponding to a Young tableau consisting of a number of boxes
  $\lambda$ which is divisible by $n$.  If $\lambda$ and $n$ have no
  common divisor, the spin chain will support deconfined spinons and
  not exhibit a Haldane gap.  If $\lambda$ and $n$ have a common
  divisor different from $n$, it will depend on the specifics of the
  model including the range of the interaction.
\end{abstract}



\maketitle

\section{Introduction}

Quantum spin chains have been a most rewarding subject of study almost
since the early days of quantum mechanics, beginning with the
invention of the Bethe ansatz in 1931~\cite{bethe31zp205} as a method
to solve the $S=\frac{1}{2}$ Heisenberg chain with nearest-neighbor
interactions.  The method led to the discovery of the Yang--Baxter
equation in 1967~\cite{yang67prl1312, baxter90}, and provides the
foundation for the field of integrable
models~\cite{KorepinBogoliubovIzergin93}.  Faddeev and
Takhtajan~\cite{faddeev-81pla375} discovered in 1981 that the
elementary excitations (now called spinons) of the spin-$1/2$
Heisenberg chain carry spin $1/2$ while the Hilbert space is spanned
by spin flips, which carry spin 1.  The fractional quantization of
spin in spin chains is conceptually similar to the fractional
quantization of charge in quantized Hall
liquids~\cite{laughlin83prl1395, stone92}.  In 1982,
Haldane~\cite{haldane83pla464, haldane83prl1153} identified the O(3)
nonlinear sigma model as the effective low-energy field theory of
SU(2) spin chains, and argued that chains with integer spin possess a
gap in the excitation spectrum, while a topological term renders
half-integer spin chains gapless~\cite{affleck90proc,Fradkin91}.

The general methods---the Bethe ansatz method and the use of effective
field theories including
bosonization~\cite{GogolinNersesyanTsvelik98,Giamarchi04}---are
complemented by a number of exactly solvable models, most prominently
among them the Majumdar--Ghosh (MG) Hamiltonian for the
$S=\frac{1}{2}$ dimer chain~\cite{majumdar-69jmp1399}, the AKLT model
as a paradigm of the gapped $S=1$ chain~\cite{affleck-87prl799,
  affleck-88cmp477}, and the Haldane--Shastry model
(HSM)~\cite{haldane88prl635, shastry88prl639, haldane91prl1529,
  haldane-92prl2021}.  The HSM is by far the most sophisticated among
these three, as it is not only solvable for the ground state, but
fully integrable due to its Yangian symmetry~\cite{haldane-92prl2021}.
The wave functions for the ground state and single-spinon excitations
are of a simple Jastrow form, elevating the conceptual similarity to
quantized Hall states to a formal equivalence.  Another unique feature
of the HSM is that the spinons are free in the sense that they only
interact through their half-Fermi statistics~\cite{haldane91prl937,
  ha-93prb12459, essler95prb13357, greiter-05prb224424,greiter06prl},
which renders the model an ideal starting point for any perturbative
description of spin systems in terms of interacting
spinons~\cite{greiter-06prl}.  The HSM has been generalized from SU(2)
to SU($n$)~\cite{kawakami92prb1005, kawakami92prbr3191, ha-92prb9359, 
  bouwknegt-96npb345, schuricht-05epl987, schuricht-06prb235105}.

For the MG and the AKLT model, only the ground states are known
exactly.  Nonetheless, these models have amply contributed to our
understanding of many aspects of spin chains, each of them through the
specific concepts captured in its ground state~\cite{
  jullien-83baps344, affleck89jpcm3047, okamoto-92pla433, eggert96prb9612, 
  white-96prb9862,sen-07prb104411,
  knabe88jsp627,fannes-89epl633,
  freitag-91zpb381,
  kluemper-91jpal955,
  kennedy-92prb304,
  kluemper-92zpb281,
  kluemper-93epl293,batchelor-94ijmpb3645,
  schollwock-96prb3304, kolezhuk-96prl5142, kolezhuk-02prb100401,
  normand-02prb104411,lauchli-06prb144426}.
The models are specific to SU(2) spin chains.  We will review both 
models below.

In the past, the motivation to study SU($n$) spin systems with $n>2$
has been mainly formal.  The Bethe ansatz method has been generalized
to multiple component systems by
Sutherland~\cite{sutherland75prb3795}, yielding the so-called nested
Bethe ansatz. In particular, this has led to a deeper understanding
of quantum integrability and the applicability of the Bethe
ansatz~\cite{choi-82pla83}.  Furthermore, the nested Bethe ansatz was
used to study the spectrum of the SU($n$)
HSM~\cite{kawakami92prb1005,ha-93prb12459}.  It has also been applied
to SU(2) spin chains with orbital degeneracy at the SU(4) symmetric
point~\cite{Li-99prb12781,Gu-02prb092404}.
Most recently, Damerau and Kl\"umper obtained highly accurate
numerical results for the thermodynamic properties of the SU(4)
spin--orbital model~\cite{damerau-06jsm12014}.
SU($n$) Heisenberg models have been studied recently by Kawashima and
Tanabe~\cite{kawashima-07prl057202} with quantum Monte Carlo, and by
Paramekanti and Marston~\cite{Paramekanti-06cm0608691} using
variational wave functions.

The effective field theory description of SU(2) spin chains by Haldane
yielding the distinction between gapless half-integer spin chains with
deconfined spinons and gapped integer spin chains with confined
spinons cannot be directly generalized to SU($n$), as there is no
direct equivalent of the CP$^1$ representation used in Haldane's
analysis.
The critical behavior of SU($n$) spin chains, however, has been
analyzed by Affleck in the framework of effective field
theories~\cite{affleck86npb409, affleck88npb582}.

An experimental realization of an SU(3) spin system, and in particular
an antiferromagnetic SU(3) spin chain, however, might be possible in
an optical lattice of ultracold atoms in the not-too-distant future.
The ``spin'' in these systems would of course not relate to the
rotation group of our physical space, but rather relate to SU(3)
rotations in an internal space spanned by three degenerate ``colors''
the atom may assume, subject to the requirement that the number of
atoms of each color is conserved.  A possible way to realize such a
system experimentally is described in Appendix~\ref{sec:exp}.
Moreover, it has been suggested recently that an SU(3) trimer state
might be realized approximately in a spin tetrahedron
chain~\cite{chen-05prb214428,chen-06prb174424}.

Motivated by both this prospect as well as the mathematical challenges
inherent to the problem, we propose several exact models for SU(3)
spin chains in this article.  The models are similar in spirit to the
MG or the AKLT model for SU(2), and consist of parent Hamiltonians and
their exact ground states.  There is no reason to expect any of these
models to be integrable, and none of the excited states are known
exactly.  We generalize several of our models to SU($n$), and use the
emerging rules to investigate and motivate which SU($n$) spin chains
exhibit spinon confinement and a Haldane gap.

The article is organized as follows.  Following a brief review of the
MG model in Sec.~\ref{sec:mg}, we introduce the trimer model for SU(3)
spin chains in Sec.~\ref{sec:trimer}.  This model consists of a
translationally invariant Hamiltonian involving four-site
interactions, with a three-fold degenerate ground state, in which
triples of neighboring sites form SU(3) singlets (or trimers).  In
Sec.~\ref{sec:su3rep}, we review the representations of SU(3), which
we use to verify the trimer model in Sec.~\ref{sec:trimercont}.  In
this section, we further provide numerical evidence that the
elementary excitations of this model transform under representation
$\bs{\bar 3}$ of SU(3) if the original spins of the model transform
under rep.\ $\bs{3}$.  We proceed by introducing Schwinger bosons in
Sec.~\ref{sec:schwinger} and a review of the AKLT model in
Sec.~\ref{sec:aklt}.  In Sec.~\ref{sec:vbs}, we formulate a family of
parent Hamiltonians for valence bond solids of SU(3) chains with spin
reps.\ $\bs{6}$, $\bs{10}$, and $\bs{8}$ on each lattice site, and
proof their validity.  We argue that only the rep.\ $\bs{10}$ and the
rep.\ $\bs{8}$ model, which are in a wider sense generalizations of
the AKLT model to SU(3), exhibit spinon confinement and hence a
Haldane-type gap in the excitation spectrum.  In Sec.~\ref{sec:sun},
we generalize three of our models from SU(3) to SU($n$).
In Sec.~\ref{sec:conf}, we use the rules emerging from the numerous
VBS models we studied to investigate which models of SU($n$) spin
chains in general exhibit spinon confinement and a Haldane gap.  In
this context, we first review a rigorous theorem due to Affleck and
Lieb~\cite{affleck-86lmp57} in Sec.\ \ref{sec:afflecklieb}.  In Sec.\
\ref{sec:generalcriterion}, we argue that the spinons in SU($n$) spin
chains with spins transforming under reps.\ with Young tableaux
consisting of a number of boxes $\lambda$ which is divisible by $n$
are always confined.  In Sec.\ \ref{sec:examples}, we construct
several specific examples to argue that if $\lambda$ and $n$ have a
common divisor different from $n$, the model will be confining only if
the interactions are sufficiently long ranged.  Specifically, the
models we study suggest that if $q$ is the largest common divisor of
$\lambda$ and $n$, the model will exhibit spinon confinement only if
the interactions extends at least to the $\frac{n}{q}$-th neighbor on
the chain.  If $\lambda$ and $n$ have no common divisor, the spinons
will be free and chain will not exhibit a Haldane gap.  We briefly
summarize the different categories of models in Sec.\ \ref{sec:sum},
and present a counter-example to the general rules in Sec.\
\ref{sec:counter}.  We conclude with a brief summary of the results 
obtained in this article in 
Sec.\ \ref{sec:conclusion}.

A brief and concise account of the SU(3) VBS models we elaborate here 
has been given previously~\cite{greiter-07prb060401}.

\section{The Majumdar--Ghosh model}
\label{sec:mg}

Majumdar and Ghosh~\cite{majumdar-69jmp1399} noticed in 1969 that on a
linear spin $S=\frac{1}{2}$ chain with an even number of sites, the
two valence bond solid or dimer states
\begin{eqnarray}
\big|\psi_{\text{MG}}^{\textrm{even}\rule{0pt}{5pt}\atop 
\textrm{(odd)}}\big\rangle 
&=&
\prod_{i\ \textrm{even}\atop (i\ \textrm{odd})}  
\Bigl(c^\dagger_{i\up} c^\dagger_{i+1\dw} - c^\dagger_{i\dw} c^\dagger_{i+1\up}
\Bigr) \vac =\nonumber \\
&=&\left\{ \begin{array}{lc}
\ket{\hbox{\begin{picture}(102,8)(-4,-3)
\put(13,0){\makebox(0,0){\rule{10.pt}{ 0.8pt}}}
\put(41,0){\makebox(0,0){\rule{10.pt}{ 0.8pt}}}
\put(69,0){\makebox(0,0){\rule{10.pt}{ 0.8pt}}}
\put(6,0){\circle{4}}
\put(20,0){\circle{4}}
\put(34,0){\circle{4}}
\put(48,0){\circle{4}}
\put(62,0){\circle{4}}
\put(76,0){\circle{4}}
\put(90,0){\circle{4}}
\end{picture}}} 
&\quad\textrm{``even''}\quad\\
\ket{\hbox{\begin{picture}(102,8)(-4,-3)
\put(27,0){\makebox(0,0){\rule{10.pt}{ 0.8pt}}}
\put(55,0){\makebox(0,0){\rule{10.pt}{ 0.8pt}}}
\put(83,0){\makebox(0,0){\rule{10.pt}{ 0.8pt}}}
\put(6,0){\circle{4}}
\put(20,0){\circle{4}}
\put(34,0){\circle{4}}
\put(48,0){\circle{4}}
\put(62,0){\circle{4}}
\put(76,0){\circle{4}}
\put(90,0){\circle{4}}
\end{picture}}} 
&\quad\textrm{``odd''}\quad\rule{0pt}{15pt}
\end{array}\right.\rule{0pt}{25pt}
\label{eq:psimg}
\end{eqnarray}
where the product runs over all even sites $i$ for one state and over
all odd sites for the other, are exact zero-energy ground
states~\cite{broek80pla261} of the parent Hamiltonian
\begin{equation}
H_{\text{MG}} = 
\sum_i \left(\boldsymbol{S}_i \boldsymbol{S}_{i+1} +
\frac{1}{2}\boldsymbol{S}_i \boldsymbol{S}_{i+2} +\frac{3}{8}\right),
 \label{eq:hmg}
\end{equation}
where
\begin{equation}
\boldsymbol{S}_i=\frac{1}{2}\,\sum_{\tau,\tau'=\up,\dw}\,
c_{i\tau}^{\dagger} \boldsymbol{\sigma}_{\tau\tau'}^{\phantom{\dagger}}
c_{i\tau'}^{\phantom{\dagger}}, 
\label{eq:s}
\end{equation}
and $\boldsymbol{\sigma}=(\sigma_x,\sigma_y,\sigma_z)$ is the vector
consisting of the three Pauli matrices.

The proof is exceedingly simple.  We rewrite
\begin{equation}
H_{\textrm{MG}}=\frac{1}{4}\sum_i H_i,\ \
H_i=\bigl(\boldsymbol{S}_i +\boldsymbol{S}_{i+1} 
+\boldsymbol{S}_{i+2}\bigr)^2 -\frac{3}{4}.
\end{equation}
Clearly, any state in which the total spin of three neighboring spins
is $\frac{1}{2}$ will be annihilated by $H_i$.  (The total spin can
only be $\frac{1}{2}$ or $\frac{3}{2}$, as
$\bf\frac{1}{2}\otimes\frac{1}{2}\otimes\frac{1}{2}=
\frac{1}{2}\oplus\frac{1}{2}\oplus\frac{3}{2}$.)  In the dimer states
above, this is always the case as two of the three neighboring spins
are in a singlet configuration, and 
$\bf 0\otimes\frac{1}{2}=\frac{1}{2}$.  Graphically, we may express
this as
\begin{equation}
H_i \ket{\hbox{\begin{picture}(40,8)(0,-3)
\put(13,0){\makebox(0,0){\rule{10.pt}{ 0.8pt}}}
\put(6,0){\circle{4}}
\put(20,0){\circle{4}}
\put(34,0){\circle{4}}
\end{picture}}}
= H_i \ket{\hbox{\begin{picture}(40,8)(0,-3)
\put(27,0){\makebox(0,0){\rule{10.pt}{ 0.8pt}}}
\put(6,0){\circle{4}}
\put(20,0){\circle{4}}
\put(34,0){\circle{4}}
\end{picture}}}=0.
\end{equation}
 As $H_i$ is positive definite, the two zero-energy eigenstates of
$H_{\textrm{MG}}$ are also ground states. 

Is the Majumdar--Ghosh or dimer state in the universality class
generic to one-dimensional spin-\half liquids, and hence a
useful paradigm to understand, say, the nearest-neighbor Heisenberg
chain?  The answer is clearly no, as the dimer states (\ref{eq:psimg})
violate translational symmetry modulo translations by two lattice
spacings, while the generic liquid is invariant.  

Nonetheless, the dimer chain shares some important properties of this
generic liquid.  First, the spinon excitations---here domain walls
between ``even'' and ``odd'' ground states---are deconfined.  (To
construct approximate eigenstates of
$H_{\textrm{MG}}$, we take momentum superpositions
of localized domain walls.)  Second, there are (modulo the overall
two-fold degeneracy) only $M+1$ orbitals available for an individual
spinon if $2M$ spins are condensed into dimers or valence bond
singlets.  This is to say, if there are only a few spinons in a long
chain, the number of orbitals available to them is roughly half the
number of sites.  This can easily be seen graphically:
\begin{center}
\begin{picture}(222,25)(-26,-17)
\put(-12,0){\makebox(0,0)[r]{\rule{10.pt}{ 0.8pt}}}
\put(13,0){\makebox(0,0){\rule{10.pt}{ 0.8pt}}}
\put(41,0){\makebox(0,0){\rule{10.pt}{ 0.8pt}}}
\put(83,0){\makebox(0,0){\rule{10.pt}{ 0.8pt}}}
\put(111,0){\makebox(0,0){\rule{10.pt}{ 0.8pt}}}
\put(139,0){\makebox(0,0){\rule{10.pt}{ 0.8pt}}}
\put(181,0){\makebox(0,0){\rule{10.pt}{ 0.8pt}}}
\put(-24,0){\circle{4}}
\put(-10,0){\circle{4}}
\put(6,0){\circle{4}}
\put(20,0){\circle{4}}
\put(34,0){\circle{4}}
\put(48,0){\circle{4}}
\put(62,0){\circle{4}}
\put(76,0){\circle{4}}
\put(90,0){\circle{4}}
\put(104,0){\circle{4}}
\put(118,0){\circle{4}}
\put(132,0){\circle{4}}
\put(146,0){\circle{4}}
\put(160,0){\circle{4}}
\put(174,0){\circle{4}}
\put(188,0){\circle{4}}
\put(62,1){\makebox(0,0){\vector(0,1){14}}}
\put(160,1){\makebox(0,0){\vector(0,1){14}}}
\put(14,-12){\makebox(0,0){\small even}}
\put(111,-12){\makebox(0,0){\small odd}}
\end{picture}
\end{center}
If we start with an even ground state on the left, the spinon to its
right must occupy an even lattice site and vice versa.  The resulting
state counting is precisely what one finds in the Haldane--Shastry
model, where it is directly linked to the half-Fermi statistics of the
spinons~\cite{haldane91prl937}.

The dimer chain is further meaningful as a piece of a general
paradigm.  The two degenerate dimer states (\ref{eq:psimg}) can be
combined into an $S=1$ chain, the AKLT chain, which serves as a
generic paradigm for $S=1$ chains which exhibit the Haldane
gap~\cite{haldane83pla464, haldane83prl1153, affleck89jpcm3047},
and provides the intellectual background for several of the exact models
we introduce further below.  Before doing so, however, we will now
introduce the trimer model, which constitutes an SU(3) analog of the MG 
model.

\section{The trimer model}
\label{sec:trimer}

\subsection{The Hamiltonian and its ground states} 
\label{sec:trimerstated}

Consider a chain with $N$ lattice sites, where $N$ has to be
divisible by three, and periodic boundary conditions (PBCs). On each
lattice site we place an SU(3) spin which transforms under the
fundamental representation $\bs{3}$, \ie the spin can take
the values (or colors) blue (b), red (r), or green (g). The trimer
states are obtained by requiring the spins on each three neighboring
sites to form an SU(3) singlet, which we call a trimer and sketch it
by
\begin{picture}(38,8)(1,-2)\linethickness{0.8pt}
\put(6,0){\circle{4}}\put(8,0){\line(1,0){10}}\put(20,0){\circle{4}}
\put(22,0){\line(1,0){10}}\put(34,0){\circle{4}}
\end{picture}.
The three 
linearly independent trimer 
states on the chain are given by
\begin{eqnarray}\label{trimerstates.3}
\ket{\psi_{\text{trimer}}^{(\mu)}}
\!&\!=\!&\!\left\{
\begin{array}{l}
|\hbox{
\setlength{\unitlength}{1pt}
\begin{picture}(80,8)(4,-2)\linethickness{0.8pt}
\put(6,0){\circle{4}}\put(8,0){\line(1,0){10}}\put(20,0){\circle{4}}
\put(22,0){\line(1,0){10}}\put(34,0){\circle{4}}\put(48,0){\circle{4}}
\put(50,0){\line(1,0){10}}\put(62,0){\circle{4}}\put(64,0){\line(1,0){10}}
\put(76,0){\circle{4}}
\end{picture}
}\rangle\,\equiv\,\ket{\psi_{\text{trimer}}^{(1)}},\\[3mm]
|\hbox{
\setlength{\unitlength}{1pt}
\begin{picture}(80,8)(4,-2)\linethickness{0.8pt}
\put(6,0){\circle{4}}\put(20,0){\circle{4}}
\put(22,0){\line(1,0){10}}\put(34,0){\circle{4}}\put(36,0){\line(1,0){10}}
\put(48,0){\circle{4}}\put(62,0){\circle{4}}\put(64,0){\line(1,0){10}}
\put(76,0){\circle{4}}
\end{picture}
}\rangle\,\equiv\,\ket{\psi_{\text{trimer}}^{(2)}},\\[3mm]
|\hbox{
\setlength{\unitlength}{1pt}
\begin{picture}(80,8)(4,-2)\linethickness{0.8pt}
\put(6,0){\circle{4}}\put(8,0){\line(1,0){10}}\put(20,0){\circle{4}}
\put(34,0){\circle{4}}\put(36,0){\line(1,0){10}}\put(48,0){\circle{4}}
\put(50,0){\line(1,0){10}}\put(62,0){\circle{4}}\put(76,0){\circle{4}}
\end{picture}
}\rangle\,\equiv\,\ket{\psi_{\text{trimer}}^{(3)}}\,.
\end{array}\right.\\[-2mm]\nn
\end{eqnarray}
Introducing operators $c_{i\sigma}^{\dagger}$ which create a fermion
of color $\sigma$ ($\sigma=\b,\r,\g$) at lattice site $i$, the trimer
states can be written as
\begin{equation}
  \label{eq:trimer}
  \ket{\psi_{\text{trimer}}^{(\mu)}\!}=\hspace{-15pt}\prod_{\scriptstyle{i} \atop 
    \left(\scriptstyle{\frac{i-\mu}{3}\,{\rm integer}}\right)}
  \hspace{-5pt}\Bigl(
  \sum_{\scriptstyle{(\alpha,\beta,\gamma)}=\atop\scriptstyle{{\pi}(\b,\r,\g)}} 
  \hspace{-5pt}\hbox{sign}({\pi})\,
  c^{\dagger}_{i\,\alpha}\,c^{\dagger}_{i+1\,\beta}\,c^{\dagger}_{i+2\,\gamma}
  \Bigr)\!\vac\!,
\end{equation}
where $\mu=1,2,3$ labels the three degenerate ground states, and $i$ runs 
over the lattice sites subject to the constraint that $\frac{i-\mu}{3}$
is integer.  The sum extends over all six permutations $\pi$ of 
the three colors b, r, and g, \ie
\begin{equation}
  \label{eq:permsum}
  \begin{split}  
  &\sum_{\scriptstyle{(\alpha,\beta,\gamma)}=\scriptstyle{{\pi}(\b,\r,\g)}} 
  \hspace{-5pt}\hbox{sign}({\pi})\,
  c^{\dagger}_{i\,\alpha}\,c^{\dagger}_{i+1\,\beta}\,c^{\dagger}_{i+2\,\gamma}\\
  &\hspace{18pt}=c^{\dagger}_{i\,\b}c^{\dagger}_{i+1\,\r}c^{\dagger}_{i+2\,\g}
   +c^{\dagger}_{i\,\r}c^{\dagger}_{i+1\,\g}c^{\dagger}_{i+2\,\b}
   +c^{\dagger}_{i\,\g}c^{\dagger}_{i+1\,\b}c^{\dagger}_{i+2\,\r}\\
  &\hspace{18pt}-c^{\dagger}_{i\,\b}c^{\dagger}_{i+1\,\g}c^{\dagger}_{i+2\,\r}
   -c^{\dagger}_{i\,\g}c^{\dagger}_{i+1\,\r}c^{\dagger}_{i+2\,\b}
   -c^{\dagger}_{i\,\r}c^{\dagger}_{i+1\,\b}c^{\dagger}_{i+2\,\g}.\hspace{5pt}
\end{split}
\end{equation}

The SU(3) generators at each lattice site $i$ are in analogy to
\eqref{eq:s} defined as
\begin{equation}
J^a_i=\frac{1}{2}\,\sum_{\sigma,\sigma'=\b,\r,\g}\,
c_{i\sigma}^{\dagger} \lambda^a_{\sigma\sigma'}
c_{i\sigma'}^{\phantom{\dagger}}, \quad a=1,\ldots,8,
\label{eq:J_a}
\end{equation}
where the $\lambda^a$ are the Gell-Mann matrices (see
App.~\ref{app:conventions}).  The operators \eqref{eq:J_a} satisfy the
commutation relations
\begin{equation}
\comm{J^a_{i}}{J^b_{j}}=\delta_{ij}\;f^{abc}J^c_{i},
\quad a,b,c=1,\ldots,8,
\label{eq:su3-spincommforsites}
\end{equation}
(we use the Einstein summation convention) with $f^{abc}$ the
structure constants of SU(3) (see App.~\ref{app:conventions}).  We
further introduce the total SU(3) spin of $\nu$ neighboring sites
$i,\ldots,i+\nu-1$,
\begin{equation}\label{eq:defJnu}
\bs{J}^{(\nu)}_i = \sum_{j=i}^{i+\nu-1}\,\bs{J}_{j},
\end{equation}
where $\bs{J}_i$ is the eight-dimensional vector formed by its
components \eqref{eq:J_a}.  The parent Hamiltonian for the trimer
states \eqref{eq:trimer} is given by
\begin{equation}
  \label{ham.trimer}
  H_{\textrm{trimer}}\, =\, \sum_{i=1}^N
  \left(\,\left(\bs{J}^{(4)}_i\right)^{\!4}\,
    -\,\frac{14}{3}\left(\bs{J}^{(4)}_i\right)^{\!2}\,+
    \,\frac{40}{9}\,\right).
\end{equation}
The $\bs{J}_i\bs{J}_j$ terms appear complicated in terms
of Gell-Mann matrices, but are rather simply when written out
using the operator $P_{ij}$, which permutes the SU($n$) spins
(here $n=3$) on sites $i$ and $j$:
\begin{equation}
  \label{eq:permut}
  \bs{J}_i\bs{J}_j = \frac{1}{2}\left( P_{ij} - \frac{1}{n} \right).
\end{equation}

To verify the trimer Hamiltonian \eqref{ham.trimer}, as well as for
the valence bond solid (VBS) models we propose below, we will need a
few higher-dimensional representations of SU(3).  We review these in
the following section.

\section{Representations of SU(3)}
\label{sec:su3rep}

\subsection{Young tableaux and representations of SU(2) }

\begin{figure}[tb]
\setlength{\unitlength}{8pt}
\begin{picture}(28,10)(0,0)
\linethickness{0.3pt}
\multiput(0.5,9)(1,0){2}{\line(0,1){1}}
\multiput(0.5,9)(0,1){2}{\line(1,0){1}}
\put(0.5,9){\makebox(1,1){1}}
\put(1,1){\line(0,1){4}}
\multiput(1,2)(0,2){2}{\circle*{0.5}}
\put(0,5){\makebox(1,1){$S^z$}}
\put(1.5,3.5){\makebox(2,1){$\ket{\up}$}}
\put(1.5,1.5){\makebox(2,1){$\ket{\dw}$}}

\put(3.5,9){\makebox(1,1){$\otimes$}}

\multiput(6.5,9)(1,0){2}{\line(0,1){1}}
\multiput(6.5,9)(0,1){2}{\line(1,0){1}}
\put(6.5,9){\makebox(1,1){2}}
\put(7,1){\line(0,1){4}}
\multiput(7,2)(0,2){2}{\circle*{0.5}}
\put(6,5){\makebox(1,1){$S^z$}}
\put(7.5,3.5){\makebox(2,1){$\ket{\up}$}}
\put(7.5,1.5){\makebox(2,1){$\ket{\dw}$}}

\put(9,9){\makebox(1,1){=}}

\multiput(11.5,8)(1,0){2}{\line(0,1){2}}
\multiput(11.5,8)(0,1){3}{\line(1,0){1}}
\put(11.5,9){\makebox(1,1){1}}
\put(11.5,8){\makebox(1,1){2}}
\put(12,2){\line(0,1){2}}
\put(12,3){\circle*{0.5}}
\put(11,4){\makebox(1,1){$S^z$}}
\put(13.8,2.5){\makebox(5,1){$\tfrac{1}{\sqrt{2}}\bigl(
\ket{\up\dw}-\ket{\dw\up}\bigr)$}}

\put(15.75,9){\makebox(1,1){$\oplus$}}

\multiput(20,9)(1,0){3}{\line(0,1){1}}
\multiput(20,9)(0,1){2}{\line(1,0){2}}
\put(20,9){\makebox(1,1){1}}
\put(21,9){\makebox(1,1){2}}
\put(21,0){\line(0,1){6}}
\multiput(21,1)(0,2){3}{\circle*{0.5}}
\put(20,6){\makebox(1,1){$S^z$}}
\put(21.6,0.5){\makebox(2,1){$\ket{\dw\dw}$}}
\put(22.8,2.5){\makebox(5,1){$\tfrac{1}{\sqrt{2}}\bigl(
\ket{\up\dw}+\ket{\dw\up}\bigr)$}}
\put(21.6,4.5){\makebox(2,1){$\ket{\up\up}$}}
\end{picture}
\caption{Tensor product of two $S=\frac{1}{2}$ spins with Young
  tableaux and weight diagrams of the occurring SU(2) representations.
  $S^z$ is the diagonal generator.}
\label{fig:su2-1}
\end{figure}
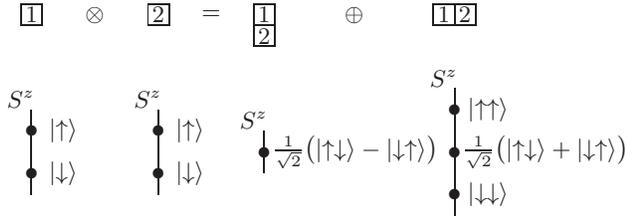
\begin{figure}[tb]
\setlength{\unitlength}{8pt}
\begin{picture}(32,8)(2,-1)
\linethickness{0.3pt}
\put(3,6){\line(1,0){1}}
\put(3,5){\line(1,0){1}}
\put(3,5){\line(0,1){1}}
\put(4,5){\line(0,1){1}}
\put(3,5){\makebox(1,1){1}}
\put(4.6,5.3){$\otimes$}
\put(6,6){\line(1,0){1}}
\put(6,5){\line(1,0){1}}
\put(6,5){\line(0,1){1}}
\put(7,5){\line(0,1){1}}
\put(6,5){\makebox(1,1){2}}
\put(3,4){$\underbrace{\phantom{iiiiiiiiii}}$}
\put(3,2.5){\line(1,0){1}}
\put(3,1.5){\line(1,0){1}}
\put(3,0.5){\line(1,0){1}}
\put(3,0.5){\line(0,1){2}}
\put(4,0.5){\line(0,1){2}}
\put(3,1.5){\makebox(1,1){1}}
\put(3,0.5){\makebox(1,1){2}}
\put(2.1,-1){$S=0$}
\put(5,1.8){$\oplus$}
\put(7,2.5){\line(1,0){2}}
\put(7,1.5){\line(1,0){2}}
\put(7,1.5){\line(0,1){1}}
\put(8,1.5){\line(0,1){1}}
\put(9,1.5){\line(0,1){1}}
\put(7,1.5){\makebox(1,1){1}}
\put(8,1.5){\makebox(1,1){2}}
\put(6.8,-0.7){$S=1$}
\put(7.6,5.3){$\otimes$}
\put(9,6){\line(1,0){1}}
\put(9,5){\line(1,0){1}}
\put(9,5){\line(0,1){1}}
\put(10,5){\line(0,1){1}}
\put(9,5){\makebox(1,1){3}}
\put(11.6,5.3){$=$}
\put(14,6){\line(1,0){1}}
\put(14,5){\line(1,0){1}}
\put(14,4){\line(1,0){1}}
\put(14,3){\line(1,0){1}}
\put(14,3){\line(0,1){3}}
\put(15,3){\line(0,1){3}}
\put(14,5){\makebox(1,1){1}}
\put(14,4){\makebox(1,1){2}}
\put(14,3){\makebox(1,1){3}}
\put(16.1,5.3){$\oplus$}
\put(18,6){\line(1,0){2}}
\put(18,5){\line(1,0){2}}
\put(18,4){\line(1,0){1}}
\put(18,4){\line(0,1){2}}
\put(19,4){\line(0,1){2}}
\put(20,5){\line(0,1){1}}
\put(18,5){\makebox(1,1){1}}
\put(19,5){\makebox(1,1){3}}
\put(18,4){\makebox(1,1){2}}
\put(17.5,2.2){$S=\frac{1}{2}$}
\put(21.1,5.3){$\oplus$}
\put(23,6){\line(1,0){2}}
\put(23,5){\line(1,0){2}}
\put(23,4){\line(1,0){1}}
\put(23,4){\line(0,1){2}}
\put(24,4){\line(0,1){2}}
\put(25,5){\line(0,1){1}}
\put(23,5){\makebox(1,1){1}}
\put(23,4){\makebox(1,1){3}}
\put(24,5){\makebox(1,1){2}}
\put(22.5,2.2){$S=\frac{1}{2}$}
\put(26.1,5.3){$\oplus$}
\put(28,6){\line(1,0){3}}
\put(28,5){\line(1,0){3}}
\put(28,5){\line(0,1){1}}
\put(29,5){\line(0,1){1}}
\put(30,5){\line(0,1){1}}
\put(31,5){\line(0,1){1}}
\put(28,5){\makebox(1,1){1}}
\put(29,5){\makebox(1,1){2}}
\put(30,5){\makebox(1,1){3}}
\put(28,2.2){$S=\frac{3}{2}$}
\put(14.5,4.5){\makebox(0,0){\line(1,2){1.8}}}
\put(14.5,4.5){\makebox(0,0){\line(1,-2){1.8}}}
\end{picture}
\caption{Tensor product of three $S=\frac{1}{2}$ spins with Young
  tableaux.  For SU($n$) with $n>2$, the tableau with three boxes on
  top of each other exists as well.}
\label{fig:su2-2}
\end{figure}
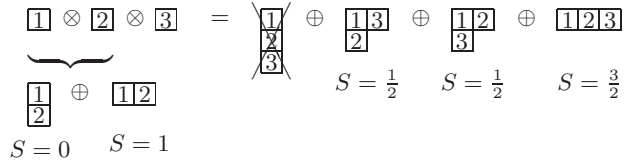
Let us begin with a review of Young tableaux and the representations
of SU(2).  The group SU(2) has three generators $S^a$, $a=1,2,3$,
which obey the algebra
\begin{equation}
\comm{S^a}{S^b}=i\epsilon^{abc}S^c,
\label{eq:su2algebra}
\end{equation}
where repeated indices are summed over and $\epsilon^{abc}$ is the
totally antisymmetric tensor.  The representations of SU(2) are
classified by the spin $S$, which takes integer or half-integer
values.  The fundamental representation of SU(2) has spin
$S=\frac{1}{2}$, it contains the two states $\ket{\up}$ and
$\ket{\dw}$.  Higher-dimensional representations can be constructed as
tensor products of fundamental representations, which is conveniently
accomplished using Young tableaux (see
\emph{e.g.}~\cite{InuiTanabeOnodera96}).  These tableaux are
constructed as follows (see Figs.~\ref{fig:su2-1} and~\ref{fig:su2-2}
for examples).  For each of the $N$ spins, draw a box numbered
consecutively from left to right.  The representations of SU(2) are
obtained by putting the boxes together such that the numbers assigned
to them increase in each row from left to right and in each column
from top to bottom.  Each tableau indicates symmetrization over all
boxes in the same row, and antisymmetrization over all boxes in the
same column.  This implies that we cannot have more than two boxes on
top of each other.  If $\kappa_i$ denotes the number of boxes in the
$i$th row, the spin is given by $S=\frac{1}{2}(\kappa_1-\kappa_2)$.

To be more explicit, let us consider the tensor product
$\bf\frac{1}{2}\otimes\frac{1}{2}\otimes\frac{1}{2}$ depicted in
Fig.~\ref{fig:su2-2} in detail.  We start with the state
$\ket{\tfrac{3}{2},\tfrac{3}{2}}=\ket{\up\up\up}$, and hence find
\begin{equation}
\ket{\tfrac{3}{2},\tfrac{1}{2}}
=\frac{1}{\sqrt{3}}\,S^-\ket{\tfrac{3}{2},\tfrac{3}{2}}=
\frac{1}{\sqrt{3}}\bigl(
\ket{\up\up\dw}+\ket{\up\dw\up}+\ket{\dw\up\up}\bigr).
\label{eq:s=3/2state}
\end{equation}
The two states with $S=S^z=\frac{1}{2}$ must be orthogonal to
\eqref{eq:s=3/2state}.  A convenient choice of basis is
\begin{equation}
\label{chiralbasis}
\begin{split}
\ket{\tfrac{1}{2},\tfrac{1}{2},+}&=
\frac{1}{\sqrt{3}}\bigl(
\ket{\up\up\dw}+\omega\ket{\up\dw\up}+\omega^2\ket{\dw\up\up}\bigr),\\[2mm]
\ket{\tfrac{1}{2},\tfrac{1}{2},-}&=
\frac{1}{\sqrt{3}}\bigl(
\ket{\up\up\dw}+\omega^2\ket{\up\dw\up}+\omega\ket{\dw\up\up}\bigr),
\end{split}
\end{equation}
where $\omega=\exp\bigl(i\frac{2\pi}{3}\bigr)$. The tableaux tell us
primarily that two such basis states exist, not what a convenient choice
of orthonormal basis states may be.

The irreducible representations of SU(2) can be classified through the
eigenvalues of the Casimir operator given by the square of the total
spin $\bs{S}^2$.  The special feature of $\bs{S}^2$ is that it
commutes with all generators $S^a$ and is hence by Schur's
lemma~\cite{Cornwell84vol2} proportional to the identity for any
finite-dimensional irreducible representation.
The eigenvalues are given by
\begin{displaymath}
  \bs{S}^2=\mathcal{C}^2_{\text{SU}(2)}=S(S+1).
\end{displaymath}

\begin{figure}[t]
\includegraphics[scale=0.18]{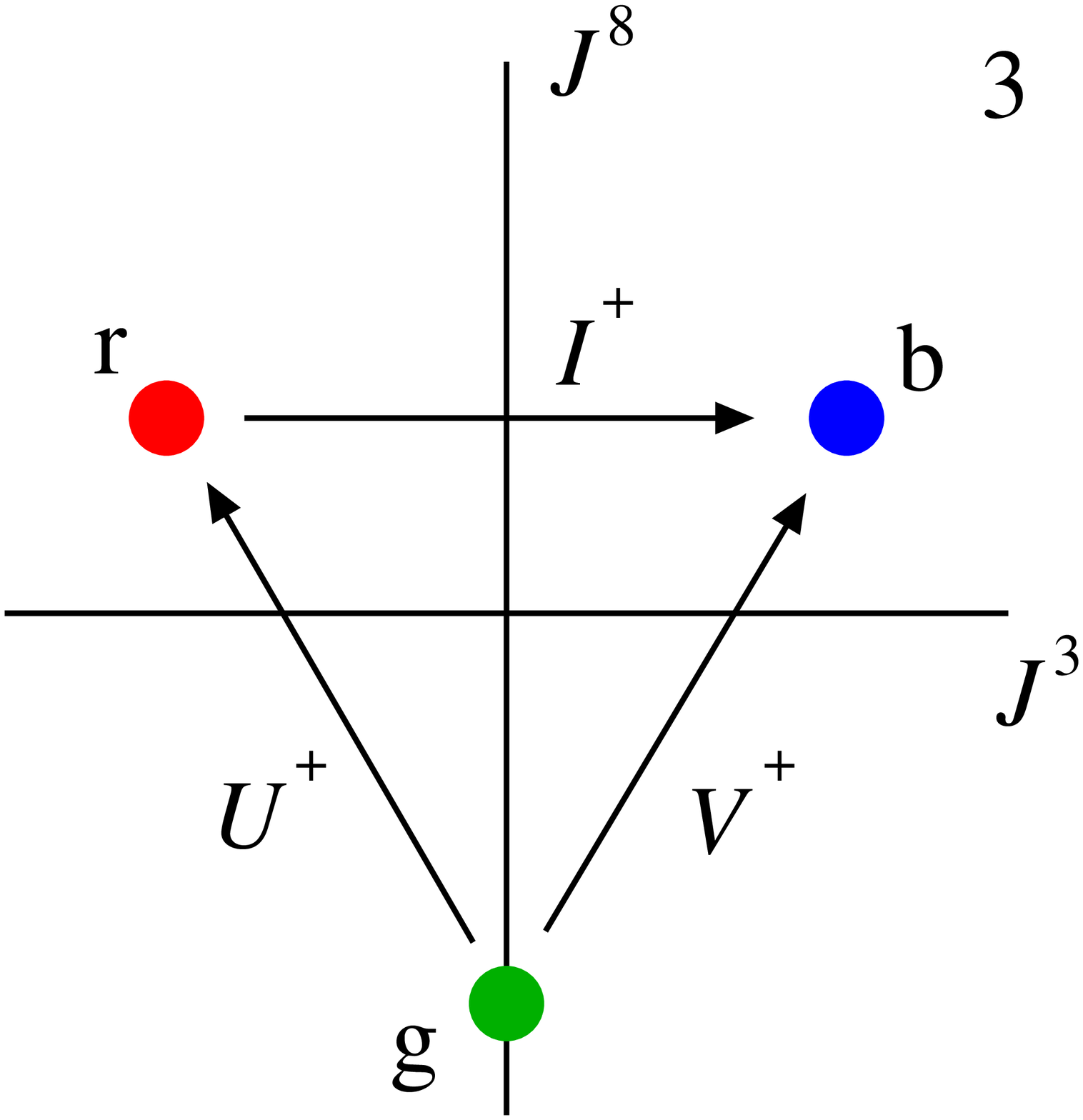}\hspace{5mm}
\includegraphics[scale=0.18]{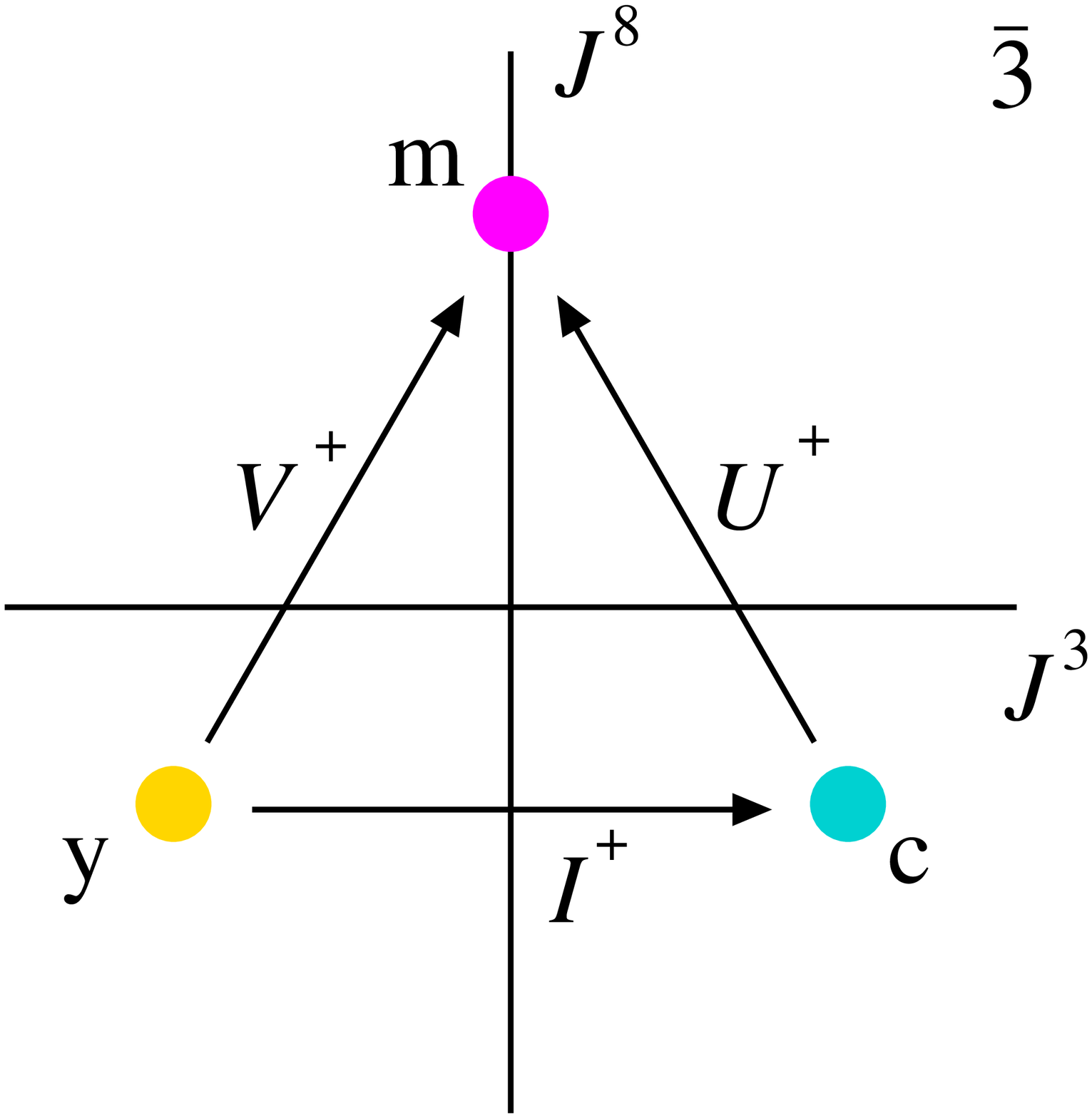}
\caption{(Color online) a) Weight diagram of the fundamental SU(3)
  representation $\bs{3}=(1,0)$. b) Weight diagram of the complex conjugate
  representation $\bs{\bar{3}}=(0,1)$.
  $J^3$ and $J^8$ denote the diagonal generators, $I^+$, $U^+$, and
  $V^+$ the raising operators. }
\label{fig:33barweight}
\end{figure}
\begin{figure}[tb]
\setlength{\unitlength}{8pt}
\begin{picture}(32,8)(2,-1)
\linethickness{0.5pt}
\put(3,6){\line(1,0){1}}
\put(3,5){\line(1,0){1}}
\put(3,5){\line(0,1){1}}
\put(4,5){\line(0,1){1}}
\put(3,5){\makebox(1,1){1}}
\put(4.6,5.3){$\otimes$}
\put(6,6){\line(1,0){1}}
\put(6,5){\line(1,0){1}}
\put(6,5){\line(0,1){1}}
\put(7,5){\line(0,1){1}}
\put(6,5){\makebox(1,1){2}}
\put(3,4.4){$\underbrace{\phantom{iiiiiiiiii}}$}
\put(3,2.9){\line(1,0){1}}
\put(3,1.9){\line(1,0){1}}
\put(3,0.9){\line(1,0){1}}
\put(3,0.9){\line(0,1){2}}
\put(4,0.9){\line(0,1){2}}
\put(3,1.9){\makebox(1,1){1}}
\put(3,0.9){\makebox(1,1){2}}
\put(3,-1){\makebox(1,1){$\bs{\bar{3}}$}}
\put(4.6,1.5){$\oplus$}
\put(6,2.4){\line(1,0){2}}
\put(6,1.4){\line(1,0){2}}
\put(6,1.4){\line(0,1){1}}
\put(7,1.4){\line(0,1){1}}
\put(8,1.4){\line(0,1){1}}
\put(6,1.4){\makebox(1,1){1}}
\put(7,1.4){\makebox(1,1){2}}
\put(6,-1){\makebox(2,1){$\bs{6}$}}
\put(7.6,5.3){$\otimes$}
\put(9,6){\line(1,0){1}}
\put(9,5){\line(1,0){1}}
\put(9,5){\line(0,1){1}}
\put(10,5){\line(0,1){1}}
\put(9,5){\makebox(1,1){3}}
\put(11.6,5.3){$=$}
\put(14,6){\line(1,0){1}}
\put(14,5){\line(1,0){1}}
\put(14,4){\line(1,0){1}}
\put(14,3){\line(1,0){1}}
\put(14,3){\line(0,1){3}}
\put(15,3){\line(0,1){3}}
\put(14,5){\makebox(1,1){1}}
\put(14,4){\makebox(1,1){2}}
\put(14,3){\makebox(1,1){3}}
\put(14,1){\makebox(1,1){$\bs{1}$}}
\put(16.1,5.3){$\oplus$}
\put(18,6){\line(1,0){2}}
\put(18,5){\line(1,0){2}}
\put(18,4){\line(1,0){1}}
\put(18,4){\line(0,1){2}}
\put(19,4){\line(0,1){2}}
\put(20,5){\line(0,1){1}}
\put(18,5){\makebox(1,1){1}}
\put(19,5){\makebox(1,1){3}}
\put(18,4){\makebox(1,1){2}}
\put(18,1){\makebox(2,1){$\bs{8}$}}
\put(21.1,5.3){$\oplus$}
\put(23,6){\line(1,0){2}}
\put(23,5){\line(1,0){2}}
\put(23,4){\line(1,0){1}}
\put(23,4){\line(0,1){2}}
\put(24,4){\line(0,1){2}}
\put(25,5){\line(0,1){1}}
\put(23,5){\makebox(1,1){1}}
\put(23,4){\makebox(1,1){3}}
\put(24,5){\makebox(1,1){2}}
\put(23,1){\makebox(2,1){$\bs{8}$}}
\put(26.1,5.3){$\oplus$}
\put(28,6){\line(1,0){3}}
\put(28,5){\line(1,0){3}}
\put(28,5){\line(0,1){1}}
\put(29,5){\line(0,1){1}}
\put(30,5){\line(0,1){1}}
\put(31,5){\line(0,1){1}}
\put(28,5){\makebox(1,1){1}}
\put(29,5){\makebox(1,1){2}}
\put(30,5){\makebox(1,1){3}}
\put(28,1){\makebox(3,1){$\bs{10}$}}
\end{picture}
\caption{Tensor product $\bs{3}\otimes\bs{3}\otimes\bs{3}$ with Young
  tableaux.}
\label{fig:youngdiagram}
\end{figure}

\subsection{Representation theory of SU(3) }

The group SU(3) has eight generators $J^a$, $a=1,\ldots,8$, which
obey the algebra
\begin{equation}
\comm{J^a}{J^b}=f^{abc}J^c,
\label{eq:su3algebra}
\end{equation}
where the structure constants $f^{abc}$ are given in
App.~\ref{app:conventions}.  For SU(3) we have two diagonal
generators, usually chosen to be $J^3$ and $J^8$, and six generators
which define the ladder operators $I^\pm=J^1\pm iJ^2$, $U^\pm=J^6\pm
iJ^7$, and $V^\pm=J^4\pm iJ^5$, respectively.  An explicit realization
of \eqref{eq:su3algebra} is, for example, given by the $J^a$'s as
expressed in terms of Gell-Mann matrices in \eqref{eq:J_a}.  This
realization defines the fundamental representation $\bs{3}$ of SU(3)
illustrated in Fig.~\ref{fig:33barweight}a.  It is three-dimensional,
and we have chosen to label the basis states by the colors blue (b),
red (r), and green (g).  The weight diagram depicted in
Fig.~\ref{fig:33barweight}a instructs us about the eigenvalues of the
diagonal generators as well as the actions of the ladder operators on
the basis states.

\begin{figure}[tb]
\setlength{\unitlength}{8pt}
\begin{picture}(16,4)(0,0)
\linethickness{0.3pt}
\multiput(0,4)(0,1){2}{\line(1,0){16}}
\put(0,3){\line(1,0){11}}
\put(0,2){\line(1,0){5}}
\multiput(0,2)(1,0){2}{\line(0,1){3}}
\multiput(4,2)(1,0){2}{\line(0,1){3}}
\put(6,3){\line(0,1){2}}
\multiput(10,3)(1,0){2}{\line(0,1){2}}
\put(12,4){\line(0,1){1}}
\multiput(15,4)(1,0){2}{\line(0,1){1}}
\put(11,0.7){\makebox(5,1){$\underbrace{\hspace{38.8pt}}$}}
\put(5,0.7){\makebox(6,1){$\underbrace{\hspace{46.7pt}}$}}
\put(11,-0.4){\makebox(5,1){$\mu_1\, $boxes}}
\put(5,-0.4){\makebox(6,1){$\mu_2\, $columns}}
\end{picture}
\caption{Dynkin coordinates $(\mu_1,\mu_2)$ for a given Young tableau.
  The columns containing three boxes represent additional SU(3)
  singlet factors, which yield equivalent representations and hence
  leave the Dynkin coordinates $(\mu_1,\mu_2)$ unchanged.}
\label{fig:dynkin}
\end{figure}
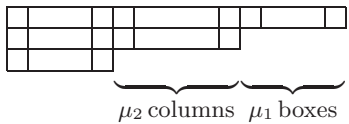

All other representations of SU(3) can be constructed by taking tensor
products of reps.\ $\bs{3}$, which is again most conveniently
accomplished using Young tableaux (see Fig.~\ref{fig:youngdiagram} for
an example).  The antisymmetrization over all boxes in the same column
implies that we cannot have more than three boxes on top of each
other.  Each tableaux stands for an irreducible representation of
SU(3), which can be uniquely labeled by their highest weight or Dynkin
coordinates $(\mu_1,\mu_2)$~\cite{Cornwell84vol2,Georgi82}
(see Fig.~\ref{fig:dynkin}).  For example, the fundamental
representation $\bs{3}$ has Dynkin coordinates (1,0).  Note that all
columns containing three boxes are superfluous, as the
antisymmetrization of three colors yields only one state.  In
particular, the SU(3) singlet has Dynkin coordinates (0,0).  In
general, the dimension of a representation $(\mu_1,\mu)$ is given by
$\frac{1}{2}(\mu_1+1)(\mu_2+1)(\mu_1+\mu_2+2)$. The labeling using
bold numbers refers to the dimensions of the representations alone.
Although this labeling is not unique, it will mostly be sufficient for
our purposes.  A representation $\bs{m}$ and its conjugated
counterpart $\bs{\overline{m}}$ are related to each other by
interchange of their Dynkin coordinates.

\subsection{Examples of representations of SU(3)}

We now consider some specific representations of SU(3) in detail.  As
starting point we use the fundamental representation $\bs{3}$ spanned
by the states $\ket{\b}$, $\ket{\r}$, and $\ket{\g}$.  The second
three-dimensional representation $\bs{\bar{3}}$ is obtained by
antisymmetrically coupling two reps.\ $\bs{3}$. The Dynkin coordinates
of the rep.\ $\bs{\bar{3}}$ are (0,1), \ie the reps.\ $\bs{3}$ and
$\bs{\bar{3}}$ are complex conjugate of each other.  An explicit
basis of the rep.\ $\bs{\bar{3}}$ is given by the colors yellow (y),
cyan (c), and magenta (m),
\begin{eqnarray}
\ket{\y}\!\!&=&\!\!\frac{1}{\sqrt{2}}\bigl(\ket{\r\g}-\ket{\g\r}\bigr),
\nonumber\\
\ket{\c}\!\!&=&\!\!\frac{1}{\sqrt{2}}\bigl(\ket{\g\b}-\ket{\b\g}\bigr),\\
\ket{\m}\!\!&=&\!\!\frac{1}{\sqrt{2}}\bigl(\ket{\b\r}-\ket{\r\b}\bigr).
\nonumber
\end{eqnarray}
The weight diagram is shown in Fig.~\ref{fig:33barweight}.b.  The
generators are given by \eqref{eq:J_a} with $\lambda^a$ replaced by
$-(\lambda^a)^*$, where $^*$ denotes complex conjugation of the matrix
elements~\cite{Georgi82}. In particular, we find
$I^+\ket{\y}=-\ket{\c}$, $U^+\ket{\c}=-\ket{\m}$, and
$V^+\ket{\y}=-\ket{\m}$.

\begin{figure}[t]
\setlength{\unitlength}{0.4pt}
\begin{picture}(320,320)(0,0)
\linethickness{0.3pt}
\put(0,166){\line(1,0){300}}
\put(150,0){\line(0,1){294}}
\multiput(50,224)(100,0){3}{\circle*{10}}
\multiput(100,137)(100,0){2}{\circle*{10}}
\put(150,50){\circle*{10}}
\put(320,300){\large\bf 6}
\put(155,290){$J^8$}
\put(295,150){$J^3$}
\put(262,225){$\ket{\b\b}$}
\put(5,225){$\ket{\r\r}$}
\put(68,245){$\tfrac{1}{\sqrt{2}}\bigl(\ket{\b\r}+\ket{\r\b}\bigr)$}
\put(210,115){$\tfrac{1}{\sqrt{2}}\bigl(\ket{\b\g}+\ket{\g\b}\bigr)$}
\put(-55,115){$\tfrac{1}{\sqrt{2}}\bigl(\ket{\r\g}+\ket{\g\r}\bigr)$}
\put(160,35){$\ket{\g\g}$}
\end{picture}
\caption{Weight diagram of the representation $\bs{6}=(2,0)$.
  The weight diagram of the conjugate representation
  $\bs{\bar{6}}=(0,2)$ is obtained by reflection at the
  origin~\cite{Cornwell84vol2}.}
\label{fig:6weight}
\end{figure}

The six-dimensional representation $\bs{6}$ has Dynkin coordinates
(2,0), and can hence be constructed by symmetrically coupling two
reps.\ $\bs{3}$. The basis states of the rep.\ $\bs{6}$ are shown in
Fig.~\ref{fig:6weight}.  The conjugate representation $\bs{\bar{6}}$
can be constructed by symmetrically coupling two reps.\
$\bs{\bar{3}}$.

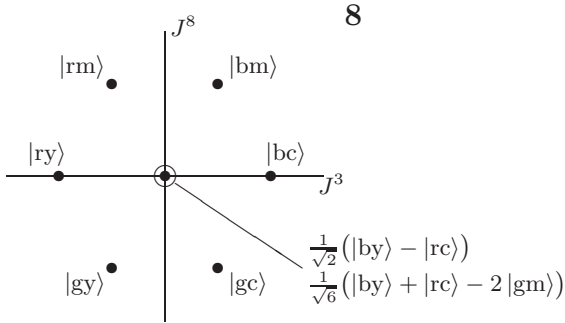
\begin{figure}[t]
\setlength{\unitlength}{0.4pt}
\begin{picture}(460,300)(0,0)
\linethickness{0.3pt}
\put(0,137){\line(1,0){300}}
\put(150,0){\line(0,1){274}}
\multiput(100,224)(100,0){2}{\circle*{10}}
\multiput(50,137)(100,0){3}{\circle*{10}}
\put(150,137){\circle{20}}
\multiput(100,50)(100,0){2}{\circle*{10}}
\put(160,130){\line(3,-2){120}}
\put(320,280){\large\bf 8}
\put(155,270){$J^8$}
\put(295,120){$J^3$}
\put(210,235){$\ket{\b\m}$}
\put(50,235){$\ket{\r\m}$}
\put(263,158){\makebox(3,1){$\ket{\b\c}$}}
\put(37,158){\makebox(3,1){$\ket{\r\y}$}}
\put(210,30){$\ket{\g\c}$}
\put(55,30){$\ket{\g\y}$}
\put(285,63){$\tfrac{1}{\sqrt{2}}\bigl(\ket{\b\y}-\ket{\r\c}\bigr)$}
\put(285,28){$\tfrac{1}{\sqrt{6}}\bigl(\ket{\b\y}+\ket{\r\c}-
2\ket{\g\m}\bigr)$}
\end{picture}
\caption{Weight diagram of the adjoint representation $\bs{8}=(1,1)$.
  The state with $J^3=J^8=0$ is doubly
  degenerate~\cite{Cornwell84vol2}.  Note that two reps.\ $\bs{8}$ can
  be constructed by combining three fundamental reps.\ $\bs{3}$
  (colors), just as two reps.\ $\bs{\frac{1}{2}}$ can be constructed
  by combining three SU(2) spins (cf. \eqref{chiralbasis}).  The
  states in the diagram span a basis for one of these
  representations.}
\label{fig:8weight}
\end{figure}

Let us now consider the tensor product
$\bs{3}\otimes\bs{\bar{3}}=\bs{1}\oplus\bs{8}$.
The weight diagram of the so-called adjoint representation
$\bs{8}=(1,1)$ is shown in Fig.~\ref{fig:8weight}. The states can be
constructed starting from the highest weight state $\ket{\b\m}$,
yielding $I^-\ket{\b\m}=\ket{\r\m}$, $U^-\ket{\b\m}=-\ket{\b\c}$,
$V^-\ket{\b\m}=\ket{\g\m}-\ket{\b\y}$, and so on. This procedure
yields two linearly independent states with $J^3=J^8=0$.  The
representation $\bs{8}$ can also be obtained by coupling of the reps.\
$\bs{6}$ and $\bs{3}$, as can be seen from the Young tableaux in
Fig.~\ref{fig:youngdiagram}.  On a more abstract level, the adjoint
representation is the representation we obtain if we consider the
generators $J^a$ themselves basis vectors.
In the weight diagram shown in Fig.~\ref{fig:8weight}, the generators
$J^3$ and $J^8$ correspond to the two states at the origin, whereas
the ladder operators $I^\pm$, $U^\pm$, and $V^\pm$ correspond to the
states at the six surrounding points.  In the notation of
Fig.~\ref{fig:8weight}, the singlet orthogonal to $\bs{8}$ is given by
$\tfrac{1}{\sqrt{3}}\bigl(\ket{\b\y}+\ket{\r\c}+\ket{\g\m}\bigr)$.

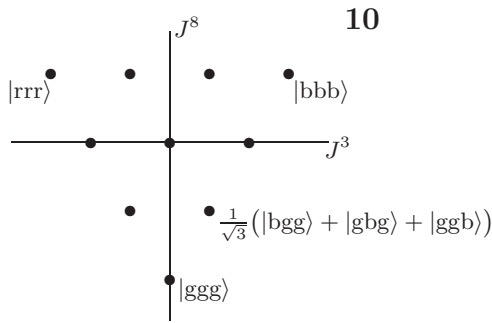
\begin{figure}[t]
\setlength{\unitlength}{0.3pt}
\begin{picture}(420,390)(50,0)
\linethickness{0.3pt}
\put(0,224){\line(1,0){400}}
\put(200,0){\line(0,1){364}}
\multiput(50,311)(100,0){4}{\circle*{13}}
\multiput(100,224)(100,0){3}{\circle*{13}}
\multiput(150,137)(100,0){2}{\circle*{13}}
\put(200,50){\circle*{13}}
\put(420,367){\large\bf 10}
\put(205,357){$J^8$}
\put(395,205){$J^3$}
\put(355,280){$\ket{\b\b\b}$}
\put(-5,280){$\ket{\r\r\r}$}
\put(260,115){$\tfrac{1}{\sqrt{3}}\bigl(\ket{\b\g\g}+\ket{\g\b\g}+
\ket{\g\g\b}\bigr)$}
\put(210,30){$\ket{\g\g\g}$}
\end{picture}
\caption{Weight diagram of the representation $\bs{10}=(3,0)$.
  The weight diagram of the conjugate representation
  $\bs{\overline{10}}=(0,3)$ is obtained by reflection at the
  origin~\cite{Cornwell84vol2}.}
  \label{fig:10weight}
\end{figure}

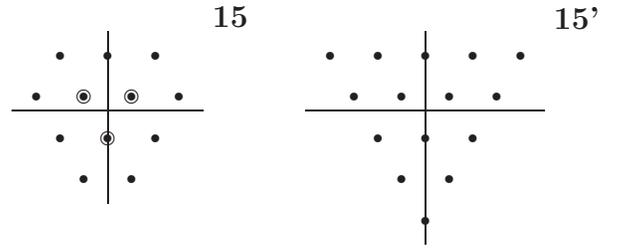
\begin{figure}[t]
\setlength{\unitlength}{0.18pt}
\begin{picture}(550,470)(0,-87)
\linethickness{0.3pt}
\put(0,195){\line(1,0){400}}
\put(200,0){\line(0,1){361}}
\multiput(100,311)(100,0){3}{\circle*{15}}
\multiput(50,224)(100,0){4}{\circle*{15}}
\multiput(150,224)(100,0){2}{\circle{30}}
\multiput(100,137)(100,0){3}{\circle*{15}}
\put(200,137){\circle{30}}
\multiput(150,50)(100,0){2}{\circle*{15}}
\put(420,371){\large\bf 15}
\end{picture}
\begin{picture}(550,550)(-50,0)
\linethickness{0.3pt}
\put(0,282){\line(1,0){500}}
\put(250,0){\line(0,1){448}}
\multiput(50,398)(100,0){5}{\circle*{15}}
\multiput(100,311)(100,0){4}{\circle*{15}}
\multiput(150,224)(100,0){3}{\circle*{15}}
\multiput(200,137)(100,0){2}{\circle*{15}}
\put(250,50){\circle*{15}}
\put(520,454){\large\bf 15'}
\end{picture}
\caption{Weight diagram of the representations $\bs{15}=(2,1)$ and
  $\bs{15'}=(4,0)$.  
  }
  \label{fig:15weight}
\end{figure}

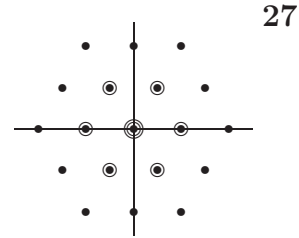
\begin{figure}[t]
\setlength{\unitlength}{0.18pt}
\begin{picture}(550,460)(50,0)
\linethickness{0.3pt}
\put(0,224){\line(1,0){500}}
\put(250,0){\line(0,1){448}}
\multiput(150,398)(100,0){3}{\circle*{15}}
\multiput(100,311)(100,0){4}{\circle*{15}}
\multiput(50,224)(100,0){5}{\circle*{15}}
\multiput(100,137)(100,0){4}{\circle*{15}}
\multiput(150,50)(100,0){3}{\circle*{15}}
\multiput(200,311)(100,0){2}{\circle{30}}
\multiput(150,224)(100,0){3}{\circle{30}}
\multiput(200,137)(100,0){2}{\circle{30}}
\put(250,224){\circle{40}}
\put(520,440){\large\bf 27}
\end{picture}
\caption{Weight diagram of the self-conjugate representation
  $\bs{27}=(2,2)$. The state with $J^3=J^8=0$ is three-fold
  degenerate~\cite{Cornwell84vol2}.}
  \label{fig:27weight}
\end{figure}

The weight diagrams of four other representations relevant to our
purposes below are shown in Figs.~\ref{fig:10weight}
to~\ref{fig:27weight}.

It is known that the physical properties of SU(2) spin chains
crucially depend on whether on the lattice sites are integer or
half-integer spins.  A similar distinction can be made for SU(3)
chains, as elaborated in Sec.~\ref{sec:conf}.  The distinction integer
or half-integer spin for SU(2) is replaced by a distinction between
three families of irreducible representations of SU(3): either the
number of boxes in the Young tableau is divisible by three without
remainder (\eg $\bs{1}$, $\bs{8}$, $\bs{10}$, $\bs{27}$), 
with remainder one (\eg $\bs{3}$, $\bs{\bar 6}$, $\bs{15}$, $\bs{15'}$),
or with remainder two 
(\eg $\bs{\bar 3}$, $\bs{6}$, $\bs{\overline{15}}$, $\bs{\overline{15}'}$). 

While SU(2) has only one Casimir operator, SU(3) has two.  The
quadratic Casimir operator is defined as
\begin{equation}
\bs{J}^2=\sum_{a=1}^8 J^aJ^a,
\label{eq:casimir}
\end{equation}
where the $J^a$'s are the generators of the representation. As
$\bs{J}^2$ commutes with all generators $J^a$ it is proportional to
the identity for any finite-dimensional irreducible representation.
The eigenvalue in a representation with Dynkin
coordinates $(\mu_1,\mu_2)$ is~\cite{Cornwell84vol2}
\begin{equation}
\bs{J}^2=\mathcal{C}^2_{\text{SU($3$)}}(\mu_1,\mu_2)=
\frac{1}{3}\bigl(\mu_1^2 + \mu_1\mu_2 + \mu_2^2 + 3\mu_1 + 3\mu_2\bigr).
\label{eq:casimirev}
\end{equation}
We have chosen the normalization in \eqref{eq:casimir} according to
the convention
\begin{displaymath}
  \mathcal{C}^2_{\text{SU($n$)}}(\text{adjoint representation})=n,
\end{displaymath}
which yields $\mathcal{C}^2_{\text{SU($3$)}}(1,1)=3$ for the
representation $\bs{8}$.  Note that the quadratic Casimir operator
cannot be used to distinguish between a representation and its
conjugate.  This distinction would require the cubic Casimir 
operator~\cite{Cornwell84vol2}, which we will not need for any of
the models we propose below.

\section{The trimer model (continued)}
\label{sec:trimercont}
\addtocounter{subsection}{1}

\subsection{Verification of the model}
\label{sec:trimerverified}

We will now proceed with the verification of the trimer Hamiltonian
\eqref{ham.trimer}.  Since the spins on the individual sites transform
under the fundamental representation $\bs{3}$, the SU(3) content of
four sites is
\begin{equation}\label{rep3power4}
\mb{3}\otimes\mb{3}\otimes\mb{3}\otimes\mb{3}\,=\,
3\cdot\mb{3}\,\oplus\,2\cdot\mb{\bar 6}\,\oplus\,3\cdot\mb{15}\,
\oplus\,\mb{15'},
\end{equation}
\ie we obtain representations $\bs{3}$, $\bs{\bar 6}$, and two
non-equivalent 15-dimensional representations with Dynkin coordinates
$(2,1)$ and $(4,0)$, respectively.  All these representations can be
distinguished by their eigenvalues of the quadratic Casimir operator,
which is given by $\left(\bs{J}^{(4)}_i\right)^2$ if the four spins
reside on the four neighboring lattice sites \hbox{$i,\ldots,i+3$.}

For the trimer states \eqref{trimerstates.3}, the situation
simplifies as we only have the two possibilities 
\begin{displaymath}
\begin{array}{rcl}
&
\begin{picture}(55,8)(0,-2)\linethickness{0.8pt}
\put(8,0){\line(1,0){10}}\put(6,0){\circle{4}}\put(20,0){\circle{4}}
\put(22,0){\line(1,0){10}}\put(34,0){\circle{4}}\put(48,0){\circle{4}}
\end{picture}
~&~\hat =~~~\mb{1}\,\otimes\,\mb{3}~=~\mb{3},\label{eq:one}\\[1mm]
&
\begin{picture}(55,8)(0,-2)\linethickness{0.8pt}
\put(8,0){\line(1,0){10}}\put(6,0){\circle{4}}
\put(20,0){\circle{4}}\put(36,0){\line(1,0){10}}\put(34,0){\circle{4}}
\put(48,0){\circle{4}}
\end{picture}~&~
\hat =~~~\mb{\bar 3} \,\otimes\,\mb{\bar 3}~=~\mb{3}\,\oplus\,\mb{\bar 6},
\end{array}
\end{displaymath}
which implies that the total SU(3) spin on four neighboring sites can
only transform under representations $\mb{3}$ or $\mb{\bar{6}}$. The
eigenvalues of the quadratic Casimir operator for these
representations are $4/3$ and $10/3$, respectively.  The auxiliary
operators
\begin{equation}
  \label{eq:auxop}
  H_i=\left(\Bigl(\bs{J}^{(4)}_i\Bigr)^2-\frac{4}{3}\right)
  \left(\Bigl(\bs{J}^{(4)}_i\Bigr)^2-\frac{10}{3}\right)
\end{equation} 
hence annihilate the trimer states for all values of $i$, while they
yield positive eigenvalues for $\bs{15}$ or $\bs{15'}$, \ie all other
states.  Summing $H_i$ over all lattice sites $i$ yields
\eqref{ham.trimer}.  We have numerically confirmed by exact
diagonalization of \eqref{ham.trimer} for chains with $N=9$ and 12
lattice sites that the three states \eqref{eq:trimer} are the only
ground states.

Note that the representation content of five neighboring sites in 
the trimer chains is just the conjugate of the above, as
\begin{displaymath}
\begin{array}{rcl}
&
\begin{picture}(65,8)(0,-2)\linethickness{0.8pt}
\put(8,0){\line(1,0){10}}\put(6,0){\circle{4}}\put(20,0){\circle{4}}
\put(22,0){\line(1,0){10}}\put(34,0){\circle{4}}\put(48,0){\circle{4}}
\put(50,0){\line(1,0){10}}\put(62,0){\circle{4}}
\end{picture}
~&~\hat =~~\mb{1}\,\otimes\,\mb{\bar 3}~=~\mb{\bar 3}\,,\\[1mm]
&
\begin{picture}(65,8)(0,-2)\linethickness{0,8pt}
\put(6,0){\circle{4}}\put(20,0){\circle{4}}
\put(36,0){\line(1,0){10}}\put(34,0){\circle{4}}\put(48,0){\circle{4}}
\put(22,0){\line(1,0){10}}\put(62,0){\circle{4}}
\end{picture}
~&~\hat =~~\mb{3}\,\otimes\,\mb{1}\,\otimes\,\mb{3}~=~\mb{\bar 3}\,\oplus\,
\mb{6}.\hspace{12pt}
\end{array}
\end{displaymath}
Since the quadratic Casimirs of conjugate representations have
identical eigenvalues, $\mathcal{C}^2_{\text{SU($3$)}}(\mu_1,\mu_2)=
\mathcal{C}^2_{\text{SU($3$)}}(\mu_2,\mu_1)$, we can construct another
parent Hamiltonian for the trimer states \eqref{eq:trimer} by simply
replacing $\bs{J}^{(4)}_i$ with $\bs{J}^{(5)}_i$ in
\eqref{ham.trimer}.  This Hamiltonian will have a different spectrum.
In comparison to the four-site interaction Hamiltonian
\eqref{ham.trimer}, however, it is more complicated while bearing no
advantages.  We will not consider it further.  

\subsection{Elementary excitations}
\label{sec:trimerexcitations}

\begin{figure}[t]
\begin{minipage}[b]{.45\linewidth}
\includegraphics[width=\linewidth]{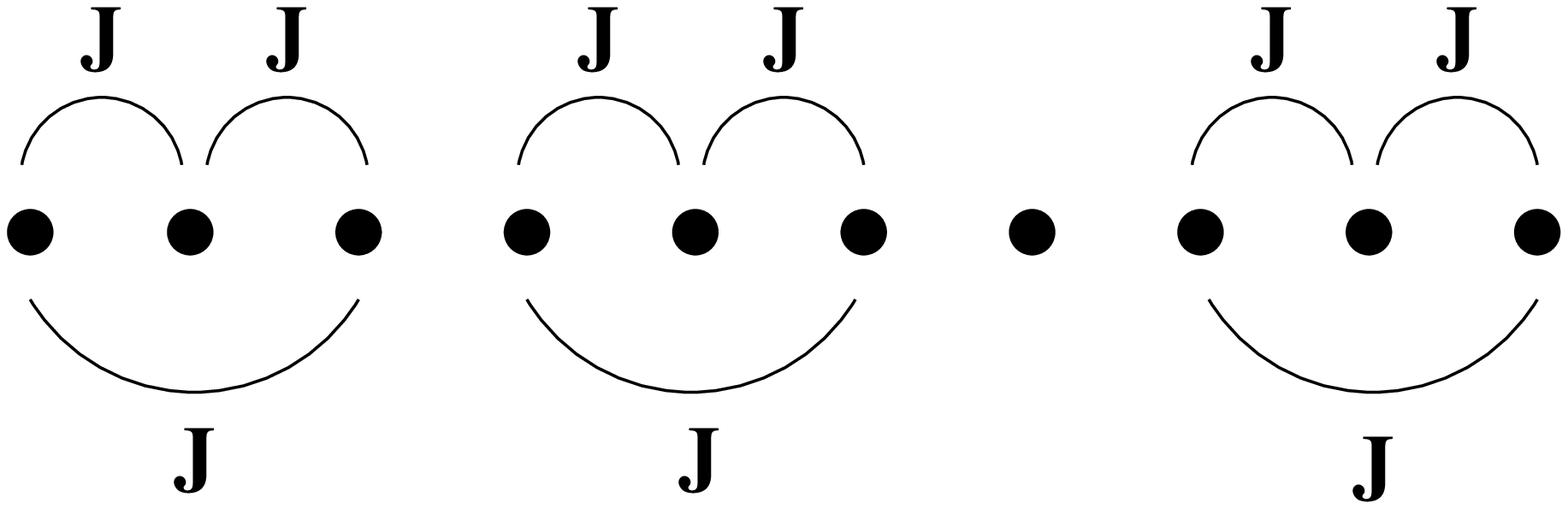}
\\ (a)
\end{minipage}\hfill
\begin{minipage}[b]{.495\linewidth}
\includegraphics[width=\linewidth]{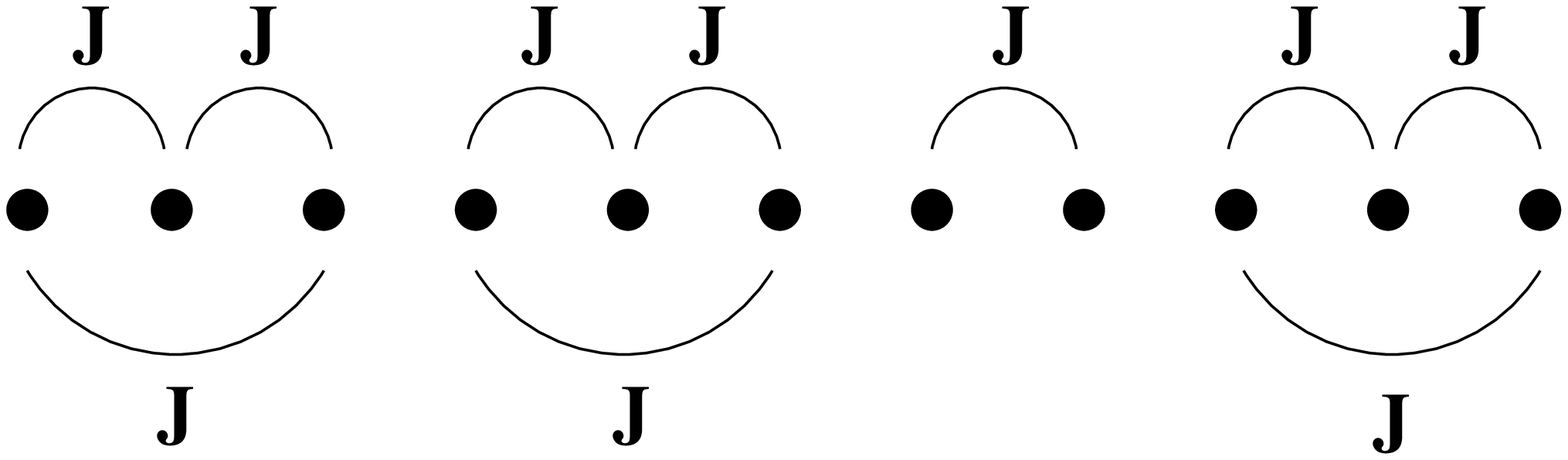}
\\ (b)
\end{minipage}
\caption{Couplings used in the numerical studies to create (a) the
  localized rep.\ $\bs{3}$ trial state and (b) the localized rep.\
  $\bs{\bar{3}}$ trial state.}
\label{fig:loc}
\end{figure}

Let us now turn to the low-lying excitations of \eqref{ham.trimer}.
In analogy with the MG model, it is evident that the SU(3) spinon
or ``coloron'' excitations correspond to domain walls between the
degenerate ground states.  For the trimer model, however, there
are two different kinds of domain walls, as illustrated by:
\begin{eqnarray}
\setlength{\unitlength}{2pt}
&\begin{picture}(90,12)(4,-2)
\linethickness{0.8pt}
\put(6,0){\circle{4}}
\put(8,0){\line(1,0){10}}
\put(20,0){\circle{4}}
\put(22,0){\line(1,0){10}}
\put(34,0){\circle{4}}
\put(48,0){\circle{4}}
\put(47.5,8){\makebox(1,1){$\mb{3}$}}
\put(62,0){\circle{4}}
\put(64,0){\line(1,0){10}}
\put(76,0){\circle{4}}
\put(78,0){\line(1,0){10}}
\put(90,0){\circle{4}}
\end{picture}&
\label{eq:trial3}\\[3mm]
&\begin{picture}(104,12)(4,-2)
\linethickness{0.8pt}
\put(6,0){\circle{4}}
\put(8,0){\line(1,0){10}}
\put(20,0){\circle{4}}
\put(22,0){\line(1,0){10}}
\put(34,0){\circle{4}}
\put(48,0){\circle{4}}
\put(50,0){\line(1,0){10}}
\put(54.5,8){\makebox(1,1){$\mb{\bar 3}$}}
\put(62,0){\circle{4}}
\put(76,0){\circle{4}}
\put(78,0){\line(1,0){10}}
\put(90,0){\circle{4}}
\put(92,0){\line(1,0){10}}
\put(104,0){\circle{4}}
\end{picture}&
\label{eq:trial3bar}
\end{eqnarray}
The first domain wall \eqref{eq:trial3} connects ground state $\mu$ to
the left to ground state $\mu+1$ to the right, where $\mu$ is defined
modulo 3 (see \eqref{eq:trimer}), and consists of an individual SU(3)
spin, which transforms under representation $\bs{3}$.  The second
domain wall \eqref{eq:trial3bar} connects ground state $\mu$ with 
ground state $\mu+2$.  It consists of two antisymmetrically coupled
spins on two neighboring sites, and hence transforms under
representation $\bs{\bar{3}}$.  As we take momentum superpositions of
the localized domain walls illustrated above, we expect one of them,
but not both, to constitute an approximate eigenstate of the trimer
model.  The reason we do not expect both of them to yield a valid
excitation is that they can decay into each other, \ie if the rep.\
$\bs{3}$ excitation is valid the rep.\ $\bs{\bar{3}}$ domain wall
would decay into two rep.\ $\bs{3}$ excitations, and vice versa.  The
question which of the two excitations is the valid one, \ie whether
the elementary excitations transform under $\bs{3}$ or $\bs{\bar{3}}$
under SU(3) rotations, can be resolved through numerical studies.  We
will discuss the results of these studies now.

\begin{table}[t]
\begin{center}
\begin{tabular}{c@{\hspace{10pt}}|@{\hspace{10pt}}c@{\hspace{10pt}}c@{\hspace{10pt}}c@{\hspace{10pt}}c}
\hline\hline
mom&$E_\text{tot}$&&\%&~over-~\\
$[2\pi/N]$&exact&trial&~~off~~&lap\\
\hline
0    &2.9735& 4.5860&54.2&0.9221\\
1, 12&6.0345&10.2804&70.4&0.5845\\
2, 11&9.0164&17.2991&91.9&0.0\\
3, 10&6.6863&13.1536&96.7&0.0\\
4, 9 &3.0896& 5.0529&63.5&0.8864\\
5, 8 &4.8744& 7.5033&53.9&0.8625\\
6, 7 &8.5618&16.6841&94.9&0.1095\\
\hline\hline
\end{tabular}
\end{center}
\caption{Energies of the rep.\ $\bs{3}$ trial states \eqref{eq:trial3}
  in comparison to the 
  exact excitation energies of the trimer model \eqref{ham.trimer} and
  their overlaps for an SU(3) spin chain with $N=13$ sites.}
\label{tab:rep3}
\end{table}

\begin{figure}[t]
\begin{center}
\includegraphics[scale=0.35,angle=270]{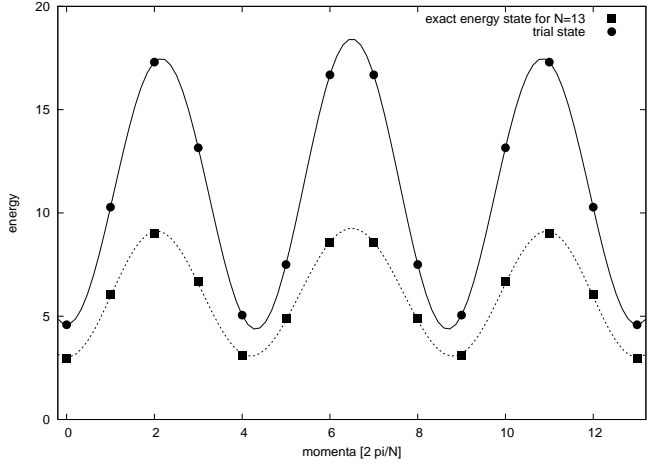}
\caption{Dispersion of the rep.\ $\bs{3}$ trial states
  \eqref{eq:trial3} in comparison to the exact excitation energies of
  \eqref{ham.trimer} for a chain with $N=13$.  The lines are a guide
  to the eye.}
\label{fig:spectrum3}
\end{center}
\end{figure}

\begin{figure}[t]
\begin{center}
\includegraphics[scale=0.35,angle=270]{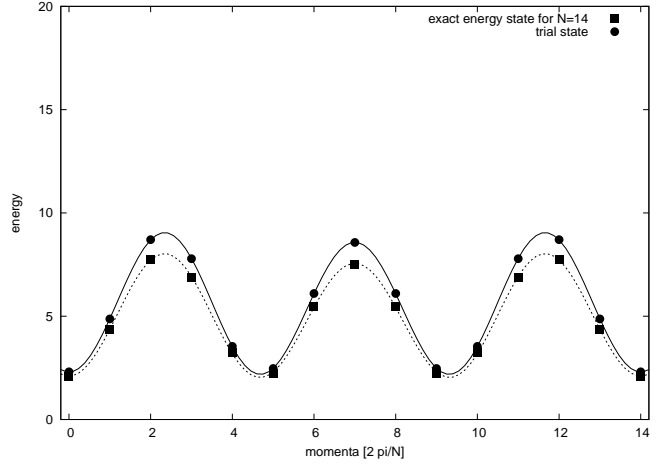}
\caption{Dispersion of the rep.\ $\bs{\bar{3}}$ trial states
  \eqref{eq:trial3bar} in comparison to the exact excitation energies
  of \eqref{ham.trimer} for a chain with $N=14$.  The lines are a
  guide to the eye.}
\label{fig:spectrum3bar}
\end{center}
\end{figure}

The rep.\ $\bs{3}$ and the rep.\ $\bs{\bar{3}}$ trial states require
chains with \mbox{$N=3\cdot\text{integer}+1$} and
\mbox{$N=3\cdot\text{integer}+2$} sites, respectively; we chose $N=13$
and $N=14$ for our numerical studies.  To create the localized domain
walls \eqref{eq:trial3} and \eqref{eq:trial3bar}, we numerically
diagonalized auxiliary Hamiltonians with appropriate couplings, as
illustrated in Fig.~\ref{fig:loc}.  From these localized excitations,
we constructed momentum eigenstates by superposition, and compared them
to the exact eigenstates of our model Hamiltonian \eqref{ham.trimer}
for chains with the same number of sites.  The results are shown in
Tab.~\ref{tab:rep3} and Fig.~\ref{fig:spectrum3} for the rep.\
$\bs{3}$ trial state, and in Tab.~\ref{tab:rep3bar} and
Fig.~\ref{fig:spectrum3bar} for the rep.\ $\bs{\bar 3}$ trial state.

\begin{table}[t]
\begin{center}
\begin{tabular}{c@{\hspace{10pt}}|@{\hspace{10pt}}c@{\hspace{10pt}}c@{\hspace{10pt}}c@{\hspace{10pt}}c}
\hline\hline
mom&$E_\text{tot}$&&\%&~over-~\\
$[2\pi/N]$&exact&trial&~~off~~&lap\\
\hline
0   &2.1013&2.3077&9.8&0.9953\\
1, 13&4.3677&4.8683&11.5&0.9864\\
2, 12&7.7322&8.7072&12.6&0.9716\\
3, 11&6.8964&7.7858&12.9&0.9696\\
4, 10&3.2244&3.5415&9.8&0.9934\\
5, 9 &2.2494&2.4690&9.7&0.9950\\
6, 8 &5.4903&6.1016&11.1&0.9827\\
7   &7.4965&8.5714&14.3&0.9562\\
\hline\hline
\end{tabular}
\caption{Energies of the rep.\ $\bs{\bar 3}$ trial states \eqref{eq:trial3bar}
  in comparison to the 
  exact excitation energies of the trimer model \eqref{ham.trimer} and
  their overlaps for an SU(3) spin chain with $N=14$ sites.}
\label{tab:rep3bar}
\end{center}
\end{table}

The numerical results clearly indicate that the rep.\ $\bs{\bar 3}$
trial states \eqref{eq:trial3bar} are valid approximations to the
elementary excitations of the trimer chain, while the rep.\ $\bs{3}$
trial states \eqref{eq:trial3} are not.  We deduce that the elementary
excitations of the trimer chain \eqref{ham.trimer} transform under
$\bs{\bar{3}}$, that is, under the representation conjugated to the
original SU(3) spins localized at the sites of the chain.  Using the
language of colors, one may say that if a basis for the original spins
is spanned by blue, red, and green, a basis for the excitations is
spanned by the complementary colors yellow, cyan, and magenta.  This
result appears to be a general feature of SU(3) spin chains, as it was
recently shown explicitly to hold for the Haldane--Shastry model as
well~\cite{ bouwknegt-96npb345, schuricht-05epl987, schuricht-06prb235105}.

Note that the elementary excitations of the trimer chain are
deconfined, meaning that the energy of two localized representation
$\bs{\bar{3}}$ domain walls or colorons 
\eqref{eq:trial3bar} does not depend on the distance between them.
The reason is simply that domain walls connect one ground state with
another, without introducing costly correlations in the region between
the domain walls.  In the case of the MG model and the trimer model
introduced here, however, there is still an energy gap associated with
the creation of each coloron, 
which is simply the energy cost associated with the domain wall.

In most of the remainder of this article, we will introduce a family
of exactly soluble valence bond models for SU(3) chains of various
spin representations of the SU(3) spins at each lattice site.  To
formulate these models, we will first review Schwinger bosons for
both SU(2) and SU(3) and the AKLT model.

\section{Schwinger Bosons}
\label{sec:schwinger}

Schwinger bosons~\cite{schwinger65proc,Auerbach94} constitute a way
to formulate spin-$S$ representations of an SU(2) algebra.  The spin
operators
\begin{equation}
\begin{array}{rcccl}
S^x + iS^y &=& S^+ &=& a^\dagger b, \\\rule{0pt}{12pt} 
S^x - iS^y &=& S^- &=& b^\dagger a, \\\rule{0pt}{12pt}
&& S^z  &=& \frac{1}{2}(a^\dagger a - b^\dagger b), 
\label{eq:schw}
\end{array}
\end{equation}
are given in terms of boson creation and annihilation
operators which obey the usual commutation relations
\begin{equation}
\begin{array}{c}
\comm{a}{a^\dagger}=\comm{b}{b^\dagger}=1, \\
\comm{a^{\phantom{\dagger}}\!\!}{b}=\comm{a}{b^\dagger}
=\comm{a^\dagger}{b}=\comm{a^\dagger}{b^\dagger}=0\rule{0pt}{16pt}.
\end{array}
\label{eq:schwb}
\end{equation}
It is readily verified with (\ref{eq:schwb}) that $S^x$, $S^y$, and
$S^z$ satisfy \eqref{eq:su2algebra}.
The spin quantum number $S$ is given by half the number of bosons,  
\begin{equation}
2S=a^\dagger a + b^\dagger b,
\label{eq:schwt}
\end{equation}
and the usual spin states (simultaneous eigenstates of $\boldsymbol{S}^2$
and $S^z$) are given by
\begin{equation}
\ket{S,m} = \frac{(a^\dagger)^{S+m}}{\sqrt{(S+m)!}} \frac{(b^\dagger)^{S-m}}
{\sqrt{(S-m)!}} \vac.
\label{eq:schs}
\end{equation}
In particular, the spin-\half states are given by
\begin{equation}
\ket{\up}=c_{\up}^\dagger \vac =a^\dagger \vac ,\qquad 
\ket{\dw}=c_{\dw}^\dagger \vac =b^\dagger \vac ,
\label{eq:schwfun}
\end{equation}
\ie $a^\dagger$ and $b^\dagger$ act just like the fermion creation
operators $c^\dagger_\up$ and $c^\dagger_\dw$ in this case.  The
difference shows up only when two (or more) creation operators act on
the same site or orbital.  The fermion operators create an
antisymmetric or singlet configuration (in accordance with the Pauli
principle), 
\begin{equation}
\ket{0,0} = c_{\up}^\dagger c_{\dw}^\dagger \vac ,
\label{eq:schwfer}
\end{equation}
while the Schwinger bosons create a totally symmetric or
triplet (or higher spin if we create more than two bosons) configuration,
\begin{eqnarray}
\ket{1,1} &=& \textstyle{\frac{1}{\sqrt{2}}} (a^\dagger)^{2}\vac 
,\nonumber\\[1mm]
\ket{1,0} &=& a^\dagger b^\dagger \vac , \\[1mm]
\ket{1,-1} &=& \textstyle{\frac{1}{\sqrt{2}}} (b^\dagger)^{2}\vac 
.\nonumber
\end{eqnarray}

The generalization to SU($n$) proceeds without incident.  We content
ourselves here by writing the formalism out explicitly for SU(3).  In
analogy to \eqref{eq:schw}, we write the SU(3) spin operators
\eqref{eq:J_a} 
\begin{equation}
\begin{array}{rcccl}
J^1 + iJ^2 &=& I^+ &=& b^\dagger r, \\\rule{0pt}{12pt} 
J^1 - iJ^2 &=& I^- &=& r^\dagger b, \\\rule{0pt}{12pt} 
&&            J^3 &=& \frac{1}{2}(b^\dagger b - r^\dagger r),\\\rule{0pt}{12pt}
J^4 + iJ^5 &=& V^+ &=& b^\dagger g, \\\rule{0pt}{12pt} 
J^4 - iJ^5 &=& V^- &=& g^\dagger b, \\\rule{0pt}{12pt} 
J^6 + iJ^7 &=& U^+ &=& r^\dagger g, \\\rule{0pt}{12pt} 
J^6 - iJ^7 &=& U^- &=& g^\dagger r, \\\rule{0pt}{12pt} 
&&             J^8 &=& 
\frac{1}{2\sqrt{3}}(b^\dagger b + r^\dagger r - 2g^\dagger g),
\label{eq:schwsu3}
\end{array}
\end{equation}
in terms of the boson annihilation and creation operators 
$b,b^\dagger$ (blue), $r,r^\dagger$ (red), and $g,g^\dagger$ (green)
satisfying
\begin{equation}
\comm{b}{b^\dagger}=\comm{r}{r^\dagger}=\comm{g}{g^\dagger}=1 
\label{eq:schwbsu3}
\end{equation}
while all other commutators vanish.  Again, it is readily verified
with \eqref{eq:schwbsu3} that the operators $J^a$ satisfy
\eqref{eq:su3algebra}.  The basis states spanning the fundamental
representation $\bs{3}$ may in analogy to \eqref{eq:schwfun} be
written using either fermion or boson creation operators:
\begin{equation}
\begin{array}{c}
\ket{\b}=c_{\b}^\dagger \vac=b^\dagger \vac ,\\\rule{0pt}{12pt} 
\ket{\r}=c_{\r}^\dagger \vac=r^\dagger \vac ,\\\rule{0pt}{12pt}
\ket{\g}=c_{\g}^\dagger \vac=g^\dagger \vac .
\end{array}
\label{eq:schwfunsu3}
\end{equation}
We write this abbreviated
\begin{equation}
\begin{array}{rcccccccl}
\bs{3}&=&(1,0)&=&
\setlength{\unitlength}{8pt}
\begin{picture}(1,1)(0,0.1)
\put(0,1){\line(1,0){1}}
\put(0,0){\line(1,0){1}}
\put(0,0){\line(0,1){1}}
\put(1,0){\line(0,1){1}}
\end{picture}
&\hat =&c_{\alpha}^\dagger\vac &=& \alpha^\dagger \vac. 
\end{array}
\end{equation}
The fermion operators can be used to combine spins transforming under
the fundamental representation $\bs{3}$ antisymmetrically, and hence 
to construct the representations 
\begin{equation}
\begin{array}{rcccccl}
\bs{\bar 3}&=&(0,1)&=&
\setlength{\unitlength}{8pt}
\begin{picture}(1,2)(0,0.5)
\put(0,2){\line(1,0){1}}
\put(0,1){\line(1,0){1}}
\put(0,0){\line(1,0){1}}
\put(0,0){\line(0,1){2}}
\put(1,0){\line(0,1){2}}
\end{picture}
&\hat =&c_{\alpha}^\dagger c_{\beta}^\dagger \vac,\\\rule{0pt}{12pt}
\bs{1}&=&(0,0)&=&
\setlength{\unitlength}{8pt}
\begin{picture}(1,3)(0,1)
\put(0,3){\line(1,0){1}}
\put(0,2){\line(1,0){1}}
\put(0,1){\line(1,0){1}}
\put(0,0){\line(1,0){1}}
\put(0,0){\line(0,1){3}}
\put(1,0){\line(0,1){3}}
\end{picture}
&\hat =&c_{\b}^\dagger c_{\r}^\dagger c_{\g}^\dagger \vac.
\end{array}
\vspace{6pt}
\label{eq:schwketsu3}
\end{equation}
The Schwinger bosons, by contrast, combine fundamental representations
$\bs{3}$ symmetrically, and hence yield representations labeled by
Young tableaux in which the boxes are arranged in a horizontal row,
like
\begin{equation}
\begin{array}{rcccccl}
\bs{6}&=&(2,0)&=&
\setlength{\unitlength}{8pt}
\begin{picture}(2,1)(0,0.1)
\put(0,1){\line(1,0){2}}
\put(0,0){\line(1,0){2}}
\put(0,0){\line(0,1){1}}
\put(1,0){\line(0,1){1}}
\put(2,0){\line(0,1){1}}
\end{picture}
&\hat =&\alpha^\dagger \beta^\dagger \vac,\\\rule{0pt}{14pt}
\bs{10}&=&(3,0)&=&
\setlength{\unitlength}{8pt}
\begin{picture}(3,1)(0,0.1)
\put(0,1){\line(1,0){3}}
\put(0,0){\line(1,0){3}}
\put(0,0){\line(0,1){1}}
\put(1,0){\line(0,1){1}}
\put(2,0){\line(0,1){1}}
\put(3,0){\line(0,1){1}}
\end{picture}
&\hat =&\alpha^\dagger \beta^\dagger \gamma^\dagger \vac,\\\rule{0pt}{14pt}
\bs{15'}&=&(4,0)&=&
\setlength{\unitlength}{8pt}
\begin{picture}(4,1)(0,0.1)
\put(0,1){\line(1,0){4}}
\put(0,0){\line(1,0){4}}
\put(0,0){\line(0,1){1}}
\put(1,0){\line(0,1){1}}
\put(2,0){\line(0,1){1}}
\put(3,0){\line(0,1){1}}
\put(4,0){\line(0,1){1}}
\end{picture}
&\hat =&\alpha^\dagger \beta^\dagger \gamma^\dagger \delta^\dagger \vac, 
\end{array}
\label{eq:schwbossu3}
\end{equation}
where $\alpha, \beta, \gamma, \ldots \in \{b,r,g\}$.
Unfortunately, it is not possible to construct representations
like 
\begin{displaymath}
\bs{8}\ =\ (1,1)\ =\
\setlength{\unitlength}{8pt}
\begin{picture}(2,1.2)(0,0.7)
\put(0,2){\line(1,0){2}}
\put(0,1){\line(1,0){2}}
\put(0,0){\line(1,0){1}}
\put(0,0){\line(0,1){2}}
\put(1,0){\line(0,1){2}}
\put(2,1){\line(0,1){1}}
\end{picture}
\vspace{2pt}
\end{displaymath}
by simply taking products of anti-commuting or commuting creation 
or annihilation operators.

\section{The AKLT model}
\label{sec:aklt}

Using the SU(2) Schwinger bosons introduced in the previous section, we
may rewrite the Majumdar--Ghosh states \eqref{eq:psimg} as
\begin{equation}
\big|\psi_{
\text{MG}}^{\textrm{even}\rule{0pt}{5pt}\atop 
\textrm{(odd)}} \big\rangle \, =
\underbrace{\displaystyle
\prod_{i\ \textrm{even}\atop (i\ \textrm{odd})}  
\Bigl(a^\dagger_{i} b^\dagger_{i+1} - b^\dagger_{i} a^\dagger_{i+1}\Bigr)
}_{\displaystyle
\equiv\Psi_{
\text{MG}}^{\textrm{even}\rule{0pt}{5pt}\atop 
\textrm{(odd)}} \big[a^\dagger,b^\dagger\big]
} \,\vac
\label{eq:mgschw}
\end{equation}
This formulation was used by Affleck, Kennedy, Lieb, and
Tasaki~\cite{affleck-88cmp477, affleck-87prl799} to propose a family
of states for higher spin representations of SU(2).  In particular, they 
showed that the valance bond solid (VBS) state
\begin{eqnarray}
\ket{\psi_{
\text{AKLT}}} \, &=&
\prod_i\,\Bigl(a^\dagger_{i} b^\dagger_{i+1} - b^\dagger_{i} a^\dagger_{i+1}\Bigr)
\,\vac =\nonumber\\
&=&
\Psi_{
\textrm{MG}}^{\scriptscriptstyle\textrm{even}}
\big[a^\dagger,b^\dagger\big] \,\cdot\, 
\Psi_{
\textrm{MG}}^{\scriptscriptstyle\textrm{odd}}
\big[a^\dagger,b^\dagger\big] \,\vac =\nonumber\\ 
\label{eq:psiaklt}
&=&
\setlength{\unitlength}{1pt}
\Big|
\begin{picture}(128,20)(-7,4)
\linethickness{0.8pt}
\multiput(0,10)(14,0){9}{\circle{4}}
\multiput(0,2)(14,0){9}{\circle{4}}
\thicklines
\multiput(2,10)(28,0){4}{\line(1,0){10}}
\multiput(16,2)(28,0){4}{\line(1,0){10}}
\thinlines
\put(37,-3){\line(1,0){10}}
\put(37,-3){\line(0,1){18}}
\put(37,15){\line(1,0){10}}
\put(47,-3){\line(0,1){18}}
\put(42,-3){\line(0,-1){9}}
\put(42,-20){\makebox(0,0)[c]{\small projection onto spin $S=1$}}
\end{picture}\Big\rangle\\[5mm]
& &\nonumber
\end{eqnarray}
is the exact zero-energy ground state of the spin-1 extended 
Heisenberg Hamiltonian
\begin{equation}
H_{\text{AKLT}} = 
\sum_i \left(\boldsymbol{S}_i \boldsymbol{S}_{i+1} +
\frac{1}{3}\bigl(\boldsymbol{S}_i \boldsymbol{S}_{i+1}\bigr)^2 
+\frac{2}{3}\right) 
\label{eq:haklt}
\end{equation}
with periodic boundary conditions.  Each term in the sum
(\ref{eq:haklt}) projects onto the subspace in which the total spin of
a pair of neighboring sites is $S=2$.  The Hamiltonian
(\ref{eq:haklt}) thereby lifts all states except (\ref{eq:psiaklt}) to
positive energies.  The VBS 
state (\ref{eq:psiaklt}) is a generic paradigm as it shares all the
symmetries, but in particular the Haldane spin
gap~\cite{haldane83pla464, haldane83prl1153, affleck89jpcm3047}, of
the spin-1 Heisenberg chain.  It even offers a particularly simple
understanding of this gap, or of the linear confinement potential
between spinons responsible for it, as illustrated by the cartoon:
\begin{center}
\setlength{\unitlength}{1pt}
\begin{picture}(196,45)(0,0)
\linethickness{0.8pt}
\multiput(0,35)(14,0){14}{\circle{4}}
\multiput(14,25)(14,0){14}{\circle{4}}
\thicklines
\multiput(2,35)(28,0){2}{\line(1,0){10}}
\multiput(72,35)(28,0){2}{\line(1,0){10}}
\multiput(142,35)(28,0){2}{\line(1,0){10}}
\multiput(16,25)(28,0){6}{\line(1,0){10}}
\put(184,25){\line(1,0){10}}
\thinlines
\put(56,36){\makebox(0,0){\vector(0,1){14}}}
\put(126,36){\makebox(0,0){\vector(0,1){14}}}
\put(91,11){\vector(-1,0){35}}
\put(91,11){\vector(1,0){35}}
\put(91,0){\makebox(0,0){\small energy cost $\propto$ distance}}
\end{picture}\\[5mm]
\end{center}
Our understanding~\cite{greiter02prb134443,greiter02prb054505} of the
connection between the confinement force and the Haldane gap is that
the confinement effectively imposes an oscillator potential for the
relative motion of the spinons.  We then interpret the zero-point
energy of this oscillator as the Haldane gap in the excitation
spectrum.

The AKLT state can also be written as a matrix
product~\cite{kluemper-91jpal955,kluemper-92zpb281,kluemper-93epl293}.
We first rewrite the valence bonds
\begin{displaymath}
\Bigl(a^\dagger_{i} b^\dagger_{i+1} - b^\dagger_{i} a^\dagger_{i+1}\Bigr)=
\Bigl(a^\dagger_{i}, b^\dagger_{i}\Bigr)
\left(\!\begin{array}{c} b^\dagger_{i+1}\\-a^\dagger_{i+1}\end{array}\!\right),
\end{displaymath}
and then use the outer product to combine the two vectors at each site
into a matrix
\begin{equation}
  \label{eq:akltmatrix}
  \begin{array}{rcl}
    M_i &\equiv& 
  \left(\!\begin{array}{c} b^\dagger_{i}\\-a^\dagger_{i}\end{array}\!\right)
  \Bigl(a^\dagger_{i}, b^\dagger_{i}\Bigr) \;\vac_i 
  \\[18pt] &=&\left(\!\begin{array}{cc} 
       \ket{1,0}_i& \sqrt{2}\ket{1,-1}_i\\[4pt]
      -\sqrt{2}\ket{1,1}_i &-\ket{1,0}_i 
    \end{array}\!\right).\\
  \end{array}
\end{equation}
Assuming PBCs, \eqref{eq:psiaklt} may then be written as the trace of
the matrix product
\begin{equation}
  \label{eq:akltm}
  \ket{\psi_{\text{AKLT}}}=\text{tr}\biggl( \prod_i M_i \biggr).
\end{equation}

In the following section, we will propose several exact models of
VBSs for SU(3).

\section{SU(3) valence bond solids}
\label{sec:vbs}

To begin with, we use SU(3) Schwinger bosons
introduced in Sec.~\ref{sec:schwinger} to rewrite the trimer states
\eqref{eq:trimer} as
\begin{eqnarray}
  \ket{\psi_\text{trimer}^{(\mu)}\!}
  &=&\hspace{-5pt}\prod_{\scriptstyle{i} \atop 
    \left(\scriptstyle{\frac{i-\mu}{3}\,{\rm integer}}\right)}
  \hspace{-5pt}\Bigl(
  \sum_{\scriptstyle{(\alpha,\beta,\gamma)}=\atop\scriptstyle{{\pi}(b,r,g)}} 
  \hspace{-5pt}\hbox{sign}({\pi})\,
  \alpha^{\dagger}_{i}\,\beta^{\dagger}_{i+1}\gamma^{\dagger}_{i+2}
  \Bigr)\vac \nonumber \\[2pt]
  &\equiv&\,\Psi^{\mu}\!\left[
    b^{\dagger},r^{\dagger},g^{\dagger}\right]\,\vac,
  \label{eq:trimerschw}
\end{eqnarray}
where, as in \eqref{eq:trimer}, $\mu=1,2,3$ labels the three
degenerate ground states, $i$ runs over the lattice sites subject to
the constraint that $\frac{i-\mu}{3}$ is integer, and the sum extends
over all six permutations $\pi$ of the three colors b, r, and g.
This formulation can be used directly to construct VBSs for SU(3) 
spin chains with spins transforming under representations $\bs{6}$ 
and $\bs{10}$ on each site.  

\subsection{The representation $\bs{6}$ VBS}

We obtain a representations $\bs{6}$ VBS from two trimer states by
projecting the tensor product of two fundamental representations
$\bs{3}$ onto the symmetric subspace, \ie onto the $\bs{6}$ in the
decomposition $\bs{3}\,\otimes\,\bs{3}=\bs{\bar{3}}\,\oplus\,\bs{6}$.
Graphically, this is illustrated as follows:
\begin{equation}
\setlength{\unitlength}{1pt}
\begin{picture}(128,48)(4,-22)
\linethickness{0.8pt}
\multiput(0,10)(14,0){9}{\circle{4}}
\multiput(14,0)(14,0){9}{\circle{4}}
\thicklines
\put(2,10){\line(1,0){10}}
\put(16,10){\line(1,0){10}}
\put(44,10){\line(1,0){10}}
\put(58,10){\line(1,0){10}}
\put(86,10){\line(1,0){10}}
\put(100,10){\line(1,0){10}}
\put(16,0){\line(1,0){10}}
\put(30,0){\line(1,0){10}}
\put(58,0){\line(1,0){10}}
\put(72,0){\line(1,0){10}}
\put(100,0){\line(1,0){10}}
\put(114,0){\line(1,0){10}}
\thinlines
\put(65,-5){\framebox(10,20)}
\put(70,-15){\line(0,1){10}}
\put(70,-22){\makebox(1,1){\small projection onto rep.\ $\bs{6}=(2,0)$}}
\put(70,25){\makebox(1,1){\small one site}}
\end{picture}
\label{fig:6state}
\end{equation}
This construction yields three linearly independent $\mb{6}$ VBS
states, as there are three ways to choose two different trimer states
out of a total of three.  These three VBS states are readily written
out using \eqref{eq:trimerschw},
\begin{equation}
\label{eq:6VBS}
\ket{\psi_{\bs{6}\, \text{VBS}}^{(\mu)}}=
\Psi^{\mu}\!\left[b^{\dagger},r^{\dagger},g^{\dagger}\right]\cdot
\Psi^{\mu+1}\!\left[b^{\dagger},r^{\dagger},g^{\dagger}\right]\,
\vac
\end{equation}
for $\mu=1$, 2, or 3.
If we pick four neighboring sites on a chain with any of these states, 
the total SU(3) spin of those may contain the representations
\begin{displaymath}
\setlength{\unitlength}{1pt}
\begin{picture}(56,12)(0,2)
\linethickness{0.8pt}
\multiput(0,10)(14,0){4}{\circle{4}}
\multiput(0,0)(14,0){4}{\circle{4}}
\thicklines
\put(2,10){\line(1,0){10}}
\put(16,10){\line(1,0){10}}
\put(16,0){\line(1,0){10}}
\put(30,0){\line(1,0){10}}
\end{picture}
\hat =~~~\bs{3}\,\otimes\,\bs{3}~=~\bs{\bar 3}\,\oplus\,\bs{6}
\end{displaymath}
or the representations
\begin{displaymath}
  \setlength{\unitlength}{1pt}
\begin{picture}(56,12)(0,2)
\linethickness{0.8pt}
\multiput(0,10)(14,0){4}{\circle{4}}
\multiput(0,0)(14,0){4}{\circle{4}}
\thicklines
\put(2,10){\line(1,0){10}}
\put(16,10){\line(1,0){10}}
\put(2,0){\line(1,0){10}}
\put(30,0){\line(1,0){10}}
\end{picture}
\hat =~~~\bs{\bar 3}\,\otimes\,\bs{\bar 3}\,\otimes\,\bs{3}~
=~2\cdot\bs{\bar 3}\,\oplus\,\bs{6}\,\oplus\,\bs{\overline{15}},
\end{displaymath}
\ie the total spin transforms under $\bs{\bar{3}}$, $\bs{6}$, or
$\bs{\overline{15}}=(1,2)$, all of which are contained in the product
\begin{multline}
  \label{eq:6666}
  \bs{6}\,\otimes\,\bs{6}\,\otimes\,\bs{6}\,\otimes\,\bs{6}=\\
  3\cdot\bs{\overline{3}}\,\oplus\,6\cdot\bs{6}\,\oplus\,
  7\cdot\bs{\overline{15}}\,\oplus\,3\cdot\bs{\overline{15}'}\,\oplus\,
  3\cdot\bs{21}\,\\ \oplus\,8\cdot\bs{24}\,\oplus\,
  6\cdot\bs{\overline{42}}\,\oplus\,\bs{45}\,\oplus\,
  6\cdot\bs{60}\,\oplus\,3\cdot\bs{63}
\end{multline}
and hence possible for a representation $\bs{6}$ spin chain in
general.  The corresponding Casimirs are given by
$\casi{0}{1}=\frac{4}{3}$, $\casi{2}{0}=\frac{10}{3}$, and
$\casi{1}{2}=\frac{16}{3}$.  This leads us to propose the parent
Hamiltonian
\begin{equation}
  \label{eq:ham6}
  H_{\bs{6}\,\text{VBS}}=\sum_{i=1}^N H_i
\end{equation}
with 
\begin{equation}
  \label{eq:hi6}
  H_i=\left(\!\left(\!\bs{J}^{(4)}_i\!\right)^2 \! - \frac{4}{3}\right)
  \!\left(\!\left(\!\bs{J}^{(4)}_i\!\right)^2 \! - \frac{10}{3}\right)
  \!\left(\!\left(\!\bs{J}^{(4)}_i\!\right)^2 \! - \frac{16}{3}\right).
\end{equation}
Note that the operators $J_i^a$, $a=1,\ldots,8$, are now given by
$6\times 6$ matrices, as the Gell-Mann matrices only provide the
generators \eqref{eq:J_a} of the fundamental representation $\bs{3}$.
Since the representations $\bs{\bar 3}$, $\bs{6}$, and
$\bs{\overline{15}}$ possess the smallest Casimirs in the expansion
\eqref{eq:6666}, $H_i$ and hence also $H_{\bs{6}\,\text{VBS}}$ are
positive semi-definite (\ie have only non-negative eigenvalues).  The
three linearly independent states \eqref{eq:6VBS} are zero-energy
eigenstates of \eqref{eq:ham6}.  

To verify that these are the only ground states, we have numerically
diagonalized \eqref{eq:ham6} for $N=6$ and $N=9$ sites.  For $N=9$, we
find zero-energy ground states at momenta $k=0,3,\ \text{and}\ 6$ (in
units of $\frac{2\pi}{N}$ with the lattice constant set to unity).
Since the dimension of the Hilbert space required the use of a LANCZOS
algorithm, we cannot be certain that there are no further ground
states.  We therefore diagonalized \eqref{eq:ham6} for $N=6$ as well,
where we were able to obtain the full spectrum.  We obtained five zero
energy ground states, two at momentum $k=0$ and one each at
$k=2,3,4$.  One of the ground states at $k=0$ and
the $k=2,4$ ground states constitute the space of momentum eigenstates
obtained by Fourier transform of the space spanned by the three
$\bs{6}$ VBS states \eqref{eq:6VBS}.  The remaining two states at
$k=0,3$ are the momentum eigenstates formed by superposition of the
state
\begin{equation}
\setlength{\unitlength}{1pt}
\begin{picture}(120,36)(-14,-12)
\linethickness{0.8pt}
\put(0,0){\circle{4}}
\put(0,10){\circle{4}}
\put(98,0){\circle{4}}
\put(98,10){\circle{4}}
\multiput(14,10)(14,0){6}{\circle{4}}
\multiput(14,0)(14,0){6}{\circle{4}}
\thicklines
\put(2,0){\line(1,0){5}}
\put(7,0){\line(1,0){5}}
\put(30,0){\line(1,0){10}}
\put(58,0){\line(1,0){10}}
\put(16,1){\line(4,3){10}}
\put(44,1){\line(4,3){10}}
\put(72,1){\line(4,3){10}}
\put(86,0){\line(1,0){5}}
\put(91,0){\line(1,0){5}}
\put(-7,5){\line(4,3){5}}
\qbezier(15,12)(28,25)(41,12)
\qbezier(43,12)(56,25)(69,12)
\qbezier(99,12)(101,14)(103,15)
\thinlines
\put(7,-8){\dashbox{2}(84,32)}
\put(-14,-15){\makebox(0,0){\small $i=$}}
\put(0,-15){\makebox(0,0){\small $6$}}
\put(14,-15){\makebox(0,0){\small $1$}}
\put(28,-15){\makebox(0,0){\small $2$}}
\put(42,-15){\makebox(0,0){\small $3$}}
\put(56,-15){\makebox(0,0){\small $4$}}
\put(70,-15){\makebox(0,0){\small $5$}}
\put(84,-15){\makebox(0,0){\small $6$}}
\put(98,-15){\makebox(0,0){\small $1$}}
\end{picture}
\label{fig:finite6state}
\end{equation}
and the same translated by one lattice spacing.  It is readily seen
that these two states are likewise zero energy eigenstates of
\eqref{eq:ham6} for $N=6$ sites.  The crucial difference, however, is
that the $\bs{6}$ VBS states \eqref{eq:6VBS} remain zero-energy
eigenstates of \eqref{eq:ham6} for all $N$'s divisible by three, while
the equivalent of \eqref{fig:finite6state} for larger $N$ do not.  We
hence attribute these two additional ground states for $N=6$ to the
finite size, and conclude that the three states
\eqref{eq:6VBS} are the only zero-energy ground states of
\eqref{eq:ham6} for general $N$'s divisible by three.

Excitations of the $\bs{6}$ VBS model are given by domain walls
between two of the ground states \eqref{eq:6VBS}.  As in the trimer
model, two distinct types of domain walls exist, which transform
according to representations $\bs{\bar{3}}$ and $\bs{3}$:
\begin{equation}\label{eq:6VBSexc}
\setlength{\unitlength}{1pt}
\begin{picture}(220,45)(4,-15)
\linethickness{0.8pt}
\multiput(0,10)(14,0){3}{\circle{4}}
\multiput(42,10)(14,0){2}{\circle*{4}}
\multiput(70,10)(14,0){6}{\circle{4}}
\multiput(154,10)(14,0){1}{\circle*{4}}
\multiput(168,10)(14,0){3}{\circle{4}}
\multiput(14,0)(14,0){15}{\circle{4}}
\thicklines
\put(2,10){\line(1,0){10}}
\put(16,10){\line(1,0){10}}
\put(44,10){\line(1,0){10}}
\put(72,10){\line(1,0){10}}
\put(114,10){\line(1,0){10}}
\put(128,10){\line(1,0){10}}
\put(86,10){\line(1,0){10}}
\put(170,10){\line(1,0){10}}
\put(184,10){\line(1,0){10}}
\put(16,0){\line(1,0){10}}
\put(30,0){\line(1,0){10}}
\put(58,0){\line(1,0){10}}
\put(72,0){\line(1,0){10}}
\put(100,0){\line(1,0){10}}
\put(114,0){\line(1,0){10}}
\put(142,0){\line(1,0){10}}
\put(156,0){\line(1,0){10}}
\put(184,0){\line(1,0){10}}
\put(198,0){\line(1,0){10}}
\put(12,-12){\makebox(0,0){\small $\Psi^1\!\cdot\!\Psi^2$}}
\put(101,-12){\makebox(0,0){\small $\Psi^2\!\cdot\!\Psi^3$}}
\put(198,-12){\makebox(0,0){\small $\Psi^1\!\cdot\!\Psi^2$}}
\put(49,25){\makebox(0,0){$\bs{\bar 3}$}}
\put(154,25){\makebox(0,0){$\bs{3}$}}
\thinlines
\put(23,-5){\dashbox{2}(52,20)}
\put(121,-6){\dashbox{2}(52,20)}
\put(135,-4){\dashbox{1}(52,20)}
\end{picture}
\end{equation}
It is not clear which excitation has the lower energy, and it appears
likely that both of them are stable against decay.  Let us first look
at the rep.\ $\bs{\bar 3}$ excitation.  The four-site Hamiltonian
\eqref{eq:hi6} annihilates the state for all $i$'s except the four
sites in the dashed box in \eqref{eq:6VBSexc}, which contains
the representations
\begin{multline}
  \bs{\bar 3}\,\otimes\,\bs{\bar 3}\,\otimes\,\bs{\bar 3}\,
  \otimes\,\bs{3}\,\otimes\,\bs{3}\\ =
  6\cdot\bs{\bar 3}\,\oplus\,5\cdot\bs{6}\,\oplus\,6\cdot
  \bs{\overline{15}}\,\oplus\,\bs{\overline{15}'}\,\oplus\,2\cdot\bs{24}\,
  \oplus\,\bs{\overline{42}}
  \nonumber
\end{multline}
\ie the representations $\bs{\overline{15}'}=(0,4)$, $\bs{24}=(3,1)$
twice, and $\bs{\overline{42}}=(2,3)$ with Casimirs
$\frac{28}{3}$, $\frac{25}{3}$ and $\frac{34}{3}$, respectively, in
addition to representations annihilated by $H_i$.  For the
rep.\ $\bs{3}$ excitation sketched on the right in \eqref{eq:6VBSexc},
there are two sets of four neighboring sites not annihilated by $H_i$
as indicated by the dashed and the dotted box.  Each set contains the
representations
\begin{displaymath}
  \bs{\bar 3}\,\otimes\,\bs{3}\,\otimes\,\bs{3}\,\otimes\,\bs{3}\,=
  \,3\cdot\bs{\bar 3}\,\oplus\,3\cdot\bs{6}\,\oplus\,2\cdot\bs{\overline{15}}\,
  \oplus\,\bs{24}
\end{displaymath}
\ie only the rep.\ $\bs{24}$ in addition to representations
annihilated by $H_i$.  For our parent Hamiltonian \eqref{eq:ham6}, it
hence may well be that the rep.\ $\bs{3}$ anti-coloron 
has the lower energy, but it is all but clear that the rep.\ $\bs{\bar
  3}$ has sufficiently higher energy to decay.  For general
representation $\bs{6}$ spin chains, it may depend on the specifics of
the model which excitation is lower in energy and whether the
conjugate excitation decays or not.

Since the excitations of the rep.\ $\bs{6}$ VBS chain are merely
domain walls between different ground states, there is no confinement
between them.  We expect the generic antiferromagnetic rep.\ $\bs{6}$
chain to be gapless, even though the model we proposed here has a gap
associated with the energy cost of creating a domain wall.

\subsection{The representation 10 VBS}\label{sec:10VBS}

Let us now turn to the $\bs{10}$ VBS chain, which is a direct
generalization of the AKLT chain to SU(3).  By combining the three
different trimer states \eqref{eq:trimerschw} for $\mu=1,2,\
\text{and}\ 3$ symmetrically,
\begin{eqnarray}
\ket{\psi_{\bs{10}\,\text{VBS}}}\!&\!=\!&\!
\Psi^1\!\left[b^{\dagger}\!,r^{\dagger}\!,g^{\dagger}\right]
\!\cdot\!\Psi^2\!\left[b^{\dagger}\!,r^{\dagger}\!,g^{\dagger}\right]
\!\cdot\!\Psi^3\!\left[b^{\dagger}\!,r^{\dagger}\!,g^{\dagger}\right]\vac
\nonumber \\[5pt]
\!&\!=\!&\!
\prod_i\Bigl(
\sum_{\scriptstyle{(\alpha,\beta,\gamma)}=\atop\scriptstyle{{\pi}(b,r,g)}} 
\hspace{-5pt}\hbox{sign}({\pi})\,
\alpha^{\dagger}_{i}\,\beta^{\dagger}_{i+1}\gamma^{\dagger}_{i+2}
\Bigr)\vac , 
\label{state.10VBS}
\end{eqnarray}
we automatically project out the rep.\ $\bs{10}$ in the decomposition
$\bs{3}\otimes\bs{3}\otimes\bs{3}=\bs{1}\oplus2\!\cdot\bs{8}\oplus\bs{10}$
generated on each lattice site by the three trimer chains.  This
construction yields a unique state, as illustrated:
\begin{equation}
\setlength{\unitlength}{1pt}
\begin{picture}(180,56)(4,-6)
\linethickness{0.8pt}
\multiput(0,35)(14,0){12}{\circle{4}}
\multiput(14,25)(14,0){12}{\circle{4}}
\multiput(28,15)(14,0){12}{\circle{4}}
\thicklines
\put(2,35){\line(1,0){10}}
\put(16,35){\line(1,0){10}}
\put(44,35){\line(1,0){10}}
\put(58,35){\line(1,0){10}}
\put(86,35){\line(1,0){10}}
\put(100,35){\line(1,0){10}}
\put(16,25){\line(1,0){10}}
\put(30,25){\line(1,0){10}}
\put(58,25){\line(1,0){10}}
\put(72,25){\line(1,0){10}}
\put(100,25){\line(1,0){10}}
\put(114,25){\line(1,0){10}}
\put(30,15){\line(1,0){10}}
\put(44,15){\line(1,0){10}}
\put(72,15){\line(1,0){10}}
\put(86,15){\line(1,0){10}}
\put(114,15){\line(1,0){10}}
\put(128,15){\line(1,0){10}}
\put(128,35){\line(1,0){10}}
\put(142,35){\line(1,0){10}}
\put(142,25){\line(1,0){10}}
\put(156,25){\line(1,0){10}}
\put(156,15){\line(1,0){10}}
\put(170,15){\line(1,0){10}}
\thinlines
\put(51,10){\framebox(10,30){}}
\put(56,2){\line(0,1){8}}
\put(56,-5){\makebox(1,1){\small projection onto $\bs{10}=(3,0)$}}
\put(56,48){\makebox(1,1){\small one site}}
\put(121,10){\dashbox{2}(24,30)}
\end{picture}
\label{fig:10state}
\end{equation}
In order to construct a parent Hamiltonian, note first that the total
spin on two (neighboring) sites of a rep.\ $\bs{10}$ chain is given by
\begin{equation}
\bs{10}\otimes\bs{10}
=\bs{\overline{10}}\oplus\bs{27}\oplus\bs{28}\oplus\bs{35}.
\label{twosites.10}
\end{equation}
On the other hand, the total spin of two neighboring sites for
the $\bs{10}$ VBS state can contain only the representations
\begin{equation}
  \bs{\bar{3}}\otimes\bs{\bar{3}}\otimes\bs{3}\otimes\bs{3}=
  2\cdot\bs{1}\oplus 4\cdot\bs{8}
  \oplus\bs{10}\oplus\bs{\overline{10}}\oplus\bs{27},
  \label{twosites.10VBS}
\end{equation}
as can be seen easily from the dashed box in the cartoon above.  (Note
that this result is independent of how many sites we include in the
dashed box.)  After the projection onto rep.\ $\bs{10}$ on each lattice
site, we find that only reps.\ $\bs{\overline{10}}=(0,3)$ and
$\bs{27}=(2,2)$ occur for the total spin of two neighboring sites for the
$\bs{10}$ VBS state.  With the Casimirs $\casi{0}{3}=6$ and
$\casi{2}{2}=8$ we obtain the parent Hamiltonian
\begin{equation}
  \label{ham.10VBS}
  H_{\bs{10}\,\text{VBS}}
  =\sum_{i=1}^N\,\left(\bigl(\bs{J}_i\bs{J}_{i+1}\bigr)^2
    +5\,\bs{J}_i\bs{J}_{i+1}+6\right),
\end{equation}
the operators $J_i^a$, $a=1,\ldots,8$, are now $10\times 10$ matrices,
and we have used $\bs{J}^2_i=6$.  $H_{\bs{10}\,\text{VBS}}$ is
positive semi-definite and annihilates the $\bs{10}$ VBS state
\eqref{state.10VBS}.  We assume that \eqref{state.10VBS} is the only
ground state of \eqref{ham.10VBS}.

The Hamiltonian \eqref{ham.10VBS} provides the equivalent of the AKLT
model~\cite{affleck-87prl799,affleck-88cmp477}, whose unique ground state is
constructed from dimer states by projection onto spin 1, for SU(3)
spin chains.  Note that as in the case of SU(2), it is sufficient to
consider linear and quadratic powers 
of the total spin of only two neighboring
sites.  This is a general feature of the corresponding SU($n$) models,
as we will elaborate in the following section.

Since the $\bs{10}$ VBS state \eqref{state.10VBS} is unique, we cannot
have domain walls connecting different ground states.  We hence expect
the coloron and anti-coloron excitations to be confined in pairs, as
illustrated below.  The state between the excitations is no longer
annihilated by \eqref{ham.10VBS}, as there are pairs of neighboring
sites containing higher-dimensional representations, as indicated by
the dotted box below.  As the number of such pairs increases linearly
with the distance between the excitation, the confinement potential
depends linearly on this distance.
\begin{equation}
\setlength{\unitlength}{1pt}
\begin{picture}(210,70)(0,0)
\linethickness{0.8pt}
\multiput(0,45)(14,0){15}{\circle{4}}
\multiput(42,45)(14,0){2}{\circle*{4}}
\multiput(154,45)(14,0){1}{\circle*{4}}
\multiput(14,35)(14,0){15}{\circle{4}}
\multiput(28,25)(14,0){15}{\circle{4}}
\multiput(182,25)(14,0){3}{\circle{4}}
\thicklines
\put(2,45){\line(1,0){10}}\put(16,45){\line(1,0){10}}
\put(44,45){\line(1,0){10}}\put(72,45){\line(1,0){10}}
\put(86,45){\line(1,0){10}}\put(114,45){\line(1,0){10}}
\put(128,45){\line(1,0){10}}\put(170,45){\line(1,0){10}}
\put(184,45){\line(1,0){10}}
\put(16,35){\line(1,0){10}}\put(30,35){\line(1,0){10}}
\put(58,35){\line(1,0){10}}\put(72,35){\line(1,0){10}}
\put(100,35){\line(1,0){10}}\put(114,35){\line(1,0){10}}
\put(142,35){\line(1,0){10}}\put(156,35){\line(1,0){10}}
\put(184,35){\line(1,0){10}}\put(198,35){\line(1,0){10}}
\put(30,25){\line(1,0){10}}\put(44,25){\line(1,0){10}}
\put(72,25){\line(1,0){10}}\put(86,25){\line(1,0){10}}
\put(114,25){\line(1,0){10}}\put(128,25){\line(1,0){10}}
\put(156,25){\line(1,0){10}}\put(170,25){\line(1,0){10}}
\put(198,25){\line(1,0){10}}\put(212,25){\line(1,0){10}}
\thinlines
\put(105,11){\vector(-1,0){49}}
\put(105,11){\vector(1,0){49}}
\put(105,0){\makebox(0,0){\small energy cost $\propto$ distance}}
\put(49,56){\makebox(0,0){$\bs{\bar 3}$}}
\put(154,56){\makebox(0,0){$\bs{3}$}}
\put(49,67){\makebox(0,0){\small coloron}}
\put(154,67){\makebox(0,0){\small anti-coloron}}
\put(93,20){\dashbox{1}(24,30)}
\end{picture}\label{10VBSstate.exc}
\end{equation}
In principle, it would also be possible to create three colorons (or
three anti-colorons) rather than a coloron--anti-coloron pair, but as
all three excitations would feel strong confinement forces, we expect
the coloron--anti-coloron pair to constitute the dominant low energy
excitation.  The confinement force between the pair induces a
linear oscillator potential for the relative motion of the
constituents.  The zero-point energy of this oscillator gives rise to
a Haldane-type energy gap
(see~\cite{greiter02prb134443,greiter02prb054505} for a similar
discussion in the two-leg Heisenberg ladder), which is independent of
the model specifics.  We expect this gap to be a generic feature of
rep.\ $\bs{10}$ spin chains with short-range antiferromagnetic
interactions.

\subsection{The representation 8 VBS}
\label{sec:8VBS}

To construct a representation $\bs{8}$ VBS state, consider first a chain
with alternating representations $\bs{3}$ and $\bs{\bar{3}}$ on
neighboring sites, which we combine into singlets.  This can be done in
two ways, yielding the two states  
\begin{equation}
\setlength{\unitlength}{1pt}
\begin{picture}(210,22)(0,-12)
\linethickness{0.8pt}
\multiput(5,0)(30,0){3}{\circle{4}}
\multiput(20,0)(30,0){3}{\circle{6}}
\multiput(138,0)(30,0){3}{\circle{4}}
\multiput(153,0)(30,0){3}{\circle{6}}
\thicklines
\multiput(7,0)(30,0){3}{\line(1,0){10}}
\multiput(156,0)(30,0){2}{\line(1,0){10}}
\put(136,0){\line(-1,0){4}}
\put(216,0){\line(1,0){3}}
\thinlines
\put(109,0){\makebox(0,0){and}}
\put(224,-3){\makebox(0,0){.}}
\put(35,-12){\makebox(0,0){\small $\bs{3}$}}
\put(50,-12){\makebox(0,0){\small $\bs{\bar{3}}$}}
\end{picture}
\nonumber
\end{equation}
We then combine a $\bs{3}$--$\bs{\bar{3}}$ state with an identical one
shifted by one lattice spacing.  This yields representations
$\bs{3}\otimes\bs{\bar{3}}=\bs{1}\oplus\bs{8}$ at each site.  The
$\bs{8}$ VBS state is obtained by projecting onto 
representation $\bs{8}$.  Corresponding to the two
$\bs{3}$--$\bs{\bar{3}}$ states illustrated above, we obtain two
linearly independent $\bs{8}$ VBS states, $\Psi^\text{L}$ and
$\Psi^\text{R}$, which may be visualized as
\begin{equation}
\setlength{\unitlength}{1pt}
\begin{picture}(210,60)(0,-24)
\linethickness{0.8pt}
\multiput(5,10)(30,0){3}{\circle{4}}
\multiput(20,10)(30,0){3}{\circle{6}}
\multiput(138,10)(30,0){3}{\circle{4}}
\multiput(153,10)(30,0){3}{\circle{6}}
\multiput(5,0)(30,0){3}{\circle{6}}
\multiput(20,0)(30,0){3}{\circle{4}}
\multiput(138,0)(30,0){3}{\circle{6}}
\multiput(153,0)(30,0){3}{\circle{4}}
\thicklines
\multiput(7,10)(30,0){3}{\line(1,0){10}}
\multiput(156,10)(30,0){2}{\line(1,0){10}}
\multiput(22,0)(30,0){2}{\line(1,0){10}}
\multiput(141,0)(30,0){3}{\line(1,0){10}}
\put(136,10){\line(-1,0){4}}
\put(216,10){\line(1,0){3}}
\put(2,0){\line(-1,0){3}}
\put(82,0){\line(1,0){4}}
\thinlines
\put(109,5){\makebox(0,0){and}}
\put(224,2){\makebox(0,0){.}}
\put(44,-6){\framebox(12,22)}
\put(50,-6){\line(0,-1){9}}
\put(50,-22){\makebox(1,1){\small projection onto $\bs{8}=(1,1)$}}
\put(50,25){\makebox(1,1){\small one site}}
\put(162.5,-6){\dashbox{2}(26,22)}
\end{picture}
\label{fig:8VBS}
\end{equation}
These states transform into each other under space reflection or color 
conjugation (interchange of $\bs{3}$ and $\bs{\bar{3}}$).

It is convenient to formulate the corresponding state vectors as a
matrix product.  Taking (\b,\r,\g ) and (\y,\c,\m ) as bases for the
reps.\ $\bs{3}$ and $\bs{\bar{3}}$, respectively, the singlet bonds in
$\Psi^\text{L}$ above can be written
\begin{multline}
  \Bigl(\ket{\b}_i \ket{\y}_{i+1} + \ket{\r}_i \ket{\c}_{i+1}
  + \ket{\g}_i \ket{\m}_{i+1}\Bigr)\\ =
  \Bigl(\ket{\b}_i,\ket{\r}_i,\ket{\g}_i\Bigr)
  \left(\!
    \begin{array}{c} \ket{\y}_{i+1}\\\ket{\c}_{i+1}\\\ket{\m}_{i+1}\end{array}
  \!\right)\!.
\nonumber
\end{multline}
We are hence led to consider matrices composed of the outer product
of these vectors on each lattice site,
\begin{displaymath}
  M_i^{\bs{1}\oplus\bs{8}} 
  = \left(\!
    \begin{array}{c} \ket{\y}_{i}\\\ket{\c}_{i}\\\ket{\m}_{i}\end{array}
    \!\right)\!
  \Bigl(\ket{\b}_i,\ket{\r}_i,\ket{\g}_i\Bigr).
\end{displaymath}\\
In the case of the AKLT model reviewed above, the Schwinger bosons take
care of the projection automatically, and we can simply assemble
these matrices into a product state.  For the $\bs{8}$ VBS,
however, we need to enforce the projection explicitly.  This is
most elegantly accomplished using the Gell-Mann matrices, yielding the
projected matrix
\begin{equation}
  \label{eq:m8}
  M_i
  =\frac{1}{2}\sum_{a=1}^8 \lambda^a\, 
  \text{tr}\Bigl(\lambda^a M_i^{\bs{1}\oplus\bs{8}}\Bigr).
\end{equation}
Here we have simply used the fact that the eight Gell-Mann matrices,
supplemented by the unitary matrix, constitute a complete basis for
the space of all complex $3\times 3$ matrices.  By omitting the unit
matrix in the expansion \eqref{eq:m8}, we effectively project out
the singlet state.  
Written out explicitly, we obtain 
\begin{widetext}
\begin{equation}
  \label{eq:m8explictly}
  M_i
  =\left(\!\begin{array}{ccc} 
  \frac{2}{3}\ket{\b\y}_i-\frac{1}{3}\ket{\r\c}_i-\frac{1}{3}\ket{\g\m}_i
  &\ket{\r\y}_i&\ket{\g\y}_i\\[4pt]
  \ket{\b\c}_i
  &-\frac{1}{3}\ket{\b\y}_i+\frac{2}{3}\ket{\r\c}_i-\frac{1}{3}\ket{\g\m}_i
  &\ket{\g\c}_i\\[4pt]
  \ket{\b\m}_i&\ket{\r\m}_i
  &-\frac{1}{3}\ket{\b\y}_i-\frac{1}{3}\ket{\r\c}_i+\frac{2}{3}\ket{\g\m}_i 
  \\[4pt]    
    \end{array}\!\right).
\end{equation}
\end{widetext}
Assuming PBCs, the $\bs{8}$ VBS state $\Psi^\text{L}$ (illustrated in
on the left \eqref{fig:8VBS}) is hence given by the trace of the
matrix product
\begin{equation}
  \label{eq:8VBSL}
  \ket{\psi_{\bs{8}\,\text{VBS}}^\text{L}}
  =\text{tr}\biggl( \prod_i M_i
  \biggr).
\end{equation}
To obtain the state $\Psi^\text{R}$ (illustrated on the right in
\eqref{fig:8VBS}) we simply have to transpose the matrices in the
product,
\begin{equation}
  \label{eq:8VBSR}
  \ket{\psi_{\bs{8}\,\text{VBS}}^\text{R}}
  =\text{tr}\biggl( \prod_i M_i
  ^\text{T} \biggr).
\end{equation}

Let us now formulate a parent Hamiltonian for these states.
If we consider two lattice sites on an SU(3) chain with a representation
$\bs{8}$ on each lattice site in general, we find the full SU(3) content
\begin{equation}
\bs{8}\,\otimes\,\bs{8}\,=\,\bs{1}\,\oplus\,2\cdot\bs{8}\,
\oplus\,\bs{10}\,\oplus\,\bs{\overline{10}}\,\oplus\,\bs{27}
\end{equation}
with $\bs{10}=(3,0)$, $\bs{\overline{10}}=(0,3)$, and $\bs{27}=(2,2)$.
On the other hand, for the $\bs{8}$ VBS states only the
representations $\bs{3}\otimes\bs{\bar 3}\,=\,\bs{1}\oplus\bs{8}$ can
occur for the total spin of two neighboring sites, as the two sites
always contain one singlet (see dashed box in \eqref{fig:8VBS} on the
right above).  With the Casimirs $\casi{0}{0}=0$ and $\casi{1}{1}=3$
for representations $\bs{1}$ and $\bs{8}$, respectively, we construct
the parent Hamiltonian
\begin{equation}
\label{ham.8VBS}
H_{\bs{8}\,\text{VBS}}=\sum_{i=1}^N\left(\bigl(\bs{J}_i\bs{J}_{i+1}\bigr)^2+
\frac{9}{2}\,\bs{J}_i\bs{J}_{i+1} + \frac{9}{2}\right),
\end{equation}
where the operators $J_i^a$, $a=1,\ldots,8$, are now $8\times 8$
matrices, and we have used the Casimir $\bs{J}_i^2=3$ on each site.
$H_{\bs{8}\,\text{VBS}}$ is positive semi-definite, and annihilates the
states $\Psi^\text{L}$ and $\Psi^\text{R}$.  We have numerically
verified for chains with $N=3$, 4, 5, and 6 lattice sites that
$\Psi^\text{L}$ and $\Psi^\text{R}$ are the only ground states of
\eqref{ham.8VBS}.

Naively, one might assume the $\bs{8}$ VBS model to support deconfined
spinons or colorons, which correspond to domain walls between the
two ground states $\Psi^\text{L}$ and $\Psi^\text{R}$.  A closer look
at the domain walls, however, shows that this is highly unlikely,
as each domain wall is a bound state of either two anti-colorons or two
colorons, as illustrated below.
\begin{equation}
\setlength{\unitlength}{1pt}
\begin{picture}(210,64)(0,-24)
\linethickness{0.8pt}
\multiput(5,10)(30,0){7}{\circle{4}}
\multiput(20,10)(30,0){7}{\circle{6}}
\multiput(35,0)(30,0){7}{\circle{6}}
\multiput(20,0)(30,0){7}{\circle{4}}
\multiput(65,10)(30,0){1}{\circle*{4}}
\multiput(50,0)(30,0){1}{\circle*{4}}
\multiput(170,10)(30,0){1}{\circle*{6}}
\multiput(155,0)(30,0){1}{\circle*{6}}
\thicklines
\multiput(7,10)(30,0){2}{\line(1,0){10}}
\multiput(83,10)(30,0){3}{\line(1,0){10}}
\multiput(187,10)(30,0){1}{\line(1,0){10}}
\multiput(22,0)(30,0){1}{\line(1,0){10}}
\multiput(68,0)(30,0){3}{\line(1,0){10}}
\multiput(172,0)(30,0){2}{\line(1,0){10}}
\thinlines
\put(22,-15){\makebox(0,0){\small $\Psi^\text{L}$}}
\put(112,-15){\makebox(0,0){\small $\Psi^\text{R}$}}
\put(202,-15){\makebox(0,0){\small $\Psi^\text{L}$}}
\put(50,22){\makebox(0,0){\small $\bs{3}$}}
\put(65,22){\makebox(0,0){\small $\bs{3}$}}
\put(155,22){\makebox(0,0){\small $\bs{\bar{3}}$}}
\put(170,22){\makebox(0,0){\small $\bs{\bar{3}}$}}
\put(57.5,34){\makebox(0,0){\small anti-colorons}}
\put(162.5,34){\makebox(0,0){\small colorons}}
\end{picture}
\label{fig:8VBSdomainwall}
\end{equation}
There is no reason to assume that the domain wall depicted above as
two anti-colorons in fact corresponds to a single coloron, as it
appears to be the case for the trimer chain.  There we created a
domain wall corresponding to a single coloron by removing one of the
rep.\ $\bs{3}$ spins from a trimer, leaving the remaining rep.\
$\bs{3}$ spins coupled antisymmetrically as in the ground state.
If we were to combine the two reps.\ $\bs{{3}}$ into a rep.\
$\bs{\bar{3}}$ in \eqref{fig:8VBSdomainwall}, we would not reproduce a
correlation present in the ground state, but enforce a new
correlation.  The correct interpretation of the domain wall between
$\Psi^\text{L}$ and $\Psi^\text{R}$ is hence that of a bound state between
two linearly confined anti-colorons.  The origin of the confining
potential is illustrated below.
\begin{equation}
\setlength{\unitlength}{1pt}
\begin{picture}(175,70)(0,-30)
\linethickness{0.8pt}
\multiput(5,10)(30,0){6}{\circle{4}}
\multiput(20,10)(30,0){5}{\circle{6}}
\multiput(35,0)(30,0){5}{\circle{6}}
\multiput(20,0)(30,0){6}{\circle{4}}
\multiput(125,10)(30,0){1}{\circle*{4}}
\multiput(50,0)(30,0){1}{\circle*{4}}
\thicklines
\multiput(7,10)(30,0){4}{\line(1,0){10}}
\multiput(143,10)(30,0){1}{\line(1,0){10}}
\multiput(22,0)(30,0){1}{\line(1,0){10}}
\multiput(68,0)(30,0){4}{\line(1,0){10}}
\thinlines
\put(22,-15){\makebox(0,0){\small $\Psi^\text{L}$}}
\put(157,-15){\makebox(0,0){\small $\Psi^\text{R}$}}
\put(50,23){\makebox(0,0){\small $\bs{3}$}}
\put(125,23){\makebox(0,0){\small $\bs{3}$}}
\put(50,34){\makebox(0,0){\small anti-coloron}}
\put(125,34){\makebox(0,0){\small anti-coloron}}
\put(87.5,-14){\vector(-1,0){37.5}}
\put(87.5,-14){\vector(1,0){37.5}}
\put(87.5,-25){\makebox(0,0){\small energy cost $\propto$ distance}}
\end{picture}
\label{fig:8VBS3bar3bar}
\end{equation}
As in the $\bs{10}$ VBS, the confinement induces a linear oscillator
potential for the relative motion of the anti-colorons. The zero-point
energy of this oscillator corresponds to a Haldane-type gap in the
spectrum.  The ground state
wave function of the oscillator is symmetric, and hence corresponds to
a symmetric combination of $\bs{3}\otimes\bs{3}$, \ie rep.\ $\bs{6}$.
The antisymmetric combination $\bs{\bar{3}}$ corresponds to the first
excited state of the oscillator, which we expect to cost more than
twice the energy of the symmetric state~\cite{greiter02prb054505}.
This statement holds for the pair of colorons in
\eqref{fig:8VBSdomainwall} as well.

\begin{figure}[t]
\includegraphics[angle=270,scale=0.53]{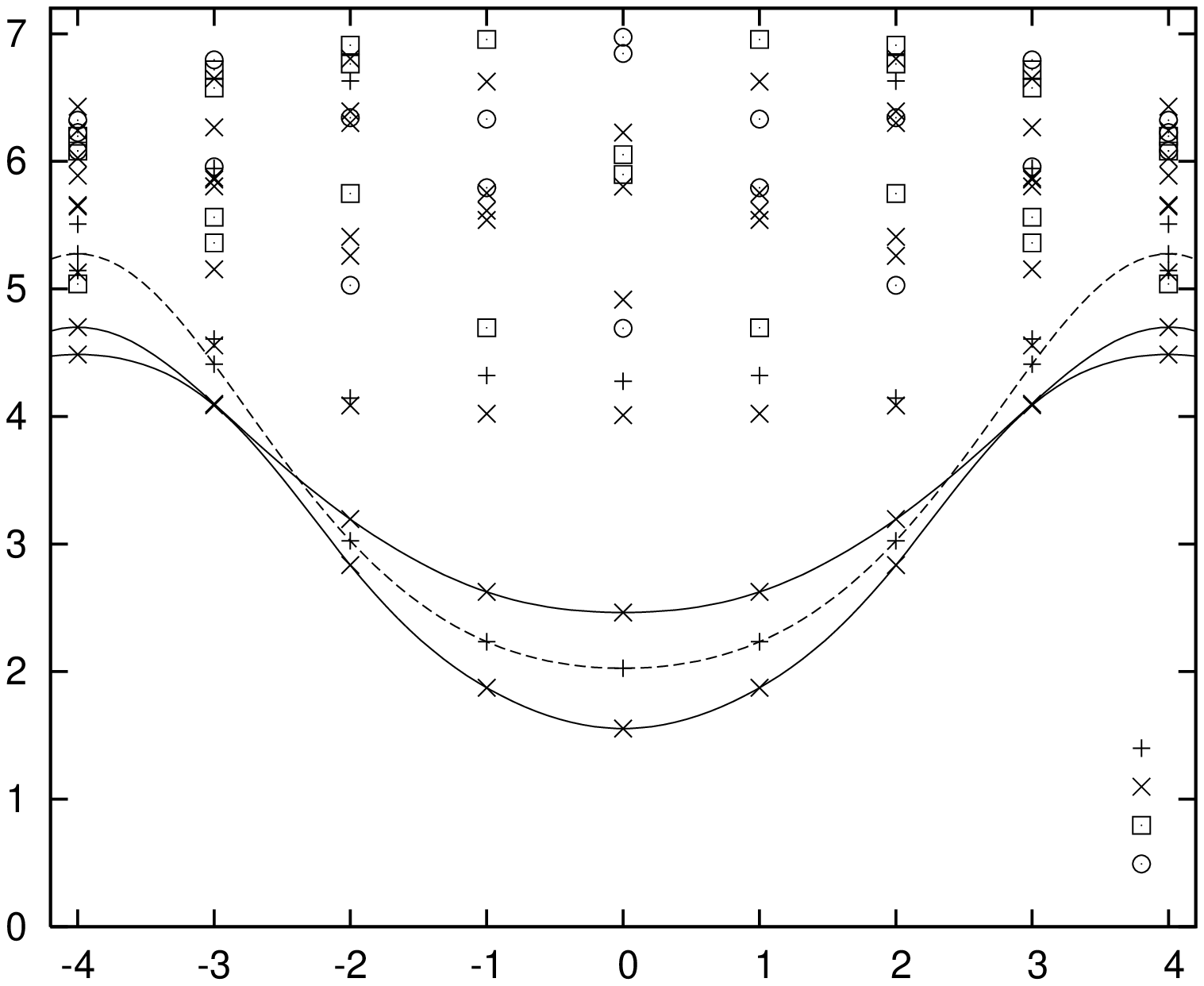}
\begin{picture}(0,0)
\put(13,5){\makebox(0,0){$k~[2\pi/N]$}}
\put(-120,110){\rotatebox{90}{$E_{\bs{8}\,\text{VBS}}$}}
\put(105,61){\makebox(0,0)[r]{{\scriptsize singlet}}}
\put(105,53){\makebox(0,0)[r]{{\scriptsize rep.\,$\bs{8}$}}}
\put(105,46){\makebox(0,0)[r]{{\scriptsize rep.\,$\bs{10}$}}}
\put(105,38){\makebox(0,0)[r]{{\scriptsize rep.\,$\bs{27}$}}}
\end{picture}
\caption{
  Spectrum of the $\bs{8}$ VBS Hamiltonian \eqref{ham.8VBS} for $N=8$
  sites obtained by exact diagonalization.  (The lines are merely
  guides to the eye.)  The ``magnon'' excitation transforming under
  rep.\ $\bs{8}$ of SU(3) has the lowest energy, followed by a singlet
  excitation, as expected from the discussion in the text.  The well
  defined modes at low energies provide strong evidence of
  coloron--anti-coloron bound states as compared to deconfined domain
  walls, and hence support our conclusion that the $\bs{8}$ VBS
  exhibits a Haldane gap due to spinon confinement.}
\label{fig:8VBSspec}
\end{figure}

The domain walls, however, are not the only low energy excitations.
In either of the ground states, we can create coloron--anti-coloron
bound states, which make no reference to the other ground state, as
illustrated below.
\begin{equation}
\setlength{\unitlength}{1pt}
\begin{picture}(175,70)(0,-30)
\linethickness{0.8pt}
\multiput(5,10)(30,0){6}{\circle{4}}
\multiput(20,10)(30,0){6}{\circle{6}}
\multiput(35,0)(30,0){5}{\circle{6}}
\multiput(20,0)(30,0){5}{\circle{4}}
\multiput(125,0)(30,0){1}{\circle*{6}}
\multiput(50,0)(30,0){1}{\circle*{4}}

\thicklines
\multiput(7,10)(30,0){6}{\line(1,0){10}}
\multiput(142,0)(30,0){1}{\line(1,0){10}}
\multiput(22,0)(30,0){1}{\line(1,0){10}}
\multiput(68,0)(30,0){2}{\line(1,0){10}}
\thinlines
\put(22,-15){\makebox(0,0){\small $\Psi^\text{L}$}}
\put(157,-15){\makebox(0,0){\small $\Psi^\text{L}$}}
\put(50,23){\makebox(0,0){\small $\bs{3}$}}
\put(125,23){\makebox(0,0){\small $\bs{\bar{3}}$}}
\put(50,34){\makebox(0,0){\small anti-coloron}}
\put(125,34){\makebox(0,0){\small coloron}}
\put(87.5,-14){\vector(-1,0){37.5}}
\put(87.5,-14){\vector(1,0){37.5}}
\put(87.5,-25){\makebox(0,0){\small energy cost $\propto$ distance}}
\end{picture}
\label{fig:8VBS33bar}
\end{equation}
The oscillator model tells us again that the ``symmetric'' combination
of $\bs{3}\otimes\bs{\bar{3}}$, \ie rep.\ $\bs{8}$, has the lowest
energy, which we expect to be comparable, if not identical, to the
energy required to create each of the domain walls above.  
The 
singlet $\bs{1}$ will have an energy comparable to that of a domain
wall transforming under either $\bs{\bar{3}}$ or $\bs{3}$.  In any
event, we expect the $\bs{8}$ VBS model to display a Haldane gap due to
coloron confinement.

The excitation spectrum of \eqref{ham.8VBS} for a chain with $N=8$
sites and PBCs is shown in Fig.~\ref{fig:8VBSspec}.  The spectrum shows
that the lowest excitation transforms under rep.~$\bs{8}$, as expected
from \eqref{fig:8VBS33bar}, with a singlet and then another rep~$\bs{8}$
following at slightly higher energies.  It is tempting to interpret 
those three levels as the lowest levels of the coloron--anti-coloron
oscillator \eqref{fig:8VBS33bar}, but then there should be another 
singlet at a comparable spacing above.  The fact that the spacings
between these excitations are significantly smaller than the energy
of the first exited state, however, would be consistent with such an 
interpretation, as the spinons in VBS models always have a local energy 
cost associated with their creation, which is specific to these models
and not related to the universal Haldane gap stemming from confinement
forces.

Most importantly, the spectrum provides strong evidence in favor of
our assumption that the domain walls are not elementary excitations,
but bound states of either two colorons or two anti-colorons, and
hence that the lowest energy excitations of finite chains are
coloron--anti-coloron bound states as illustrated in
\eqref{fig:8VBS33bar}.  The assumption is crucial for our conclusion
that the model exhibits a Haldane gap.  If the low energy sector of
the model was determined by two deconfined domain walls, we would see
a continuum of states in the spectrum, similar to the spectrum seen in
spin $S=\frac{1}{2}$ chains of SU(2).  The well defined low-energy
modes in Fig.~\ref{fig:8VBSspec}, however, look much more like the
spinon--spinon bound state excitations seen in $S=1$ chains
or two-leg $S=\frac{1}{2}$ Heisenberg ladders.  In particular, if we
assume that the individual domain walls transform under reps.\
$\bs{6}$ and $\bs{\bar{6}}$, we expect excitations transforming under
the representations contained in
$\bs{6}\otimes\bs{\bar{6}}=\bs{1}\oplus\bs{8}\oplus\bs{27}$ to be
approximately degenerate.  Fig.~\ref{fig:8VBSspec} shows clearly that
such a multiplet is not present a the lowest energies.

\section{SU($\bs{n}$) models}
\label{sec:sun}

In this section, we generalize three of the models proposed for SU(3)
spin chains, the trimer model, the symmetric representation $\bs{10}$ VBS, 
and the matrix product state $\bs{8}$ VBS to the case of 
SU($n$) spin chains.

\subsection{The $\bs{n}$-mer model}

Consider an  SU($n$) spin chain with $N$ sites, where $N$ is a multiple of $n$,
with a spin transforming according 
to the fundamental representation $\bs{n}$ of  SU($n$) at each lattice site,
\begin{equation}
\begin{array}{rcccccccl}
\bs{n}&=&(1,0,\ldots,0)&=&
\setlength{\unitlength}{8pt}
\begin{picture}(1,1)(0,0.1)
\put(0,1){\line(1,0){1}}
\put(0,0){\line(1,0){1}}
\put(0,0){\line(0,1){1}}
\put(1,0){\line(0,1){1}}
\end{picture}
&\hat =&c_{\sigma}^\dagger\vac, 
\end{array}
\end{equation}
where $\sigma$ denotes a ``flavor'', $\sigma\in\{{\rm f}_1,\ldots,{\rm
  f}_n\}$, and $c_{\sigma}^\dagger$ creates a fermion of flavor
$\sigma$.
  
The SU($n$) generators at site $i$ are in analogy to
\eqref{eq:s} and \eqref{eq:J_a} defined as
\begin{equation}
J^a_i=\frac{1}{2}\,\sum_{\sigma,\sigma'={\rm f}_1,\ldots,{\rm f}_n}\,
c_{i\sigma}^{\dagger} V^a_{\sigma\sigma'}
c_{i\sigma'}^{\phantom{\dagger}}, \quad a=1,\ldots,n^2-1,
\label{eq:J_an}
\end{equation}
where the $V^a$ denote the $n^2\!-\!1$ SU($n$) Gell-Mann
matrices~\cite{macfarlane-68cmp77}.  The generators are normalized
through the eigenvalue the quadratic Casimir operator takes in the
adjoint representation, $\bs{J}^2=\casin{1,0,\ldots,0,1}=n$.

To determine the eigenvalues of the quadratic Casimir for general
representations of SU($n$), a significant amount of representation
theory is required~\cite{Humphreys87}.  We content ourselves here
by providing the formulas up to $n=6$ in App.~\ref{quadraticcasimirs}.

In analogy to the trimer states \eqref{eq:trimer}, we construct 
the $n$-mer states of an SU($n$) spin chain 
by combining sets of $n$ neighboring spins into a singlet,
\begin{equation}
  \label{eq:nmer}
  \ket{\psi_{n\text{-mer}}^{(\mu)}\!}=\hspace{-15pt}\prod_{\scriptstyle{i} \atop 
    \left(\scriptstyle{\frac{i-\mu}{n}\,{\rm integer}}\right)}
  \hspace{-5pt}\Bigl(
  \sum_{\scriptstyle{(\sigma_1,\ldots,\sigma_n)}=
    \atop\scriptstyle{{\pi}({\rm f}_1,\ldots,{\rm f}_n)}} 
  \hspace{-10pt}\hbox{sign}({\pi})\,
  \prod_{\kappa=1}^n c^{\dagger}_{i-1+\kappa\;\sigma_\kappa}
  \Bigr)\!\vac\!,
\end{equation}
where $\mu=1,\ldots,n$ labels the $n$ degenerate ground states, and
$i$ runs over the lattice sites subject to the constraint that
$\frac{i-\mu}{n}$ is integer.  The sum extends over all $n!$
permutations $\pi$ of the $n$ flavors ${\rm f}_1,\ldots,{\rm f}_n$.

In order to identify a parent Hamiltonian, consider the total SU($n$)
spin on $n+1$ neighboring sites for the $n$-mer states.  Following the
rules of combining representations labeled by Young tableaux (see \eg
\cite{Cornwell84vol2,Georgi82}), it is not difficult to see that the
total spin will only contain representations given by tableaux with
$n+1$ boxes and two columns, \ie tableaux of the form
\begin{displaymath}
\setlength{\unitlength}{8pt}
\begin{picture}(5,7.5)
\put(0,0){\line(0,1){7.5}}
\put(1,0){\line(0,1){7.5}}
\put(2,3.5){\line(0,1){4}}
\put(0,0){\line(1,0){1}}
\put(0,1){\line(1,0){1}}
\put(0,2.5){\line(1,0){1}}
\put(0,3.5){\line(1,0){2}}
\put(0,4.5){\line(1,0){2}}
\put(0,6.5){\line(1,0){2}}
\put(0,7.5){\line(1,0){2}}
\put(4.5,5.5){\makebox(0,0)
{$\left.\begin{array}{c}~\\~\end{array}\right\}{\small ~\nu~\text{rows}}$}}
\end{picture}
\end{displaymath}
with $1\le\nu\le\frac{n+1}{2}$.  The 
eigenvalues of the quadratic Casimir operator for these
representations are
\begin{equation}
f_n(\nu)=\frac{1}{2n}\Bigl(n^2(2\nu -1) -2n(\nu -1)^2 -1 \Bigr).
\end{equation}
An educated guess for a parent Hamiltonian for the $n$-mer chain
hence appears to be
\begin{equation}
H_{\text{trial}}=\sum_{i=1}^N H_i\quad
\label{eq:MGSUni}
\end{equation}
with
\begin{equation}
H_i=\prod_{\nu=1}^{\lfloor\frac{n+1}{2}\rfloor}
\left(\!\left(\bs{J}_i^{(n+1)}\right)^2 - f_n(\nu)\right),
\label{eq:MGSUn}
\end{equation}
where $\lfloor \; \rfloor$ denotes the floor function, \ie $\lfloor
x\rfloor$ is the largest integer $l\le x$, and we use the notation
introduced in \eqref{eq:defJnu}.  

This construction yields the MG model~\cite{majumdar-69jmp1399} for
SU(2), the trimer model \eqref{ham.trimer} for SU(3), and a valid
parent Hamiltonian for the four degenerate 4-mer states for SU(4).
For $n\ge 5$, however, the decomposition of the tensor product
$\bs{n}^{\otimes (n+1)}$ contains irreducible representations
corresponding to Young tableaux with more than two columns, whose
Casimirs are equal or smaller than a number of Casimirs included in
the list $f_n(\nu)$, $\nu=1,2,\ldots,\lfloor\frac{n+1}{2}\rfloor$.  If
the Casimir of such an ``undesired'' representation not included in
the list is smaller than an odd number of Casimirs included in the
list, we obtain negative eigenvalues for $H_i$, and it is not a priori
clear any more that the Hamiltonian \eqref{eq:MGSUni} is positive
semi-definite.  An obvious cure to this problem is to write
\begin{equation}
  H_{n\text{-mer}}=\sum_{i=1}^N H_i^2,\quad
  \label{eq:hisquare}
\end{equation}
with $H_i$ as in \eqref{eq:MGSUn}.  This does, however, not cure
potential problems arising from undesired representations which 
share the eigenvalues of the Casimir with one of the
representations from the list, as it happens to be the case for $n=5$.
The Hamiltonian \eqref{eq:hisquare} likewise annihilates these
representations, giving rise to a remote possibility that the $n$-mer
states \eqref{eq:nmer} are not the only ground states of
\eqref{eq:hisquare}.  The potential relevance of these problems has to
be investigated for each $n$ separately.

\subsection{The representation $\bs{(n,0,\ldots,0)}$ VBS}

As a generalization of the AKLT model for SU(2) and the ${\bs{10}}$ VBS
model for SU(3) discussed above, we now consider a VBS for an SU($n$) chain
of spins transforming under the symmetric representation
\begin{equation}
  \label{eq:n000}
  \begin{array}{rcccl}
    (n,0,\ldots,0)&=&
    \setlength{\unitlength}{8pt}
    \begin{picture}(5,2)(0,3.15)
      \put(0,4){\line(1,0){5}}
      \put(0,3){\line(1,0){5}}
      \put(0,3){\line(0,1){1}}
      \put(1,3){\line(0,1){1}}
      \put(4,3){\line(0,1){1}}
      \put(5,3){\line(0,1){1}}
      \put(2.5,1.5){\makebox(0,0)%
        {$\underbrace{\quad\text{~~~}\qquad}_{\text{\small $n$~boxes}}$}}
   \end{picture}
    &\hat =& b_{\sigma_1}^\dagger b_{\sigma_2}^\dagger\ldots b_{\sigma_n}^\dagger
    \vac,\\[2pt] \nonumber
\end{array}
\end{equation}
where each $b_{\sigma}^\dagger$, $\sigma\in\{{\rm f}_1,\ldots,{\rm
  f}_n\}$, is an SU($n$) Schwinger boson.  The VBS state is obtained
by combining $n$ $n$-mer states \eqref{eq:nmer}, one for each
$\mu=1,\dots,n$, in that the total spin on each lattice site is
projected onto the symmetric representation $(n,0,\ldots,0)$.  This
yields
\begin{equation}
  \label{eq:n000VBS}
  \ket{\psi_{(n,0,\ldots,0)\, \text{VBS}}}=
  \prod_{i} \Bigl(\!\!
  \sum_{\scriptstyle{(\sigma_1,\ldots,\sigma_n)}=
    \atop\scriptstyle{{\pi}({\rm f}_1,\ldots,{\rm f}_n)}} 
  \hspace{-12pt}\hbox{sign}({\pi})
  \prod_{\kappa=1}^n b^{\dagger}_{i-1+\kappa,\sigma_\kappa}
  \Bigr)\!\vac\!.
\end{equation}

Let us now construct a parent Hamiltonian for the symmetric VBS
\eqref{eq:n000VBS}.  The total SU($n$) spin of two neighboring sites
of a representation $(n,0,\ldots,0)$ spin chain in general
contains all the representations corresponding to Young tableaux 
with $2n$ boxes and at most two rows, \ie all tableaux of the form
\begin{displaymath}
\setlength{\unitlength}{8pt}
\begin{picture}(9,5)(0,1.5)
\put(0,6){\line(1,0){9}}
\put(0,5){\line(1,0){9}}
\put(0,4){\line(1,0){5}}
\put(0,4){\line(0,1){2}}
\put(1,4){\line(0,1){2}}
\put(4,4){\line(0,1){2}}
\put(5,4){\line(0,1){2}}
\put(6,5){\line(0,1){1}}
\put(8,5){\line(0,1){1}}
\put(9,5){\line(0,1){1}}
\put(2.5,2.5){\makebox(0,0)%
{$\underbrace{\quad\text{~~~}\qquad}_{\text{\small $n-\nu$~columns}\qquad\ }$}}
\put(7,2.5){\makebox(0,0)%
{$\underbrace{\quad\text{~\,~}\quad}_{\quad\text{\small $2\nu$ boxes}}$}}
\end{picture}
\end{displaymath}
The eigenvalues of the quadratic Casimir operator for these
representations are given by
\begin{equation}
  \label{eva:aklt}
  g_n(\nu)\equiv \casin{2\nu,n\!-\!\nu,0,\dots,0}=2n^2-4n+\nu(\nu+1). 
\end{equation}

On the other hand, the total SU($n$) spin of two neighboring sites of
the representation $(n,0,\ldots,0)$ VBS \eqref{eq:n000VBS} has to be
contained in the product
\begin{equation}
  \label{eq:sunprod}
  \setlength{\unitlength}{8pt} 
  \begin{picture}(1.2,1.7)(0,0.6)
    \multiput(0,0)(1,0){2}{\line(0,1){2}}
    \multiput(0,0)(0,1){3}{\line(1,0){1}}
  \end{picture}
  ^{\otimes n-1}\otimes\
  \begin{picture}(1,1)(0,0.1)
    \multiput(0,0)(1,0){2}{\line(0,1){1}}
    \multiput(0,0)(0,1){2}{\line(1,0){1}}
  \end{picture}
  \ \otimes\
  \begin{picture}(1,1)(0,0.1)
    \multiput(0,0)(1,0){2}{\line(0,1){1}}
    \multiput(0,0)(0,1){2}{\line(1,0){1}}
  \end{picture}
\end{equation}
As we project the spin on each lattice site onto the representation
$(n,0,\ldots,0)$, only two these representations remain:
\begin{displaymath}
 \setlength{\unitlength}{8pt}
\begin{picture}(19,2.5)(0,.6)
\linethickness{0.3pt}
\multiput(0,0)(1,0){2}{\line(0,1){2}}
\multiput(4,0)(1,0){2}{\line(0,1){2}}
\multiput(6,1)(1,0){2}{\line(0,1){1}}
\multiput(0,1)(0,1){2}{\line(1,0){7}}
\put(0,0){\line(1,0){5}}
\put(8,0.5){\makebox(3,1){and}}
\multiput(12,0)(1,0){2}{\line(0,1){2}}
\multiput(16,0)(1,0){3}{\line(0,1){2}}
\multiput(12,0)(0,1){3}{\line(1,0){6}}
\end{picture}
\end{displaymath}
The eigenvalues of the quadratic Casimir operator are given by
$g_n(0)=2n(n-2)$ and $g_n(1)=2(n-1)^2$, respectively.  Hence, using
$\bs{J}^2_i=n(n-1)$, we obtain the parent Hamiltonian 
\begin{equation}
  \label{ham.aklt.sun}
  \begin{split}
  &H_{{\scriptstyle (n,0,\ldots,0)}\, {\scriptstyle \text{VBS}}}\\
  &=\sum_{i=1}^N
  \Bigl(\bigl(\bs{J}_i\bs{J}_{i+1}\bigr)^2 + (2n-1)\,\bs{J}_i\bs{J}_{i+1}+
  n(n-1)\Bigr).
  \end{split}
\end{equation}
Since $g_n(0)\ge 0$ for $n\ge 2$ and $g_n(\nu)$ is a strictly
increasing function of $\nu$, the Hamiltonian \eqref{ham.aklt.sun} is
positive semi-definite.  For $n=2$, we recover the AKLT model
\eqref{eq:haklt}; for $n=3$, we recover the $\bs{10}$ VBS model
\eqref{ham.10VBS}.

\subsection{An example of a matrix product state}
\label{sec:mp}

In principle, a matrix product VBS can be formulated on all SU($n$)
chains with spins transforming under the symmetric combination of any
representation $\bs{m}$ and its conjugate representation
$\bs{\overline{m}}$.  Unless the rep.~$\bs{m}$ is self-conjugate,
we obtain two inequivalent states, which transform into each other
under space reflection.  The construction of these is analogous to the
$\bs{8}$ VBS introduced above, and likewise best illustrated as
\begin{equation}
\setlength{\unitlength}{1.333pt}
\begin{picture}(170,52)(0,-24)
\linethickness{0.8pt}
\multiput(5,10)(30,0){3}{\makebox(0,0){\small $\bs{m}$}}
\multiput(20,10.6)(30,0){3}{\makebox(0,0){\small $\bs{\overline{m}}$}}
\multiput(5,0.6)(30,0){3}{\makebox(0,0){\small $\bs{\overline{m}}$}}
\multiput(20,0)(30,0){3}{\makebox(0,0){\small $\bs{m}$}}
\multiput(128,10)(30,0){2}{\makebox(0,0){\small $\bs{m}$}}
\multiput(143,10.6)(30,0){1}{\makebox(0,0){\small $\bs{\overline{m}}$}}
\multiput(128,0.6)(30,0){2}{\makebox(0,0){\small $\bs{\overline{m}}$}}
\multiput(143,0)(30,0){1}{\makebox(0,0){\small $\bs{m}$}}
\thicklines
\multiput(9,10)(30,0){3}{\line(1,0){7}}
\multiput(24,0)(30,0){2}{\line(1,0){7}}
\multiput(147,10)(30,0){1}{\line(1,0){7}}
\multiput(132,0)(30,0){1}{\line(1,0){7}}
\put(1,0){\line(-1,0){3}}
\put(84,0){\line(1,0){3}}
\put(124,10){\line(-1,0){3}}
\put(162,0){\line(1,0){3}}
\thinlines
\put(104,5){\makebox(0,0){and}}
\put(169,2){\makebox(0,0){.}}
\put(44,-6){\framebox(12,22)}
\put(50,-6){\line(0,-1){7}}
\put(50,-20){\makebox(1,1){\small projection onto the symmetric}}
\put(50,-28){\makebox(1,1){\small combination in 
$\bs{m}\otimes\bs{\overline{m}}$}}
\put(50,23){\makebox(1,1){\small one site}}
\put(122.5,-6){\dashbox{2}(26,22)}
\end{picture}
\vspace{10pt}
\label{fig:mmbarVBS}
\end{equation}
The thick lines indicate that we combine pairs of neighboring
representations $\bs{m}$ and $\bs{\overline{m}}$ into singlets.  On
each lattice site, we project onto the symmetric combination of
$\bs{m}$ and $\bs{\overline{m}}$, as indicated.  By ``symmetric
combination'' we mean that if representations $\bs{m}$ and
$\bs{\overline{m}}$ of SU($n$) have Dynkin coordinates
$(\mu_1,\mu_2,\ldots,\mu_{n-1})$ and
$(\mu_{n-1},\mu_{n-2},\ldots,\mu_1)$, respectively, we combine them
into the representation with Dynkin coordinates
$(\mu_1\!+\!\mu_{n-1},\mu_2\!-\!\mu_{n-2},\ldots,\mu_{n-1}\!+\!\mu_1)$.
In other words, we align the columns of both tableaux horizontally,
and hence obtain a tableau with twice the width, without ever adding a
single box vertically to a column of the tableaux we started with.
The states \eqref{fig:mmbarVBS} we obtain are translationally
invariant and we expect the parent Hamiltonians to involve
nearest-neighbor interactions only.

In this subsection, we will formulate the simplest SU($n$) model
of this general family.  We take $\bs{m}$ to be the representation
formed by antisymmetrically combining $\kappa\le \frac{n}{2}$ fundamental 
representations,
\begin{displaymath}
\bs{m}\;=\;[\kappa]\;\equiv\; (0,\ldots,0,1,0,\ldots,0)\;=\;
\setlength{\unitlength}{8pt}
\begin{picture}(7,3)(0.7,5.2)
\put(1,3.5){\line(0,1){4}}
\put(2,3.5){\line(0,1){4}}
\put(1,3.5){\line(1,0){1}}
\put(1,4.5){\line(1,0){1}}
\put(1,6.5){\line(1,0){1}}
\put(1,7.5){\line(1,0){1}}
\put(5,5.5){\makebox(0,0){$\left.\begin{array}{c}~\\~\end{array}\right\}
{\small ~\kappa~\text{boxes}}$\ ,}}
\put(-7,3.6){\line(0,1){0.8}}
\put(-7,2.6){\makebox(0,0){\small $\kappa$-th entry}}
\end{picture}\\[20pt]
\end{displaymath}
which implies that we consider a model with the representation
corresponding to a Young tableaux with a column with $n-\kappa$ boxes
to the left of a column with $\kappa$ boxes at each lattice site:\\
\setlength{\unitlength}{8pt}
\begin{picture}(8,8)(-14,0)
\put(-7,3.8){$[\kappa ,n\!-\!\kappa]\;\equiv$}
\put(0,0){\line(0,1){7.5}}
\put(1,0){\line(0,1){7.5}}
\put(2,3.5){\line(0,1){4}}
\put(0,0){\line(1,0){1}}
\put(0,1){\line(1,0){1}}
\put(0,2.5){\line(1,0){1}}
\put(0,3.5){\line(1,0){2}}
\put(0,4.5){\line(1,0){2}}
\put(0,6.5){\line(1,0){2}}
\put(0,7.5){\line(1,0){2}}
\put(4.5,5.5){\makebox(0,0){$\left.\begin{array}{c}~\\~\end{array}\right\}
{\small ~\kappa~\text{rows}}$}}
\end{picture}

The construction of the parent Hamiltonian is similar to the $n$-mer
model above.  The total spin on two neighboring lattice sites can only
assume representations contained in $\bs{m}\otimes\bs{\overline{m}}$,
\ie representations corresponding to tableaux of the form
\begin{displaymath}
[\nu ,n-\nu]\;=\;
\setlength{\unitlength}{8pt}
\begin{picture}(8.5,4)(-0.3,3.9)
\put(0,0){\line(0,1){7.5}}
\put(1,0){\line(0,1){7.5}}
\put(2,4){\line(0,1){3.5}}
\put(0,0){\line(1,0){1}}
\put(0,1){\line(1,0){1}}
\put(0,3){\line(1,0){1}}
\put(0,4){\line(1,0){2}}
\put(0,5){\line(1,0){2}}
\put(0,6.5){\line(1,0){2}}
\put(0,7.5){\line(1,0){2}}
\put(4.5,5.75){\makebox(0,0)
{$\left.\begin{array}{c}~\\[-0.6pt]~\end{array}\right\}
{\small ~\nu~\text{rows}}$}}
\end{picture}\\[26pt] 
\end{displaymath}
with $0\le\nu\le\kappa$.  The eigenvalues of the quadratic
Casimir operator for these representations are
\begin{equation}
  h_n(\nu)=\nu\,(n-\nu+1).
\end{equation}
The obvious proposal for a parent Hamiltonian is hence 
\begin{equation}
H=\sum_{i=1}^N H_i, \quad 
H_i=\prod_{\nu=0}^{\lfloor\frac{n}{2}\rfloor}
\left(\!\left(\bs{J}_i^{(2)}\right)^2 - h_n(\nu)\right),
\label{eq:mmbarham}
\end{equation}
where $\lfloor x\rfloor$ denotes again the floor function.  This
Hamiltonian singles out the matrix product state \eqref{fig:mmbarVBS}
as unique ground states for $n\le 5$, but suffers from the same
shortcomings as \eqref{eq:MGSUni} with \eqref{eq:MGSUn} for $n\ge 6$.

\section{Spinon confinement and the Haldane gap}
\label{sec:conf}

\subsection{The Affleck--Lieb theorem}
\label{sec:afflecklieb}

For the generic SU(2) spin chain, Haldane~\cite{haldane83pla464,
  haldane83prl1153} identified the O(3) nonlinear sigma model as the
effective low energy field theory of SU(2) spin chains, and argued
that chains with integer spin possess a gap in the excitation spectrum,
while a topological term renders
half-integer spin chains gapless~\cite{affleck90proc,Fradkin91}.  The exact
models for SU(2) spin chains we reviewed above, the MG and the AKLT
chain, serve a paradigms to illustrate the general principle.
Unfortunately, the effective field theory description of Haldane
yielding the distinction between gapless half integer spin chains with
deconfined spinons and gapped integer spin chains with confined
spinons cannot be directly generalized to SU($n$) chains, as there is
no direct equivalent of the CP$^1$ representation used in Haldane's
analysis.

Nonetheless, there is a rigorous and rather general result for
antiferromagnetic chains of SU($n$) spins transforming under a
representation corresponding to a Young tableau with a number of boxes
not divisible by $n$: Affleck and Lieb~\cite{affleck-86lmp57} showed
that if the ground state is non-degenerate, and the Hamiltonian
consists nearest-neighbor interactions only, than the gap in the
excitation spectrum vanishes as $1/N$ (where $N$ is the number of
sites) in the thermodynamic limit.  This result is fully consistent
with the picture suggested by the models introduced above.  Like the
MG model, the trimer model and the representation $\bs{6}$ VBS have
degenerate ground states and interactions which extend beyond the
nearest neighbor, which implies that the theorem is not directly
applicable.

On physical grounds, however, the statement that a given model is
gapless (\ie the excitation gap vanishes in the thermodynamic limit)
implies that the spinons are deconfined.  The reason is simply that if
there was a confinement force between them, the zero-point energy
associated with the quantum mechanical oscillator of the relative
motion between the spinons would inevitably yield an energy gap.  The
MG, the trimer and the $\bs{6}$ VBS model constitute pedagogically valuable
paradigms of deconfined spinons.  Since the excitations in these
models are literally domain walls between different ground states, no
long-range forces can exist between them.

\subsection{A general criterion for spinon confinement}
\label{sec:generalcriterion}

More importantly, however, the exact models we introduced above
provide information about the models of SU($n$) spin chains with
representations corresponding to Young tableaux with a number of boxes
divisible by $n$, \ie models for which the Affleck--Lieb theorem is
not applicable.  We have studied two SU(3) models belonging to this
family 
in Sec.~\ref{sec:vbs}, the rep.~$\bs{10}$ VBS
and the rep.~$\bs{8}$ VBS, and found that both have confined
spinons or colorons and hence display a Haldane-type gap in the
spectrum.  

In this section, we will argue that 
models of antiferromagnetic chains of SU($n$) spins transforming under
a representation corresponding to a Young tableau consisting of a
number of boxes $\lambda$ divisible by $n$ generally possess a
Haldane-type gap due to spinon confinement forces.

We should caution immediately that our argument is based on several
assumptions, which we consider reasonable, but which we are ultimately
unable to prove.

The first, and also the most crucial, is the assumption that the
question of whether a given model supports free spinon excitations can
be resolved through study of the corresponding VBS state.  This
assumption definitely holds for SU(2) spin chains, where the MG model
for $S=\frac{1}{2}$ indicates that the spinons are free, while the
AKLT model for $S=1$ serves as a paradigm for spinon confinement and
hence the Haldane gap.  The general conclusions we derived from our
studies of the SU(3) VBSs above rely on this assumption.  The
numerical results we reported on the rep.~$\bs{8}$ VBS provide
evidence that this assumption holds at least for this model.

Let us consider an SU($n$) spin chain with spins transforming under
a representation corresponding to Young tableau consisting of $L$ columns
with $\kappa_1\le\kappa_2\le\ldots\le\kappa_L<n$ boxes each,
\begin{equation}
  \label{eq:rep}
  [\kappa_1,\kappa_2,\ldots,\kappa_L]\;\equiv\;
\setlength{\unitlength}{8pt}
\begin{picture}(9,4)(0,3)
\put(0,0){\line(1,0){1}}
\multiput(0,1)(0,1){2}{\line(1,0){2}}
\put(2,2.45){\line(1,0){2.75}}
\put(4.75,3.75){\line(1,0){1.75}}
\put(6.5,5){\line(1,0){2.25}}
\put(7.75,6){\line(1,0){3}}
\put(0,7){\line(1,0){10.75}}
\put(0,6){\line(1,0){2}}
\multiput(0,0)(1,0){2}{\line(0,1){7}}
\put(2,1){\line(0,1){6}}
\put(4.75,3.75){\line(0,-1){1.3}}
\put(6.5,5){\line(0,-1){1.25}}
\multiput(9.75,6)(1,0){2}{\line(0,1){1}}
\multiput(7.75,5)(1,0){2}{\line(0,1){2}}
\put(0.55,-1){\makebox(0,0){\small $\kappa_L$}}
\put(8.25,4){\makebox(0,0){\small $\kappa_3$}}
\put(10,4.3){\makebox(0,0){\small $\kappa_2\kappa_1$}}
\end{picture} \\[27pt] \phantom{oooops} 
\end{equation}
with a total number of boxes
\begin{displaymath}
  \lambda=\sum_{l=1}^L \kappa_l
\end{displaymath}
divisible by $n$.  We denote the Dynkin coordinates of this
representation by $(\mu_1,\mu_2,\ldots,\mu_{n-1})$, which implies
\begin{displaymath}
  \sum_{i=1}^{n-1}\mu_i=L.
\end{displaymath}
Note that this representation is, by definition, given by the 
maximally symmetric component of the tensor product of the individual
columns,
\begin{equation}
  \label{eq:sym}
  [\kappa_1,\kappa_2,\ldots,\kappa_L]=\mathcal{S}
  \big([\kappa_1]\otimes [\kappa_2]\otimes\ldots\otimes [\kappa_L ]\big).
\end{equation}
For convenience, we denote the $\left({n \atop \kappa}\right)$
dimensional representation $[\kappa]$ in this subsection as
$\bs{\kappa}_l\equiv [\kappa_l]=
\setlength{\unitlength}{1.25pt}
\begin{picture}(10,3)(0,-2)
\put(5,0){\circle{8}}
\put(5,0){\makebox(0,0){\small $l$}}
\end{picture}$.

Since $\lambda$ is divisible by $n$, it will always be possible to
obtain a singlet from the complete sequence of representations
$\bs{\kappa}_1,\bs{\kappa}_2,\ldots,\bs{\kappa}_L$ by combining them
antisymmetrically.  To be precise, when we write that we combine
representations $\bs{\kappa_1}$ and $\bs{\kappa_2}$ antisymmetrically, we
mean we obtain a new representation $[\kappa_1+\kappa_2]$ by
stacking the two columns with $\kappa_1$ and $\kappa_2$ boxes on
top of each other if $\kappa_1+\kappa_2<n$, and a new representation 
$[\kappa_1+\kappa_2-n]$ if $\kappa_1+\kappa_2\ge n$.  In equations,
we write this as
\begin{equation}
  \label{eq:asym}
  \mathcal{A}
  \big([\kappa_1]\otimes [\kappa_2]\big)\equiv
  \begin{cases}
    [\kappa_1+\kappa_2]&\text{for}\ \kappa_1+\kappa_2<n\\
    [\kappa_1+\kappa_2-n]&\text{for}\ \kappa_1+\kappa_2\ge n
  \end{cases}
\end{equation}
Following the notation used above, we indicate the
antisymmetric combination of representations $\bs{\kappa}_l$ by a line
connecting them.  In particular, we depict the singlet formed by
combining $\bs{\kappa}_1,\bs{\kappa}_2,\ldots,\bs{\kappa}_L$ on
different lattice sites as
\begin{equation}
\label{eq:kappasinglet}
  \setlength{\unitlength}{1.25pt}
\begin{picture}(90,0)(0,-1)
\linethickness{0.8pt}
\multiput(5,0)(20,0){3}{{\circle{8}}}
\multiput(85,0)(20,0){1}{{\circle{8}}}
\put(5,0){\makebox(0,0){\small 1}}
\put(25,0){\makebox(0,0){\small 2}}
\put(45,0){\makebox(0,0){\small 3}}
\put(65,0){\makebox(0,0){\small \ldots}}
\put(85,0){\makebox(0,0){\small $L$}}
\thicklines
\multiput(15,0)(20,0){2}{\makebox(0,0){\line(2,0){12}}}
\put(49,0){\line(1,0){8}}
\put(81,0){\line(-1,0){8}}
\end{picture}
\vspace{10pt}
\end{equation}
The understanding here is that we combine them in the order indicated
by the line, \ie in \eqref{eq:kappasinglet} we first combine
$\bs{\kappa}_1$ and $\bs{\kappa}_2$, then we combine the result with
$\bs{\kappa}_3$, and so on.  Depending on the order of the
representations $\bs{\kappa}_l$ on the line, we obtain different, but
not necessarily orthogonal, singlets.  We assume that it is irrelevant
whether we combine the representations starting from the left or from
the right of the line, as the resulting state will not depend on it.

In general, it will be possible to construct a number $\le\lambda/n$
of singlets out of various combinations of the $\bs{\kappa}$'s, one
for each block of $\bs{\kappa}$'s for which the values of $\kappa$ add
up to $n$ as we combine the representation in the order described above.  
In this case, we will be able to construct one VBS for
each singlet, and subsequently combine them at each site symmetrically
to obtain the desired representation \eqref{eq:rep}.  The argument
for spinon confinement we construct below will hold for each of the
individual VBSs, and hence for the combined VBS as well.  It is hence
sufficient for our purposes to consider situations where the entire
sequence $\bs{\kappa}_1,\bs{\kappa}_2,\ldots,\bs{\kappa}_L$
is needed to construct a singlet.

A possible VBS ``ground state'' for a representation corresponding to a
Young tableau with $L=4$ columns is depicted below.
\begin{equation}
  \setlength{\unitlength}{1.25pt}
\begin{picture}(160,70)(0,-15)
\linethickness{0.8pt}
\multiput(0,40)(20,0){9}{{\circle{7}}}
\multiput(0,40)(20,0){9}{\makebox(0,0){\small 1}}
\multiput(0,30)(20,0){9}{{\circle{7}}}
\multiput(0,30)(20,0){9}{\makebox(0,0){\small 2}}
\multiput(0,20)(20,0){9}{{\circle{7}}}
\multiput(0,20)(20,0){9}{\makebox(0,0){\small 3}}
\multiput(0,10)(20,0){9}{{\circle{7}}}
\multiput(0,10)(20,0){9}{\makebox(0,0){\small 4}}
\thicklines
\multiput(3.5,38.25)(20,0){8}{\line(2,-1){13}}
\multiput(3.5,28.25)(20,0){8}{\line(2,-1){13}}
\multiput(3.5,18.25)(20,0){8}{\line(2,-1){13}}
\thinlines
\put(54,4){\framebox(12,42)}
\put(60,4){\line(0,-1){7}}
\put(60,-10){\makebox(1,1){\small projection onto representation}}
\put(60,-18){\makebox(1,1){\small $[\kappa_1,\kappa_2,\kappa_3,\kappa_4]$}}
\put(60,55){\makebox(1,1){\small one site}}
\end{picture}
\label{eq:VBS1}
\end{equation}
In general, there are as many inequivalent VBS ``ground states'' as
there are inequivalent ways to order the representations
$\bs{\kappa}_1,\bs{\kappa}_2,\ldots,\bs{\kappa}_L$, \ie
the number of inequivalent VBS ``ground states'' is given by
\begin{displaymath}
  \frac{L!}{\mu_1!\cdot\mu_2!\cdot\ldots\cdot\mu_{n-1}!}.
\end{displaymath}
To give an example, the following VBS
\begin{equation}
  \setlength{\unitlength}{1.25pt}
\begin{picture}(160,70)(0,-15)
\linethickness{0.8pt}
\multiput(0,40)(20,0){9}{{\circle{7}}}
\multiput(0,40)(20,0){9}{\makebox(0,0){\small 2}}
\multiput(0,30)(20,0){9}{{\circle{7}}}
\multiput(0,30)(20,0){9}{\makebox(0,0){\small 1}}
\multiput(0,20)(20,0){9}{{\circle{7}}}
\multiput(0,20)(20,0){9}{\makebox(0,0){\small 3}}
\multiput(0,10)(20,0){9}{{\circle{7}}}
\multiput(0,10)(20,0){9}{\makebox(0,0){\small 4}}
\thicklines
\multiput(3.5,38.25)(20,0){8}{\line(2,-1){13}}
\multiput(3.5,28.25)(20,0){8}{\line(2,-1){13}}
\multiput(3.5,18.25)(20,0){8}{\line(2,-1){13}}
\thinlines
\put(54,4){\framebox(12,42)}
\put(60,4){\line(0,-1){7}}
\put(60,-10){\makebox(1,1){\small projection onto representation}}
\put(60,-18){\makebox(1,1){\small $[\kappa_1,\kappa_2,\kappa_3,\kappa_4]$}}
\put(60,55){\makebox(1,1){\small one site}}
\end{picture}
\label{eq:VBS2}
\end{equation}
is inequivalent to the one above if and only if $\kappa_1\ne\kappa_2$.
Note that all these VBS ``ground states'' states are translationally
invariant.  We expect some of these states, but not all of them,
degenerate in energy for the appropriate parent Hamiltonian, and have
accordingly written ``ground states'' in quotation marks.  For example,
if we form a SU(4) VBS with representation 
$[\kappa_1,\kappa_2,\kappa_3]=[1,1,2]$, the
combination 
\begin{equation}
  \setlength{\unitlength}{1.25pt}
\begin{picture}(50,0)(0,-1)
\linethickness{0.8pt}
\multiput(5,0)(20,0){3}{{\circle{8}}}
\put(5,0){\makebox(0,0){\small 1}}
\put(25,0){\makebox(0,0){\small 3}}
\put(45,0){\makebox(0,0){\small 2}}
\thicklines
\multiput(15,0)(20,0){2}{\makebox(0,0){\line(2,0){12}}}
\end{picture}\nonumber
\vspace{4pt}
\end{equation}
might yield a state with a lower energy for the appropriate
Hamiltonian than the state formed by combining
\begin{equation}
  \setlength{\unitlength}{1.25pt}
\begin{picture}(50,0)(0,-1)
\linethickness{0.8pt}
\multiput(5,0)(20,0){3}{{\circle{8}}}
\put(5,0){\makebox(0,0){\small 1}}
\put(25,0){\makebox(0,0){\small 2}}
\put(45,0){\makebox(0,0){\small 3}}
\thicklines
\multiput(15,0)(20,0){2}{\makebox(0,0){\line(2,0){12}}}
\end{picture}. \nonumber
\vspace{4pt}
\end{equation}
Simple examples where we have only two inequivalent VBS ground states
are provided by the matrix product states discussed in
Secs.~\ref{sec:8VBS} and \ref{sec:mp}.

We will now argue that the elementary excitations of the corresponding
VBS models are always confined.  To this end, we first observe that
any domain wall between translationally invariant ``ground states''
consists of a total of $m\cdot n$ representations $\bs{\kappa}_l$ ($m$
integer).  To illustrate this, consider a domain wall between the
``ground states'' depicted in \eqref{eq:VBS2} and \eqref{eq:VBS1}:
\begin{equation}
  \setlength{\unitlength}{1.25pt}
\begin{picture}(170,49)(20,-2)
\linethickness{0.8pt}
\multiput(20,40)(20,0){2}{{\circle{7}}}
\multiput(20,40)(20,0){2}{\makebox(0,0){\small 2}}
\multiput(80,40)(20,0){6}{{\circle{7}}}
\multiput(80,40)(20,0){6}{\makebox(0,0){\small 1}}
\multiput(20,30)(20,0){3}{{\circle{7}}}
\multiput(20,30)(20,0){3}{\makebox(0,0){\small 1}}
\multiput(100,30)(20,0){5}{{\circle{7}}}
\multiput(100,30)(20,0){5}{\makebox(0,0){\small 2}}
\multiput(20,20)(20,0){4}{{\circle{7}}}
\multiput(20,20)(20,0){4}{\makebox(0,0){\small 3}}
\multiput(120,20)(20,0){4}{{\circle{7}}}
\multiput(120,20)(20,0){4}{\makebox(0,0){\small 3}}
\multiput(20,10)(20,0){5}{{\circle{7}}}
\multiput(20,10)(20,0){5}{\makebox(0,0){\small 4}}
\multiput(140,10)(20,0){3}{{\circle{7}}}
\multiput(140,10)(20,0){3}{\makebox(0,0){\small 4}}
%
\put(60,40){{\circle*{7}}}
\put(80,30){{\circle*{7}}}
\put(100,20){{\circle*{7}}}
\put(120,10){{\circle*{7}}}
{\color{white}
\put(60,40){\makebox(0,0){\small\bf 2}}
\put(80,30){\makebox(0,0){\small\bf 2}}
\put(100,20){\makebox(0,0){\small\bf 3}}
\put(120,10){\makebox(0,0){\small\bf 4}}
}
\thicklines
\multiput(23.5,38.25)(20,0){2}{\line(2,-1){13}}
\multiput(23.5,28.25)(20,0){2}{\line(2,-1){13}}
\multiput(23.5,18.25)(20,0){2}{\line(2,-1){13}}
%
\put(63.5,28.25){\line(2,-1){13}}
\put(63.5,18.25){\line(2,-1){13}}
\multiput(83.5,38.25)(20,0){5}{\line(2,-1){13}}
\multiput(83.5,28.25)(20,0){5}{\line(2,-1){13}}
\multiput(83.5,18.25)(20,0){5}{\line(2,-1){13}}
\end{picture}
\nonumber
\end{equation}
In the example, the domain wall consists of reps.~$\bs{\kappa}_2$,
$\bs{\kappa}_2$, $\bs{\kappa}_3$, and $\bs{\kappa}_4$.  If the
translationally invariant states on both sites are true ground states,
the domain wall is likely to correspond to two elementary excitations:
a rep.~$\bs{\bar{\kappa}}_1$ spinon consisting of an antisymmetric
combination of a $\bs{\kappa}_2$, a $\bs{\kappa}_3$, and a
$\bs{\kappa}_4$, as indicated by the line in the drawing, and another
rep.~$\bs{\kappa}_2$ spinon.  The reason we assume that
$\bs{\kappa}_2$, $\bs{\kappa}_3$, and $\bs{\kappa}_4$ form a
rep.~$\bs{\bar{\kappa}}_1$ is simply that this combination is present
in both ground states on either side, and hence bound to be the
energetically most favorable combination.  The second
$\bs{\kappa}_2$, however, is not part of this elementary excitation,
as combining it antisymmetrically with the others (\ie the
$\bs{\bar{\kappa}}_1$) would induce correlations which are not present
in the ground state.  We hence conclude that the second
$\bs{\kappa}_2$ is an elementary excitation as well.  The domain wall
depicted above consists of a spinon transforming under
rep.~$\bs{\bar{\kappa}}_1$ and an anti-spinon transforming under
rep.~$\bs{\kappa}_2$.

The next step in our argument is to notice that the spinon and the
complementary particle created simultaneously which may either be its
anti-particle or some other spinon, are confined through a linear
potential.  To see this, we pull them apart and look at the state
inbetween (color online):
\begin{equation}
  \setlength{\unitlength}{1.25pt}
\begin{picture}(180,73)(0,-10)
\linethickness{0.8pt}
\multiput(0,40)(20,0){1}{{\circle{7}}}
\multiput(0,40)(20,0){1}{\makebox(0,0){\small 2}}
\multiput(40,40)(20,0){8}{{\circle{7}}}
\multiput(40,40)(20,0){8}{\makebox(0,0){\small 1}}
\multiput(0,30)(20,0){2}{{\circle{7}}}
\multiput(0,30)(20,0){2}{\makebox(0,0){\small 1}}
\multiput(40,30)(20,0){4}{{\circle{7}}}
\multiput(40,30)(20,0){4}{\makebox(0,0){\small 2}}
\multiput(140,30)(20,0){3}{{\circle{7}}}
\multiput(140,30)(20,0){3}{\makebox(0,0){\small 2}}
\multiput(0,20)(20,0){7}{{\circle{7}}}
\multiput(0,20)(20,0){7}{\makebox(0,0){\small 3}}
\multiput(160,20)(20,0){2}{{\circle{7}}}
\multiput(160,20)(20,0){2}{\makebox(0,0){\small 3}}
\multiput(0,10)(20,0){8}{{\circle{7}}}
\multiput(0,10)(20,0){8}{\makebox(0,0){\small 4}}
\multiput(180,10)(20,0){1}{{\circle{7}}}
\multiput(180,10)(20,0){1}{\makebox(0,0){\small 4}}
%
\put(20,40){{\circle*{7}}}
\put(120,30){{\circle*{7}}}
\put(140,20){{\circle*{7}}}
\put(160,10){{\circle*{7}}}
{\color{white}
\put(20,40){\makebox(0,0){\small\bf 2}}
\put(120,30){\makebox(0,0){\small\bf 2}}
\put(140,20){\makebox(0,0){\small\bf 3}}
\put(160,10){\makebox(0,0){\small\bf 4}}
}
%
\thicklines
\put(123.5,28.25){\line(2,-1){13}}
\put(143.5,18.25){\line(2,-1){13}}
\thicklines
\multiput(3.5,38.25)(20,0){1}{\line(2,-1){13}}
\multiput(3.5,28.25)(20,0){2}{\line(2,-1){13}}
\multiput(3.5,18.25)(20,0){2}{\line(2,-1){13}}
{\color{red}
\multiput(43.5,38.25)(20,0){3}{\line(2,-1){13}}
\multiput(43.5,28.25)(20,0){4}{\line(2,-1){13}}
\multiput(43.5,18.25)(20,0){4}{\line(2,-1){13}}
%
%
\put(90,25){\makebox(0,0){\line(2,3){15.9}}}
}
\multiput(123.5,38.25)(20,0){3}{\line(2,-1){13}}
\multiput(123.5,28.25)(20,0){3}{\line(2,-1){13}}
\multiput(123.5,18.25)(20,0){3}{\line(2,-1){13}}
\thinlines
\put(20,56){\makebox(0,0){\small $\bs{\kappa}_2$-spinon}}
\put(140,56){\makebox(0,0){\small $\bs{\bar{\kappa}}_1$-spinon}}
\put(80,-2){\vector(-1,0){60}}
\put(80,-2){\vector(1,0){60}}
\put(80,-8){\makebox(0,0){\small energy cost $\propto$ distance}}
\end{picture}
\label{eq:kappaconf}
\end{equation}
The state between spinon and anti-spinon is not translationally
invariant.  In the example, the unit cell of this state is depicted in
red and consists of two regular bonds with three ``singlet lines''
between the sites, one stronger bond with four lines (which cross in
the cartoon), and one weaker bond with only two lines.  If we assume
that the two states \eqref{eq:VBS2} and \eqref{eq:VBS1} on both sides
are true ground states, it is clear that the irregularities in the
strength of the bond correlations will cause the state between the
spinon and anti-spinon to have a higher energy than either of them.
This additional energy cost will induce a linear confinement potential
between the spinons, and hence a linear oscillator potential for the
relative motion of the particles.  As in the models studied above, the
Haldane gap corresponds to the zero-point energy of this oscillator.

If one of the ``ground states'' to the left or to the right of the
domain wall is not a true ground state, but a translationally
invariant state corresponding to a higher energy than the ground
state, there will be a confining force between this domain wall and
another domain wall which transforms the intermediate ``ground state''
with a higher energy back into a true ground state.  This force will
be sufficient to account for a Haldane gap, regardless of the strength
or existence of a confinement force between the constituent particles
of each domain wall.

The lowest-lying excitations of a representation
$[\kappa_1,\kappa_2,\ldots,\kappa_L]$ spin chain, however, will in
general not be domain walls, but spinons created by breaking up one of
the singlets \eqref{eq:kappasinglet} in a ground state.  This is 
illustrated below for the ground state \eqref{eq:VBS1}:
\begin{equation}
  \setlength{\unitlength}{1.25pt}
\begin{picture}(170,49)(20,-2)
\linethickness{0.8pt}
\multiput(20,40)(20,0){2}{{\circle{7}}}
\multiput(20,40)(20,0){2}{\makebox(0,0){\small 1}}
\multiput(80,40)(20,0){6}{{\circle{7}}}
\multiput(80,40)(20,0){6}{\makebox(0,0){\small 1}}
\multiput(20,30)(20,0){3}{{\circle{7}}}
\multiput(20,30)(20,0){3}{\makebox(0,0){\small 2}}
\multiput(100,30)(20,0){5}{{\circle{7}}}
\multiput(100,30)(20,0){5}{\makebox(0,0){\small 2}}
\multiput(20,20)(20,0){4}{{\circle{7}}}
\multiput(20,20)(20,0){4}{\makebox(0,0){\small 3}}
\multiput(120,20)(20,0){4}{{\circle{7}}}
\multiput(120,20)(20,0){4}{\makebox(0,0){\small 3}}
\multiput(20,10)(20,0){5}{{\circle{7}}}
\multiput(20,10)(20,0){5}{\makebox(0,0){\small 4}}
\multiput(140,10)(20,0){3}{{\circle{7}}}
\multiput(140,10)(20,0){3}{\makebox(0,0){\small 4}}
%
\put(60,40){{\circle*{7}}}
\put(80,30){{\circle*{7}}}
\put(100,20){{\circle*{7}}}
\put(120,10){{\circle*{7}}}
{\color{white}
\put(60,40){\makebox(0,0){\small\bf 1}}
\put(80,30){\makebox(0,0){\small\bf 2}}
\put(100,20){\makebox(0,0){\small\bf 3}}
\put(120,10){\makebox(0,0){\small\bf 4}}
}
\thicklines
\multiput(23.5,38.25)(20,0){2}{\line(2,-1){13}}
\multiput(23.5,28.25)(20,0){2}{\line(2,-1){13}}
\multiput(23.5,18.25)(20,0){2}{\line(2,-1){13}}
%
\put(63.5,28.25){\line(2,-1){13}}
\put(63.5,18.25){\line(2,-1){13}}
\multiput(83.5,38.25)(20,0){5}{\line(2,-1){13}}
\multiput(83.5,28.25)(20,0){5}{\line(2,-1){13}}
\multiput(83.5,18.25)(20,0){5}{\line(2,-1){13}}
\end{picture}
\nonumber
\end{equation}
In the example, we have created a spinon transforming under
rep.~$\bs{\bar{\kappa}}_1$ and its anti-particle, a spinon
transforming under rep.~$\bs{\kappa}_1$.  This is, however, irrelevant
to the argument---we may break the singlet in any way we like.
The important feature is that we obtain, by construction, at least two
excitations, and that these are confined.  In our specific example, the 
confining potential is equivalent to the confining potential in
\eqref{eq:kappaconf} above (color online):
\begin{equation}
  \setlength{\unitlength}{1.25pt}
\begin{picture}(180,78)(0,-15)
\linethickness{0.8pt}
\multiput(0,40)(20,0){1}{{\circle{7}}}
\multiput(0,40)(20,0){1}{\makebox(0,0){\small 1}}
\multiput(40,40)(20,0){8}{{\circle{7}}}
\multiput(40,40)(20,0){8}{\makebox(0,0){\small 1}}
\multiput(0,30)(20,0){2}{{\circle{7}}}
\multiput(0,30)(20,0){2}{\makebox(0,0){\small 2}}
\multiput(40,30)(20,0){4}{{\circle{7}}}
\multiput(40,30)(20,0){4}{\makebox(0,0){\small 2}}
\multiput(140,30)(20,0){3}{{\circle{7}}}
\multiput(140,30)(20,0){3}{\makebox(0,0){\small 2}}
\multiput(0,20)(20,0){7}{{\circle{7}}}
\multiput(0,20)(20,0){7}{\makebox(0,0){\small 3}}
\multiput(160,20)(20,0){2}{{\circle{7}}}
\multiput(160,20)(20,0){2}{\makebox(0,0){\small 3}}
\multiput(0,10)(20,0){8}{{\circle{7}}}
\multiput(0,10)(20,0){8}{\makebox(0,0){\small 4}}
\multiput(180,10)(20,0){1}{{\circle{7}}}
\multiput(180,10)(20,0){1}{\makebox(0,0){\small 4}}
%
\put(20,40){{\circle*{7}}}
\put(120,30){{\circle*{7}}}
\put(140,20){{\circle*{7}}}
\put(160,10){{\circle*{7}}}
{\color{white}
\put(20,40){\makebox(0,0){\small\bf 1}}
\put(120,30){\makebox(0,0){\small\bf 2}}
\put(140,20){\makebox(0,0){\small\bf 3}}
\put(160,10){\makebox(0,0){\small\bf 4}}
}
%
\thicklines
\put(123.5,28.25){\line(2,-1){13}}
\put(143.5,18.25){\line(2,-1){13}}
\thicklines
\multiput(3.5,38.25)(20,0){1}{\line(2,-1){13}}
\multiput(3.5,28.25)(20,0){2}{\line(2,-1){13}}
\multiput(3.5,18.25)(20,0){2}{\line(2,-1){13}}
{\color{red}
\multiput(43.5,38.25)(20,0){3}{\line(2,-1){13}}
\multiput(43.5,28.25)(20,0){4}{\line(2,-1){13}}
\multiput(43.5,18.25)(20,0){4}{\line(2,-1){13}}
%
%
\put(90,25){\makebox(0,0){\line(2,3){15.9}}}
}
\multiput(123.5,38.25)(20,0){3}{\line(2,-1){13}}
\multiput(123.5,28.25)(20,0){3}{\line(2,-1){13}}
\multiput(123.5,18.25)(20,0){3}{\line(2,-1){13}}
\thinlines
\put(20,56){\makebox(0,0){\small $\bs{\kappa}_1$-spinon}}
\put(140,56){\makebox(0,0){\small $\bs{\bar{\kappa}}_1$-spinon}}
\put(80,-2){\vector(-1,0){60}}
\put(80,-2){\vector(1,0){60}}
\put(80,-8){\makebox(0,0){\small energy cost $\propto$ distance}}
\end{picture}
\nonumber
\end{equation}
We leave it to the reader to convince him- or herself that the
conclusions regarding confinement drawn from the simple examples
studied here hold in general.

\subsection{Models with confinement through interactions 
extending beyond nearest neighbors }
\label{sec:examples}

Let us briefly summarize the results obtained.  The SU($n$) models we
have studied so far fall into two categories.  The models belonging to
the first---the trimer chain, the $\bs{6}$ VBS, and the $n$-mer
chain---have $n$ degenerate ground states, which break translational
invariance up to translations by $n$ lattice spacings.  The Young
tableaux describing the representations of SU($n$) at each site
consist of a number of boxes $\lambda$ which is smaller than $n$, and
hence obviously not divisible by $n$.  The models support deconfined
spinon excitations, and hence do not possess a Haldane gap in the
spectrum.  The Hamiltonians of these models require interactions
between $n+1$ neighboring sites along the chain.  Even though the
Affleck--Lieb theorem is not directly applicable to the models we
constructed above, it is applicable to SU($n$) spin chains with spins
transforming under the same representations.  Like the VBS models, the
theorem suggests that there is no Haldane gap in this family of
models.

The models belonging to the second category---the $\bs{10}$ VBS, the
$\bs{8}$ VBS, the representation $(n,0,\ldots,0)$ VBS, and the
$\bs{m}$-$\bs{\overline{m}}$ matrix product state---have
translationally invariant ground states.  The ground states are unique
for some models, like the $\bs{10}$ and the $(n,0,\ldots,0)$ VBS, and
degenerate for others, like the $\bs{8}$ VBS.  The Young tableaux
describing the representations of SU($n$) at each site consist of a
number of boxes $\lambda$ which is divisible by $n$.  The
Affleck--Lieb theorem is not applicable to models of this category.
The spinon excitations for this category of models are subject to
confinement forces, which give rise to a Haldane gap.  The parent
Hamiltonians for these models require interactions between
nearest-neighbor sites only.

At first glance, this classification might appear complete.  Further
possibilities arise, however, in SU($n$) spin chains where number of
boxes $\lambda$ the Young tableau consists of and $n$ have a common
divisor different from $n$, which obviously requires that $n$ is not a
prime number.  In this case, it is possible to construct VBS models in
which the ground state breaks translational invariance only up to
translations by $n/q$ lattice spacings, where $q$ denotes the largest
common divisor of $\lambda$ and $n$ such that $q<n$.  This implies
that the ground state of the appropriate, translationally invariant
Hamiltonian will be $n/q$-fold degenerate.  In the examples we will
elaborate on below, the parent Hamiltonians for these models require
interactions between $\frac{n}{q}+1$ sites, a feature we conjecture to
hold in general.  The spinon excitations of these models are confined,
even though the Affleck--Lieb theorem states that the nearest-neighbor
Heisenberg chain of SU($n$) spins transforming under these
representations is gapless.  (We implicitly assume here that the 
ground states of the SU($n$) nearest-neighbor Heisenberg chains are
non-degenerate.)  Let us illustrate the general features of
this third category of models with a few simple examples.

(1) Consider an SU(4) chain with spins transforming under the
10-dimensional representation
\begin{displaymath}
  (2,0,0)\ =\
  \setlength{\unitlength}{8pt}
  \begin{picture}(2,1)(0,0.15)
    \multiput(0,0)(1,0){3}{\line(0,1){1}}
    \multiput(0,0)(0,1){2}{\line(1,0){2}}
  \end{picture}\ \ .
\end{displaymath}
Following the construction principle of the $\bs{6}$ VBS of SU(3), we
find that the two degenerate VBS states illustrated through
\begin{equation}
\setlength{\unitlength}{1pt}
\begin{picture}(162,48)(4,-22)
\linethickness{0.8pt}
\multiput(0,10)(14,0){12}{\circle{4}}
\multiput(0,0)(14,0){12}{\circle{4}}
\thicklines
\multiput(2,10)(14,0){3}{\line(1,0){10}}
\multiput(58,10)(14,0){3}{\line(1,0){10}}
\multiput(114,10)(14,0){3}{\line(1,0){10}}
\put(-2,0){\line(-1,0){4}}
\multiput(2,0)(14,0){1}{\line(1,0){10}}
\multiput(30,0)(14,0){3}{\line(1,0){10}}
\multiput(86,0)(14,0){3}{\line(1,0){10}}
\multiput(142,0)(14,0){1}{\line(1,0){10}}
\put(156,0){\line(1,0){4}}
\thinlines
\put(37,-5){\framebox(10,20)}
\put(107,-5){\dashbox{2}(38,20)}
\put(42,-5){\line(0,-1){8}}
\put(42,-20){\makebox(1,1){\small projection onto rep.\ $(2,0,0)$}}
\put(42,23){\makebox(1,1){\small one site}}
\end{picture}
\label{eq:200state}
\end{equation}
are exact zero-energy ground states of 
\begin{equation}
  \label{eq:SU4model}
  H_{(2,0,0)\,\text{VBS}}
  =\sum_{i=1}^N
  \left(\Bigl(\bs{J}_i^{(3)}\Bigr)^4-12\Bigl(\bs{J}_i^{(3)}\Bigr)^2+
    \frac{135}{4}\right),
\end{equation}
which contains next-nearest neighbor interactions.  

The example illustrates the general rule stated above. The largest
common divisor of $n=4$ and $\lambda=2$ is $q=2$.  This implies
$\frac{n}{q}=2$ and hence two degenerate VBS states which break
translational invariance up to translations by two lattice spacings.
The parent Hamiltonian requires interaction between three neighboring
sites.

According to the Affleck--Lieb theorem, models of antiferromagnetic
SU(4) chains of representation $(2,0,0)$ with nearest-neighbor
Heisenberg interactions and non-degenerate ground states are gapless
in the thermodynamic limit, which implies that the spinons are
deconfined.  In all the models we have studied in previous sections,
the conclusions drawn from the Affleck--Lieb theorem were consistent
with those drawn from our exact models.  For the present model,
however, they are not consistent.

Specifically, we strongly conjecture that the spinons in the $(2,0,0)$
VBS are confined.  This conjecture is based on the reasonable
assumption that the elementary excitations of the model transform as
either the fundamental representation $\bs{4}=(1,0,0)$ of SU(4) or its
conjugate representation $\bs{\bar 4}=(0,0,1)$.  This assumption
implies that a single domain wall in one of the $4$-mer chains used to
construct the VBS state \eqref{eq:200state} shifts this chain by one
lattice spacing.  If we assume a ground state to the left of the
spinon, the state on to the right will have a higher energy for the
next-nearest neighbor Hamiltonian \eqref{eq:SU4model}, as illustrated
below.  
\begin{equation}
\setlength{\unitlength}{1pt}
\begin{picture}(170,60)(4,-25)
\linethickness{0.8pt}
\multiput(0,10)(14,0){13}{\circle{4}}
\multiput(0,0)(14,0){13}{\circle{4}}
\multiput(0,10)(14,0){3}{\circle*{4}}
\multiput(140,0)(14,0){3}{\circle*{4}}
\thicklines
\multiput(2,10)(14,0){2}{\line(1,0){10}}
\multiput(44,10)(14,0){3}{\line(1,0){10}}
\multiput(100,10)(14,0){3}{\line(1,0){10}}
\multiput(156,10)(14,0){1}{\line(1,0){10}}
\put(-2,0){\line(-1,0){4}}
\multiput(2,0)(14,0){1}{\line(1,0){10}}
\multiput(30,0)(14,0){3}{\line(1,0){10}}
\multiput(86,0)(14,0){3}{\line(1,0){10}}
\multiput(142,0)(14,0){2}{\line(1,0){10}}
\put(170,10){\line(1,0){4}}
\thinlines
\put(84,-14){\vector(-1,0){70}}
\put(84,-14){\vector(1,0){70}}
\put(84,-22){\makebox(0,0){\small energy cost $\propto$ distance}}
\put(14,22){\makebox(0,0){$\bs{\bar 4}$}}
\put(14,32){\makebox(0,0){\small spinon}}
\put(154,22){\makebox(0,0){$\bs{\bar 4}$}}
\put(154,32){\makebox(0,0){\small spinon}}
\put(65,-5){\dashbox{1}(38,20)}
\end{picture}
\label{eq:200spinons}
\end{equation}
To recover the ground state, a second domain wall is required
nearby, which is bound to the first by a linear potential.

Our conclusion is not in contradiction with the rigorous result of
Affleck and Lieb, as they explicitly restrict themselves to models
with nearest-neighbor interactions.  If we had only nearest-neighbor
interactions, the energy expectation value in the region between the
domain walls would not be higher than in the ground state, as one can
see easily from the cartoon above---the sequence of alternating links
would merely shift from (strong, weak, strong, weak) to (strong,
strong, weak, weak).

(2) The situation is similar for an SU(6) chain with spins transforming
under the 56-dimensional representation
\begin{displaymath}
  (3,0,0,0,0)\ =\
  \setlength{\unitlength}{8pt}
  \begin{picture}(3,1)(0,0.15)
    \multiput(0,0)(1,0){4}{\line(0,1){1}}
    \multiput(0,0)(0,1){2}{\line(1,0){3}}
  \end{picture}\ \ .
\end{displaymath}
With $n=6$ and $\lambda=3$, we have again $\frac{n}{q}=2$.
Accordingly, we find that the two VBS states illustrated through
\begin{equation}
\setlength{\unitlength}{1pt}
\begin{picture}(162,58)(4,-22)
\linethickness{0.8pt}
\multiput(0,20)(14,0){12}{\circle{4}}
\multiput(0,10)(14,0){12}{\circle{4}}
\multiput(0,0)(14,0){12}{\circle{4}}
\thicklines
\multiput(2,20)(14,0){5}{\line(1,0){10}}
\multiput(86,20)(14,0){5}{\line(1,0){10}}
\put(-2,10){\line(-1,0){4}}
\put(-2,0){\line(-1,0){4}}
\multiput(2,10)(14,0){1}{\line(1,0){10}}
\multiput(30,10)(14,0){5}{\line(1,0){10}}
\multiput(114,10)(14,0){3}{\line(1,0){10}}
\put(156,10){\line(1,0){4}}
\multiput(2,0)(14,0){3}{\line(1,0){10}}
\multiput(58,0)(14,0){5}{\line(1,0){10}}
\multiput(142,0)(14,0){1}{\line(1,0){10}}
\put(156,0){\line(1,0){4}}
\thinlines
\put(37,-5){\framebox(10,30)}
\put(107,-5){\dashbox{2}(38,30)}
\put(42,-5){\line(0,-1){8}}
\put(42,-20){\makebox(1,1){\small projection onto rep.\ $(3,0,0,0,0)$}}
\put(42,33){\makebox(1,1){\small one site}}
\end{picture}
\label{eq:30000state}
\end{equation}
are exact ground states of a parent Hamiltonian containing up to
next-nearest-neighbor interactions only, and that the spinon excitations
are confined.

(3) As a third example, consider an SU(6) spin chain with spins
transforming under the 21-dimensional representation
\begin{displaymath}
  (2,0,0,0,0)\ =\
  \setlength{\unitlength}{8pt}
  \begin{picture}(2,1)(0,0.15)
    \multiput(0,0)(1,0){3}{\line(0,1){1}}
    \multiput(0,0)(0,1){2}{\line(1,0){2}}
  \end{picture}\ \ .
\end{displaymath}
This implies $\frac{n}{q}=3$.  We find that the three VBS states
illustrated by
\begin{equation}
\setlength{\unitlength}{1pt}
\begin{picture}(162,48)(4,-22)
\linethickness{0.8pt}
\multiput(0,10)(14,0){12}{\circle{4}}
\multiput(0,0)(14,0){12}{\circle{4}}
\thicklines
\multiput(2,10)(14,0){5}{\line(1,0){10}}
\multiput(86,10)(14,0){5}{\line(1,0){10}}
\put(-2,0){\line(-1,0){4}}
\multiput(2,0)(14,0){2}{\line(1,0){10}}
\multiput(44,0)(14,0){5}{\line(1,0){10}}
\multiput(128,0)(14,0){2}{\line(1,0){10}}
\put(156,0){\line(1,0){4}}
\thinlines
\put(37,-5){\framebox(10,20)}
\put(93,-5){\dashbox{2}(52,20)}
\put(42,-5){\line(0,-1){8}}
\put(42,-20){\makebox(1,1){\small projection onto rep.\ $(2,0,0,0,0)$}}
\put(42,23){\makebox(1,1){\small one site}}
\end{picture}
\label{eq:20000state}
\end{equation}
are exact ground states of a parent Hamiltonian involving the
quadratic Casimir of total spin of four neighboring sites,
\begin{equation}
  \label{eq:ham20000}
  H_{(2,0,0,0,0)\,\text{VBS}}=\sum_{i=1}^N H_i
\end{equation}
with 
\begin{equation}
  \label{eq:hi20000}
  H_i=\left(\!\left(\!\bs{J}^{(4)}_i\!\right)^2 \! - \frac{32}{3}\right)
  \!\left(\!\left(\!\bs{J}^{(4)}_i\!\right)^2 \! - \frac{44}{3}\right)
  \!\left(\!\left(\!\bs{J}^{(4)}_i\!\right)^2 \! - \frac{50}{3}\right).
\end{equation}
These VBS states break translational symmetry only up to translations
by three lattice spacings.  The spinons of this model are again
confined through a linear potential, as illustrated below.
\begin{equation}
\setlength{\unitlength}{1pt}
\begin{picture}(176,60)(4,-25)
\linethickness{0.8pt}
\multiput(0,10)(14,0){15}{\circle{4}}
\multiput(0,0)(14,0){15}{\circle{4}}
\multiput(0,10)(14,0){5}{\circle*{4}}
\multiput(126,0)(14,0){5}{\circle*{4}}
\thicklines
\put(-2,0){\line(-1,0){4}}
\multiput(2,10)(14,0){4}{\line(1,0){10}}
\multiput(72,10)(14,0){5}{\line(1,0){10}}
\multiput(156,10)(14,0){2}{\line(1,0){10}}
\put(184,10){\line(1,0){4}}
\multiput(2,0)(14,0){2}{\line(1,0){10}}
\multiput(44,0)(14,0){5}{\line(1,0){10}}
\multiput(128,0)(14,0){4}{\line(1,0){10}}
\thinlines
\put(91,-14){\vector(-1,0){63}}
\put(91,-14){\vector(1,0){63}}
\put(91,-22){\makebox(0,0){\small energy cost $\propto$ distance}}
\put(28,22){\makebox(0,0){$\bs{\bar 6}$}}
\put(28,32){\makebox(0,0){\small spinon}}
\put(154,23){\makebox(0,0){$\bs{\bar 6}$}}
\put(154,33){\makebox(0,0){\small spinon}}
\put(107,-5){\dashbox{1}(52,20)}
\end{picture}
\label{eq:20000spinons}
\end{equation}
The conclusions we have drawn for this VBS model rest on the
assumption that the quadratic Casimirs of the representations
contained in the tensor product shown in the dashed box in
\eqref{eq:20000state} as well as in the tensor product one obtains if
one shifts this box by one lattice spacing to the left or to the right
are smaller than the largest Casimir contained in the tensor product
shown in the dotted box in \eqref{eq:20000spinons}.  We have verified the
validity of this assumption for the (2,0,0,0,0) VBS model we
considered here, but not shown it to hold for similar models with
larger $n$ or $\lambda$.

(4) Finally, consider an SU(6) spin chain
with spins transforming under the 70-dimensional representation
\begin{displaymath}
  (1,1,0,0,0)\ =\
  \setlength{\unitlength}{8pt}
  \begin{picture}(2,1)(0,-0.35)
    \multiput(0,0)(1,0){3}{\line(0,1){1}}
    \multiput(0,0)(1,0){2}{\line(0,-1){1}}
    \multiput(0,0)(0,1){2}{\line(1,0){2}}
    \multiput(0,-1)(0,1){1}{\line(1,0){1}}
  \end{picture}\ \ .
\end{displaymath}
Thus we have once again $\frac{n}{q}=2$.  In a notation
similar to the one introduced for the $\bs{8}$ VBS, 
\begin{displaymath}
  \setlength{\unitlength}{8pt}
  \begin{picture}(1.8,1)(0,0.15)
    \multiput(0,0)(1,0){2}{\line(0,1){1}}
    \multiput(0,0)(0,1){2}{\line(1,0){1}}
  \end{picture}\hat =
  \setlength{\unitlength}{1pt}
  \begin{picture}(10,10)(0,0)
    \put(6,2.5){\circle{4}}
  \end{picture},\ \
  \setlength{\unitlength}{8pt}
  \begin{picture}(1.8,1)(0,-0.35)
    \multiput(0,0)(1,0){2}{\line(0,1){1}}
    \multiput(0,0)(1,0){2}{\line(0,-1){1}}
    \multiput(0,-1)(0,1){3}{\line(1,0){1}}
  \end{picture}\hat =
  \setlength{\unitlength}{1pt}
  \begin{picture}(11,10)(0,0)
    \put(6,2.5){\circle{6}}
  \end{picture} ,
\end{displaymath}
the two degenerate VBSs are illustrated by
\begin{equation}
\setlength{\unitlength}{1pt}
\begin{picture}(162,48)(4,-22)
\linethickness{0.8pt}
\multiput(5,10)(60,0){3}{\circle{4}}
\multiput(50,10)(60,0){3}{\circle{4}}
\multiput(20,0)(60,0){3}{\circle{4}}
\multiput(35,0)(60,0){3}{\circle{4}}
\multiput(20,10)(60,0){3}{\circle{6}}
\multiput(35,10)(60,0){3}{\circle{6}}
\multiput(5,0)(60,0){3}{\circle{6}}
\multiput(50,0)(60,0){3}{\circle{6}}
\thicklines
\multiput(7,10)(60,0){3}{\line(1,0){10}}
\multiput(38,10)(60,0){3}{\line(1,0){10}}
\multiput(23,10)(60,0){3}{\line(1,0){9}}
\multiput(37,0)(60,0){3}{\line(1,0){10}}
\multiput(8,0)(60,0){3}{\line(1,0){10}}
\multiput(53,0)(60,0){2}{\line(1,0){9}}
\put(2,0){\line(-1,0){4}}
\put(173,0){\line(1,0){4}}
\thinlines
\put(44,-6){\framebox(12,22)}
\put(50,-6){\line(0,-1){9}}
\put(50,-22){\makebox(1,1){\small projection onto rep.\ $(1,1,0,0,0)$}}
\put(50,25){\makebox(1,1){\small one site}}
\put(104.5,-6){\dashbox{2}(41,22)}
\end{picture}
\label{eq:11000state}
\end{equation}
are exact ground states of a parent Hamiltonian involving the
quadratic Casimir of the total spin of three neighboring sites,
\begin{equation}
  \label{eq:ham11000}
  H_{(1,1,0,0,0)\,\text{VBS}}=\sum_{i=1}^N H_i
\end{equation}
with 
\begin{equation}
  \label{eq:hi11000}
  H_i=\left(\!\left(\!\bs{J}^{(3)}_i\!\right)^2 \! - 20\right)
  \!\left(\!\left(\!\bs{J}^{(3)}_i\!\right)^2 \! - 70\right)
  \!\left(\!\left(\!\bs{J}^{(3)}_i\!\right)^2 \! - 540\right).
\end{equation}
The states \eqref{eq:11000state} are not the only VBSs one can form.
Other possibilities like
\begin{equation}
\setlength{\unitlength}{1pt}
\begin{picture}(174,12)(4,2)
\linethickness{0.8pt}
\multiput(5,10)(60,0){3}{\circle{6}}
\multiput(50,10)(60,0){3}{\circle{6}}
\multiput(20,0)(60,0){3}{\circle{6}}
\multiput(35,0)(60,0){3}{\circle{6}}
\multiput(20,10)(60,0){3}{\circle{4}}
\multiput(35,10)(60,0){3}{\circle{4}}
\multiput(5,0)(60,0){3}{\circle{4}}
\multiput(50,0)(60,0){3}{\circle{4}}
\thicklines
\multiput(8,10)(60,0){3}{\line(1,0){10}}
\multiput(37,10)(60,0){3}{\line(1,0){10}}
\multiput(22,10)(60,0){3}{\line(1,0){11}}
\multiput(38,0)(60,0){3}{\line(1,0){10}}
\multiput(7,0)(60,0){3}{\line(1,0){10}}
\multiput(52,0)(60,0){2}{\line(1,0){11}}
\put(3,0){\line(-1,0){4}}
\put(172,0){\line(1,0){4}}
\thinlines
\put(74.5,-6){\dashbox{2}(41,22)}
\end{picture}
\nonumber
\end{equation}
or
\begin{equation}
\setlength{\unitlength}{1pt}
\begin{picture}(178,16)(4,-2)
\linethickness{0.8pt}
\multiput(5,10)(30,0){6}{\circle{4}}
\multiput(20,10)(30,0){6}{\circle{6}}
\multiput(5,0)(30,0){6}{\circle{6}}
\multiput(20,0)(30,0){6}{\circle{4}}
\thicklines
\multiput(7,10)(30,0){6}{\line(1,0){10}}
\multiput(23,10)(60,0){3}{\line(1,0){10}}
\multiput(8,0)(30,0){6}{\line(1,0){10}}
\multiput(52,0)(60,0){2}{\line(1,0){10}}
\put(2,0){\line(-1,0){4}}
\put(172,0){\line(1,0){4}}
\thinlines
\put(104.5,-6){\dashbox{2}(41,24)}
\put(89.5,-8){\dashbox{2}(41,24)}
\end{picture},
\nonumber
\end{equation}
however, contain additional representations for the total SU(6) spin
of three neighboring sites, and are hence expected to possess
higher energies.

Spinon excitations transforming under the 6-dimensional rep.\
$(0,0,0,0,1)$ are linearly confined to spinons transforming under the
15-dimensional rep.\ $(0,0,0,1,0)$:
\begin{equation}
\setlength{\unitlength}{1pt}
\begin{picture}(180,60)(4,-25)
\linethickness{0.8pt}
\multiput(5,10)(60,0){3}{\circle{4}}
\multiput(50,10)(60,0){3}{\circle{4}}
\multiput(20,0)(60,0){3}{\circle{4}}
\multiput(35,0)(60,0){3}{\circle{4}}
\multiput(20,10)(60,0){3}{\circle{6}}
\multiput(35,10)(60,0){3}{\circle{6}}
\put(185,10){\circle{6}}
\multiput(5,0)(60,0){3}{\circle{6}}
\multiput(50,0)(60,0){3}{\circle{6}}
\multiput(5,10)(30,0){1}{\circle*{4}}
\multiput(20,10)(15,0){2}{\circle*{6}}
\multiput(170,0)(15,0){1}{\circle*{6}}
\multiput(155,0)(30,0){2}{\circle*{4}}
\thicklines
\multiput(7,10)(60,0){3}{\line(1,0){10}}
\put(172,10){\line(1,0){10}}
\multiput(52,10)(60,0){2}{\line(1,0){11}}
\multiput(23,10)(60,0){3}{\line(1,0){9}}
\multiput(8,0)(60,0){3}{\line(1,0){10}}
\multiput(37,0)(60,0){3}{\line(1,0){10}}
\multiput(53,0)(60,0){2}{\line(1,0){9}}
\put(173,0){\line(1,0){10}}
\put(2,0){\line(-1,0){4}}
\put(188,10){\line(1,0){4}}
\thinlines
\put(20,25){\makebox(0,0){\small $(0,0,0,0,1)$}}
\put(20,36){\makebox(0,0){\small spinon }}
\put(170,25){\makebox(0,0){\small $(0,0,0,1,0)$}}
\put(170,36){\makebox(0,0){\small spinon}}
\put(74.5,-6){\dashbox{2}(41,22)}
\put(95,-14){\vector(-1,0){75}}
\put(95,-14){\vector(1,0){75}}
\put(95,-22){\makebox(0,0){\small energy cost $\propto$ distance}}
\end{picture}
\label{eq:11000spinons}
\end{equation}
The VBS configuration we have drawn between the two spinons in
\eqref{eq:11000spinons} constitutes just one of several possibilities.
We expect, however, that this possibility corresponds to the lowest
energy among them for the Hamiltonian \eqref{eq:ham11000}.  This
concludes our list of examples.

The models introduced in this subsection are interesting in that they
provide us with examples where spinon confinement, and with the
confinement the existence of a Haldane gap, is caused by interactions
extending beyond nearest neighbors.  The conclusion drawn from the
Affleck--Lieb theorem for SU($n$) models with spins transforming under
representations we have labeled here as the ``third category'' hence
appear to hold for models with nearest-neighbor interactions only, to
which the theorem is applicable.  For these models, the theorem states
that the spectrum is gapless, which according to our understanding
implies that the models support deconfined spinon excitations.  The
examples we have studied, by contrast, suggest that models with
longer-ranged interactions belonging to this category exhibit
confinement forces between the spinon excitations and hence possess a
Haldane gap.

It is worth noting that even though in the examples we elaborated here
$\lambda <n$, we expect our conclusions to hold for models with
$\lambda >n$ as well.  To see this, let $m>0$ be an integer such that
$n m<\lambda <n(m+1)$.  We can now construct a first VBS with spinon
confinement using $n m$ boxes of the Young tableau and combine it with
a second by projection on each side with a second VBS constructed from
the remaining $\lambda'= \lambda-n m$ boxes.  Since the spinons
of the first VBS are always confined and hence gapped, the final VBS
will support deconfined spinons if and only if the second VBS will
support them, which in turn will depend on the largest common divisor
$q'$ of $\lambda'$ and $n$ as well as the range of the interaction.
Since the largest common divisor $q$ of $n$ and $\lambda$ is
equal to $q'$, there is no need to think in terms of $\lambda'$ and $q'$.
The conclusions regarding confinement and the Haldane gap will not
depend on the distinction between $\lambda$ and $\lambda'$.

\subsection{The different categories of models}
\label{sec:sum}

In this section, we used the rules
emerging from the numerous examples we studied to argue that models of
SU($n$) spin chains in general fall into three categories.  The
classification depends on the number of boxes $\lambda$ the Young
tableau corresponding to the representation of the individual spins
consists off, as follows:
\begin{enumerate}
\item If $\lambda$ and $n$ have no common divisor, the models will
  support free spin excitations and hence not exhibit a Haldane gap.

  The general reasoning here is simply that the VBS states in this
  category break translational invariance up to translations by $n$
  lattice spacings, and that there are (at least) $n$ degenerate
  VBS ground states to each model.  Spinons transforming under
  representations of Young tableaux with an arbitrary number of boxes
  can be accommodated in domain walls between these different ground
  states.  Consequently, the spinons are deconfined.  

\item If $\lambda$ is divisible by $n$, the general argument we have
  constructed in Sec.\ \ref{sec:generalcriterion} indicates that the
  model will exhibit spinon confinement and hence a Haldane gap.

\item \label{three}
  If $\lambda$ and $n$ have a common divisor different from $n$,
  the examples studied in Sec.\ \ref{sec:examples} suggest that the
  question of whether the spinons are confined or not depends on the
  details of the interactions.  If $q$ is the largest common divisor
  of $\lambda$ and $n$, interactions ranging to the $\frac{n}{q}$-th
  neighbor were required for spinon confinement in the models we
  studied.  The Affleck-Lieb theorem~\cite{affleck-86lmp57}, on the
  other hand, tells us that SU($n$) chains with nearest-neighbor
  Heisenberg interactions belonging to this category are gapless if
  the ground states are non-degenerate.
\end{enumerate}

Note that the second category is really just the special case $q=n$
of the third:  with $\frac{n}{q}=1$, nearest-neighbor interactions
are already sufficient for spinon confinement and a Haldane gap.

The conclusion that interactions ranging to the $\frac{n}{q}$-th
neighbor are required for spinon confinement and a Haldane gap,
however, is not universally valid.  A counter-example is provided by
the extended VBSs (XVBSs) introduced by Affleck, Arovas, Marston, and
Rabson~\cite{affleck-91npb467}.  In this model, each site effectively
takes the role of two neighboring sites when a VBS is constructed, and
nearest-neighbor interactions are already sufficient to cause spinon
confinement.  We briefly review this model in the following
subsection.

\subsection{A counter-example: the SU(4) representation $\bs{6}$
extended VBS}
\label{sec:counter}

In an article devoted to quantum antiferromagnets with spins transforming under
 the self-conjugate representations of SU($2n$),
Affleck \ea~\cite{affleck-91npb467} introduced an extended VBS for the
six-dimensional SU(4) representation $(0,1,0)$.  This 
representation is, with regard to the number of boxes the corresponding
Young tableau consists of, not distinguishable from the symmetric 
representation $\bs{10}$ considered in Sec.~\ref{sec:examples}, as both
tableaux consist of two boxes:
\begin{equation}
\setlength{\unitlength}{0.8pt}  
\begin{picture}(116,36)(-6,-6)
\put(-6,10){\line(1,0){10}} 
\put(-6,20){\line(1,0){10}} 
\put(-6,10){\line(0,1){10}} 
\put(4,10){\line(0,1){10}} 
\put(14,15){\makebox(1,1){$\otimes$}}
\put(24,10){\line(1,0){10}} 
\put(24,20){\line(1,0){10}} 
\put(24,10){\line(0,1){10}} 
\put(34,10){\line(0,1){10}} 
\put(47,15){\makebox(1,1){$=$}}
\put(60,10){\line(1,0){20}} 
\put(60,20){\line(1,0){20}} 
\put(60,10){\line(0,1){10}} 
\put(70,10){\line(0,1){10}} 
\put(80,10){\line(0,1){10}} 
\put(90,15){\makebox(1,1){$\oplus$}}
\put(100,5){\line(1,0){10}} 
\put(100,15){\line(1,0){10}} 
\put(100,25){\line(1,0){10}} 
\put(100,5){\line(0,1){20}} 
\put(110,5){\line(0,1){20}} 
\put(-1,-2){\makebox(1,1){$\bs{4}$}}
\put(29,-2){\makebox(1,1){$\bs{4}$}}
\put(70,-2){\makebox(1,1){$\bs{10}$}}
\put(105,-7){\makebox(1,1){$\bs{6}$}}
\end{picture}
\end{equation}
Since the two boxes are combined antisymmetrically for rep.~$\bs{6}$,
a VBS constructed along the lines of Sec.~\ref{sec:examples} would no
longer provide a paradigm for antiferromagnetic spin chains of the
corresponding representation in general.  Affleck
\ea~\cite{affleck-91npb467} have constructed an extended VBS, 
which is illustrated in the following cartoon:
\begin{equation}
\label{gs.su4.xvbs}
\setlength{\unitlength}{1pt}
\begin{picture}(170,20)(0,0)
 \multiput(0,10)(30,0){6}{\circle{4}}
 \multiput(8,10)(30,0){6}{\circle{4}}
 \multiput(4,10)(30,0){6}{\circle{16}}
\thicklines
 \multiput(2,10)(60,0){3}{\line(1,0){4}}
 \multiput(10,10)(30,0){5}{\line(1,0){18}}
 \put(-8,10){\line(1,0){6}}
 \put(160,10){\line(1,0){7}}
\end{picture}
\end{equation}
Here each small circle represents a fundamental representation
$\bs{4}$ of SU(4) (a box in the Young tableau), and each large circle a
lattice site.  The lines connecting four dots indicate that these four
fundamental representations are combined into an SU(4) singlet.  The
total spin of two neighboring sites in this state may assume the
representations
\begin{equation}
  \bs{4}\otimes\bs{\bar 4}=\bs{1}\oplus\bs{15},
\end{equation}
while combining two reps.~$\bs{6}$ on neighboring sites in general
yields
\begin{equation}
  \bs{6}\otimes\bs{6}=\bs{1}\oplus\bs{15}\oplus\bs{20}.
\end{equation}
To construct a parent Hamiltonian for the XVBS \eqref{gs.su4.xvbs}, it
is hence sufficient to sum over projectors onto rep.~$\bs{20}$ on all
pairs of neighboring sites.  Using our conventions (Affleck \ea
~\cite{affleck-91npb467} have normalized the eigenvalues of the
quadratic Casimir operator to $\casinn{4}{\textrm{adjoint rep.}}=8$),
the parent Hamiltonian takes the form
\begin{equation}
  \label{ham.su4.xvbs}
  H_{\textrm{XVBS}}=\sum_{i=1}^N \left( \bs{J}_i\bs{J}_{i+1} +
     \frac{1}{3}\left(\bs{J}_i\bs{J}_{i+1}\right)^2+\frac{5}{12} \right),
\end{equation}
where the operators $J_i^a$, $a=1,\ldots,15$, are given by $6\times 6$
matrices.  Note that the ground state is two fold degenerate, as it 
breaks translational symmetry modulo translations by two lattice spacings.

We conjecture that the lowest lying excitation is a
bound state consisting of two fundamental reps $\bs{4}$, which most
likely are combined antisymmetrically into a rep.~$\bs{6}$:
\begin{equation*}
  \setlength{\unitlength}{1pt}
\begin{picture}(170,20)(0,0)
 \multiput(0,10)(30,0){6}{\circle{4}}
 \multiput(8,10)(30,0){6}{\circle{4}}
 \multiput(68,10)(22,0){2}{\circle*{4}}
 \multiput(4,10)(30,0){6}{\circle{16}}
\thicklines
 \multiput(32,10)(90,0){2}{\line(1,0){4}}
 \multiput(10,10)(30,0){5}{\line(1,0){18}}
 \put(-8,10){\line(1,0){6}}
 \put(160,10){\line(1,0){7}}
\end{picture}
\end{equation*}
The SU(4) rep.~$\bs{6}$ XVBS provides us with an example where a
nearest-neighbor Hamiltonian is sufficient to induce spinon
confinement and a Haldane gap, even though the largest common divisor
of $n=4$ and $\lambda=2$ is $q=2$, \ie interactions including
$\frac{n}{q}=2$ neighbors would be required following the rules
derived from the examples in Sec.~\ref{sec:examples}.  The reason for
this discrepancy is that in the XVBS model considered here, each site
effectively takes the role of two neighboring sites.  
Affleck \ea~\cite{affleck-91npb467,marston-89prb11538} conjecture
that the ground state of the SU(4) rep.~$\bs{6}$ nearest-neighbor 
Heisenberg model is, like the XVBS reviewed here, two-fold degenerate,
which implies that the Affleck-Lieb theorem is not applicable.

This example is valuable in showing that it is advisable to explicitly
construct the VBS for a given representation of SU($n$) in order to
verify the applicability of the general rules motivated above.

\section{Conclusion}
\label{sec:conclusion}

In the first part of this article, we have formulated several exact
models of SU(3) spin chains.  We introduced a trimer model and
presented evidence that the elementary excitations of the model
transform under the SU(3) representations conjugate to the
representation of the original spin on the chain.  We then introduced
three SU(3) valence bond solid chains with spins transforming under
representations $\bs{6}$, $\bs{10}$, and $\bs{8}$, respectively.  We
argued that of these four models, the coloron excitations are confined
only in the $\bs{10}$ and the $\bs{8}$ VBS models, and that only those
models with confined spinons exhibit a Haldane gap.  We subsequently
generalized three of our models to SU($n$), and investigated again
which models exhibit spinon confinement.  

Finally, we used the rules emerging from the numerous examples we
studied to argue that models of SU($n$) spin chains in general fall
into three categories with regard to spinon confinement and the
Haldane gap.  These are summarized in the previous subsection.
The results rely crucially on the assumption that the conclusions we
obtained for the VBS models we studied are of general validity.
This assumption certainly holds for the corresponding SU(2) models,
and appears reasonable on physical grounds.  Ultimately, however,
it is only an assumption, or at best a hypotheses.

On a broader perspective, we believe that the models we have studied
provide further indication that SU($n$) spin chains are an
equally rich and rewarding subject of study as SU(2) spin chains have 
been since Bethe.

\section*{ACKNOWLEDGMENTS}

We would like to thank Peter Zoller and Peter W\"olfle, but 
especially Dirk Schuricht and Ronny Thomale for many highly inspiring 
discussions of various aspects of this work.  We further wish to
thank Dirk Schuricht for valuable suggestions on the manuscript.  
One of us (SR) was supported by a Ph.D.\ scholarship of the Cusanuswerk.

\appendix
\section{A proposal for an experimental realization of SU(3) spin 
chains in an optical lattice}
\label{sec:exp}

In this appendix, we wish to describe a proposal for an experimental
realization of SU(3) spin chains. The most eligible candidate for such
experiments are ultra-cold gases in optical lattices. Recently, these
systems have become an interesting playground for the realization of
various problems of condensed matter physics, such as the phase
transition from a superfluid to a Mott
insulator~\cite{Jaksch-98prl3108,Greiner-02n39}, the fermionic Hubbard
model~\cite{Honerkamp-04prl170403,Honerkamp-04prb094521,rapp-06cm0607138},
and SU(2) spin chains~\cite{Duan-03prl090402,Garcia-04prl250405}.  In
particular, the Hamiltonians for spin lattice models may be engineered
with polar molecules stored in optical lattices, where the spin is
represented by a single-valence electron of a heteronuclear
molecule~\cite{micheli-06np341,brennen-06qp0612180}.

In a most naive approach, one might expect to realize an SU(3) spin
(at a site in an optical lattice) by using atoms with three internal
states, like an atom with spin $S=1$.  If we now were to interpret the
$S^z=+1$ state as SU(3) spin ``blue'', the $S^z=0$ state as ''red'',
and the $S^z=-1$ state as ``green'', however, the SU(3) spin would not
be conserved.  The SU(2) algebra would allow for the process
$\ket{+1,-1}\rightarrow\ket{0,0}$, which in SU(3) language corresponds
to the forbidden process $\ket{\b,\g}\rightarrow\ket{\r,\r}$.

\begin{figure}
\begin{center}
\includegraphics[width=0.38\textwidth]{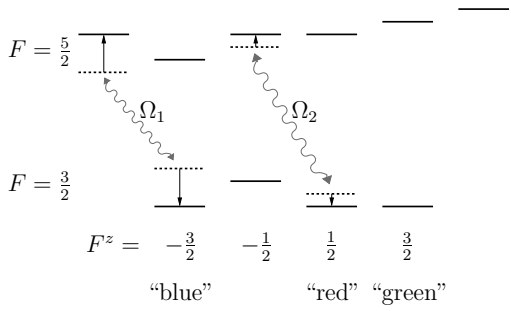}
\caption{Effective lifting of the $F^z=-\frac{1}{2}$ state.}
\label{fig:exp}
\end{center}
\end{figure}

A more sophisticated approach is hence required.  One way to obtain a
system with three internal states in which the number of particles in
each state (\ie of each color) is conserved 
is to manipulate an atomic system with total
angular momentum $F=3/2$ (where
$\bs{F}=\bs{S}_{\text{el}}+\bs{L}_{\text{orb}}+\bs{S}_{\text{nuc}}$
includes the internal spin of the electrons, the orbital angular
momentum, and the spin of the nucleus) to simulate an SU(3) spin.  The
important feature here is that the atoms have four internal states,
corresponding to
$F^z=-\frac{3}{2},-\frac{1}{2},+\frac{1}{2},+\frac{3}{2}$.  
For such atoms, one has to suppress the
occupation of one of the ``middle'' states, say the $F^z=-\frac{1}{2}$
state, by effectively lifting it to a higher energy while keeping the
other states approximately degenerate.  This can be accomplished
through a combination of an external magnetic field and two carefully
tuned lasers, which effectively push down the energies of the
$F^z=-\frac{3}{2}$ and the $F^z=+\frac{1}{2}$ states by coupling these
states to states of (say) the energetically higher $F=5/2$ multiplet
(see Fig.~\ref{fig:exp}).
At sufficiently low temperatures, we are hence left with a system with
three internal states $F^z=-\frac{3}{2},+\frac{1}{2},+\frac{3}{2}$,
which we may identify with the colors ``blue'', ``red'', and ``green''
of an SU(3) spin.  In leading order, the number of particles of each
color is now conserved, as required by SU(3) symmetry.
For example, conservation of $F^z$ forbids processes in which a
``blue'' and a ``green'' particle turn into two ``red'' ones,
$\ket{\b,\g}\rightarrow\ket{\r,\r}$.  Higher order processes of the
kind $\ket{\b,\g,\g}\rightarrow\ket{\r,\r,\r}$ are still possible, but
negligible if the experiment is conducted at sufficiently short time
scales.

If one places fermionic atoms with an artificial SU(3) spin engineered
along the lines of this or a related proposal in an optical lattice
and allows for a weak hopping of the atoms on the lattice, one has
developed an experimental realization of an SU(3) Hubbard model.
If the energy cost $U$ of having two atoms on the same lattice site is
significantly larger than the hopping $t$, and the density is one atom
per site, the system will effectively constitute an SU(3)
antiferromagnet.  The dimension of this antiferromagnet will depend on
the optical lattice, which can be one-, two-, or three-dimensional.

In principle, the above proposal can be generalized to SU($n$), even
though the experimental obstacles are likely to grow ``exponentially''
with $n$.  Besides, it is far from clear that such an endeavor is
worthwhile, as all the non-trivial properties of SU($n$) are already
present in SU(3) chains (while SU(2) constitutes a special case).


\section{Gell-Mann matrices}
\label{app:conventions}

The Gell-Mann matrices are given by~\cite{Cornwell84vol2,Georgi82}
\begin{displaymath}
\begin{array}{c@{}c@{}c@{\hspace{6pt}}c@{}c@{}c@{\hspace{6pt}}c@{}c@{}c}
\lambda^1&=&\!\left(\begin{array}{ccc}0&1&0\\1&0&0\\0&0&0\end{array}
\right)\!\!,&
\lambda^2&=&\!\left(\begin{array}{ccc}0&-i&0\\i&0&0\\0&0&0\end{array}
\right)\!\!,&
\lambda^3&=&\!\left(\begin{array}{ccc}1&0&0\\0&-1&0\\0&0&0\end{array}
\right)\!\!,\\[27pt]
\lambda^4&=&\!\left(\begin{array}{ccc}0&0&1\\0&0&0\\1&0&0\end{array}
\right)\!\!,&
\lambda^5&=&\!\left(\begin{array}{ccc}0&0&-i\\0&0&0\\i&0&0\end{array}
\right)\!\!,&
\lambda^6&=&\!\left(\begin{array}{ccc}0&0&0\\0&0&1\\0&1&0\end{array}
\right)\!\!,\\[27pt]
\lambda^7&=&\!\left(\begin{array}{ccc}0&0&0\\0&0&-i\\0&i&0\end{array}
\right)\!\!,&
\lambda^8&=&
\multicolumn{3}{l}{{\displaystyle\frac{1}{\sqrt{3}}}
\!\left(\begin{array}{ccc}1&0&0\\0&1&0\\0&0&-2\end{array}\right)\!\!.}
\end{array}
\end{displaymath}
They are normalized as
$\mathrm{tr}\left(\lambda^a\lambda^b\right)=2\delta_{ab}$ and satisfy
the commutation relations
$\comm{\lambda^a}{\lambda^b}=2f^{abc}\lambda^c.$ The structure
constants $f^{abc}$ are totally antisymmetric and obey Jacobi's
identity 
\begin{displaymath}
f^{abc}f^{cde}+f^{bdc}f^{cae}+f^{dac}f^{cbe}=0.
\end{displaymath} 
Explicitly, the non-vanishing structure constants are given by $f^{123}=i$,
$f^{147}=f^{246}=f^{257}=f^{345}=-f^{156}=-f^{367}=i/2$,
$f^{458}=f^{678}=i\sqrt{3}/2$, and 45 others obtained by permutations of 
the indices.\\[-0.1pt]

\section{Eigenvalues of the quadratic Casimir operator}
\label{quadraticcasimirs}

The eigenvalues of the quadratic Casimir operator for representations
$\casin{\mu_1,\mu_2,\ldots,\mu_{n-1}}$ of SU($n$) up to $n=6$ are given by:
\begin{itemize}
\item[] $ \casinn{2}{\mu}=\frac{1}{4}\left(\mu^2+2\mu\right)=
       \frac{\mu}{2}\left(\frac{\mu}{2}+1\right)$ 
\item[] $\casinn{3}{\mu_1,\mu_2}=\frac{1}{3} \bigl( \mu_1^2 + \mu_1\mu_2 +
       \mu_2^2 + 3\mu_1 + 3\mu_2 \bigr)$
\item[] $\casinn{4}{\mu_1,\mu_2,\mu_3} =
       \frac{1}{8}\left(3\mu_1^2+4\mu_2^2+
         3\mu_3^2+4\mu_1\mu_2\right.$\\[3pt]
       $ \phantom{\casinn{2}{\mu}} \left.
      + 2\mu_1\mu_3 +4\mu_2\mu_3 + 12\mu_1+ 16\mu_2+12\mu_3 \right)$
\item[] $\casinn{5}{\mu_1,\mu_2,\mu_3,\mu_4} = \frac{1}{5}\left( 2\mu_1^2 
      + 3\mu_2^2 + 3\mu_3^2 + 2\mu_4^2 \right.$\\[3pt] 
$\phantom{\casinn{2}{\mu}} + 3\mu_1\mu_2 + 4\mu_2\mu_3 + 3\mu_3\mu_4 + 
2\mu_1\mu_3 + \mu_1\mu_4$\\[3pt]
$\phantom{\casinn{2}{\mu}} \left. + 2\mu_2\mu_4 + 10\mu_1
+ 15\mu_2 + 15\mu_3 + 10\mu_4 \right)$
\item[] $\casinn{6}{\mu_1,\mu_2,\mu_3,\mu_4,\mu_5}= $\\[3pt]
$\phantom{\casinn{2}{\mu}}\!\!\!\frac{1}{12}\left( 
  5\mu_1^2 + 8\mu_2^2 + 9\mu_3^2 + 8\mu_4^2 + 5\mu_5^2 \right.$\\[3pt]
$\phantom{\casinn{2}{\mu}}+ 8\mu_1\mu_2 + 12\mu_2\mu_3 + 12\mu_3\mu_4 
  + 8\mu_4\mu_5  $\\[3pt]
$\phantom{\casinn{2}{\mu}} + 4\mu_1\mu_4 + 6\mu_1\mu_3  + 8\mu_2\mu_4 + 6\mu_3\mu_5$\\[3pt]
$\phantom{\casinn{2}{\mu}} + 4\mu_2\mu_5 + 2\mu_1\mu_5 + 30\mu_1 + 48\mu_2$\\[3pt]
$\phantom{\casinn{2}{\mu}}\left. + 54\mu_3 + 48\mu_4 + 30\mu_5 \right)$\\
\end{itemize}
The general method to obtain these and further eigenvalues for $n>6$
requires a discussion of representation theory~\cite{Humphreys87} at a level
which is beyond the scope of this article.

The dimensionality of a representation $(\mu_1,\mu_2,\ldots,\mu_{n-1})$
is determined by the so-called Hook formula~\cite{Cornwell84vol2} 
\begin{equation}
  \label{eq:hook-formula}
  \text{dim}=\frac{\prod_{i<j}^n\left( \lambda_i - \lambda_j + j - i  \right)}
             {\prod_{i<j}^n\left( j - i  \right)},
\end{equation}
where $\lambda_i=\sum_{j=i}^{n-1}\mu_j$ for $i=1,\ldots,n$. In particular, 
it yields for $n=2,3,4$:

\begin{itemize}
\item[] $\dimn{2}{\mu}=\mu+1$
\item[] $\dimn{3}{\mu_1,\mu_2}=\frac{1}{2}(\mu_1+1)(\mu_2+1)(\mu_1+\mu_2+2)$
\item[] \parbox{\linewidth}{$\dimn{4}{\mu_1,\mu_2,\mu_3}=\frac{1}{12}(\mu_1+1)(\mu_2+1)
         (\mu_3+1)$\\[3pt]
        $\phantom{\text{dimdim}}(\mu_1+\mu_2+2)(\mu_2+\mu_3+2)
         (\mu_1+\mu_2+\mu_3+3)$}
\end{itemize}

\end{document}